\documentclass[11pt]{article}%
\usepackage{amssymb}
\usepackage{amsfonts}
\usepackage{amsmath}
\usepackage{mathrsfs}
\usepackage{graphicx}%
\usepackage{verbatim}
\usepackage{color}
\usepackage{authblk}

\usepackage{algorithm}
\usepackage{algpseudocode}
\usepackage{subeqnarray}
\usepackage{amsmath,bm}
\usepackage{subfig}
\usepackage{array,calc}
\usepackage{bbm}
\usepackage{hyperref}
\usepackage{dsfont}
\usepackage[title]{appendix}
\newcommand{\RNum}[1]{\uppercase\expandafter{\romannumeral #1\relax}}
\makeatother

\allowdisplaybreaks[4]
\usepackage{algorithm}
\usepackage{algpseudocode}
\usepackage{setspace}
\algnewcommand{\Inputs}[1]{%
	\State \textbf{Inputs:}
	\Statex \hspace*{\algorithmicindent}\parbox[t]{.9\linewidth}{\raggedright #1}
}
\algnewcommand{\Initialize}[1]{%
	\State \textbf{Initialization:}
	\Statex \hspace*{\algorithmicindent}\parbox[t]{.9\linewidth}{\raggedright #1}
}

\providecommand{\U}[1]{\protect\rule{.1in}{.1in}}
\begin{document}

\title{Heterogeneity induced localization of traffic congestion on networks}
\author[1,2]{ Zhidong He\thanks{zhidonghe@outlook.com; Z.He is with DS Information Technology Co., Ltd. He contributed to this work during his PhD study in Delft University of Technology. }}
\affil[1]{\footnotesize Faculty of Electrical Engineering, Mathematics and Computer Science, Delft University of Technology, The Netherlands}
\affil[2]{\footnotesize DS Information Technology Co., Ltd., Shanghai, China}
\date{}
\setcounter{Maxaffil}{0}
\renewcommand\Affilfont{\itshape\small}
\maketitle

\begin{abstract}
The emergence of congestion is a critical phenomenon in transport systems.
Transport is organized along pathways abstracted by links, which connect different nodes as regions to form the network.
The modeling of traffic has so far mainly been based on one-dimensional lattices or networked queuing systems, which restricts mechanistic insights and analytical tractability for traffic dynamics on general networks.
An interpretable kinetics model is necessary for understanding the impact of the network structure on traffic congestion in large-scale networks.
Considering a generalized flow-density relation within each node, we propose and investigate a general traffic model on networks with external injections and exits.
We observe various phases in terms of node density that exhibit distinct behaviors under different injection rates. 
The onset of congestion along with the coexistence of free nodes and congested nodes, manifested by the spatial localization of traffic congestion, is induced by the asymmetry of the flow transitions between neighbors in the network. 
%The heterogeneity of both the network and the external injections highlights the existence of the localization phase.
The proposed congestion centrality allows us to determine the congestion threshold, identify congested nodes and predict congestion propagation. 
Our model and results are shown to cover a broad spectrum of implications for transport dynamics on networks as well as offer a practical prospect for traffic optimizations such as congestion alleviation and targeted diffusion. 
\end{abstract}

\section{Introduction}
Our daily lives heavily depend on the functioning of different forms of transport systems, ranging from subcellular level of molecular motors on cytoskeleton \cite{hirokawa2005molecular}, data packages delivery on Internet \cite{mieghem2006data} to urban traffic \cite{daganzo2011macroscopic}, pedestrian movement \cite{helbing2001self} and international capital flows \cite{anderson2018economy}. 
As a dynamic complex system, traffic on networks frequently incurs collective phenomena, such as transitions between free flow and congestion states, as well as the evolution of congestion propagation and dissipation.
Understanding the impact of microscopic behavior of individuals and of the network structure on the traffic dynamics is of utmost importance both for utilizing the transport facilities more efficient and for designing better traffic control strategies.

Several types of models have been applied to analyze dynamic properties of traffic systems \cite{helbing2001traffic}\cite{chowdhury2000statistical}\cite{mendes2012traffic}. Macroscopic traffic models based on kinetic gas theory and hydromechanics can describe  traffic behaviors exhaustively in one-dimensional lattices with explicit results in a mean-field approach \cite{daganzo2007urban}\cite{lighthill1955kinematic}\cite{newell1993simplified}, while microscopic models can reflect individual behavior more realistically, but involve high computing cost for analyzing large-scale systems \cite{daganzo1994cell}\cite{schreckenberg1995discrete}\cite{bressloff2013stochastic}. 
The underlying network structure for transportation is formed by nodes that represent regions, and the items (e.g., biological cells, particles or vehicles) within the nodes can move along the links that represent pathways.
%where the items can be flow mass in  fluid	mechanics, particles in gas system or vehicle in road transport.
%Transport is usually organized along link-like pathways, which connect different nodes as regions to form a network structure.
%, which introduce the network model as an excellent proxy 
A network science approach \cite{newman2018networks} to transportation incorporates the network structure and the overlying traffic dynamics. Complementary network-based studies, e.g., in terms of queuing system \cite{wu2006congestion}\cite{zhao2005onset}, percolation theory \cite{li2015percolation}, exclusion processes \cite{neri2011totally} and shockwave model \cite{mones2014shock}, have provided practical insights in the formation of congestion \cite{wu2006congestion}\cite{sole2016congestion} and percolation transitions of traffic flow \cite{zeng2019switch}, but were mainly limited to numerical simulations and empirical traffic data.
Notwithstanding the importance of a kinetic model for exploring the traffic dynamics on a general network, there are mainly two obstacles for a traffic model that is analytically tractable and provides mechanistic insights. The first obstacle is that the topological disorder \cite{moretti2013griffiths} induced by heterogeneous networks could dramatically differentiate the dynamics from almost deterministic models in homogeneous networks. The second is the generic non-linearity of regional properties due to the interaction of items among neighboring regions or within each region.

This work aims to provide an analytic approach for exploring the traffic dynamics on networks.
We propose a macroscopic traffic model that incorporates both local nonlinear features and collective interactions of traffic flows on a general heterogeneous network.
The flow network in our model defines the flows between nodes (representing regions) via weighted directed links (representing flow transition probability).
%, which can be determined by the underlying structure and a given routing mechanism. 
The external injection and the exiting flow at each node are introduced as independent processes, i.e., the items are injected at source nodes and exit the system once they arrive at their respective destination nodes.
The imposed external injection can describe an injection rate for electronic migration, injected freight loads for transportation \cite{sole2016congestion}, or Langmuir kinetics for particle dynamics \cite{parmeggiani2004totally}.	
Meanwhile, the interaction behaviour among items within each node is described by a generalized form of flow in terms of the number of items in this region. 

The model addresses the question of how the topology of a flow network and the flow-density relation affect transport characteristics and congestion onsets.
The richness of the phase diagrams with respect to the injection rates reflects the profound diversity of the traffic dynamics. Especially, we observe and explore the phase coexistence state with a bimodal distribution of the node density in a system under specific external injections, manifested by the spatial localization of traffic congestion. 
%The frequency observed hysteresis in real-world of the congestion threshold could also arise from this density localization. 
Moreover, we propose the congestion centrality, that merely depends upon the flow network and the external injections and plays a critical role for the onset of congestion, identifying the congested regions and predicting congestion propagation.
Finally, we discuss how our findings can be applied for traffic optimization among various domains including vehicle transportation, targeted medicine and nervous systems.

%Transport of matter and energy is essential for maintaining complex structure.
%The present papers explores the surprising richness of non-linearity induced spatial localization that can occur in the particular case of driven diffusive systems.

%the collective motion of motor proteins along cytoskeletal filaments

%So   far  most   theoretical   treatments   of   single-file   diffusion   have   focused   onindistinguishable particles.  The reference model for this process is the one-dimensionalsymmetric  exclusion  processes  (SEP),  a  lattice  model  where  hard-core  particles  hoprandomly to nearest neighbour sites, provided the target is empty [14,15]. 

\section{Traffic dynamics on networks}\label{sec:traffic_model}
\subsection{Flow networks}
\begin{figure}[!htp]
	\centering
	\subfloat[Process illustration]
	{\includegraphics[width=0.28\textheight]{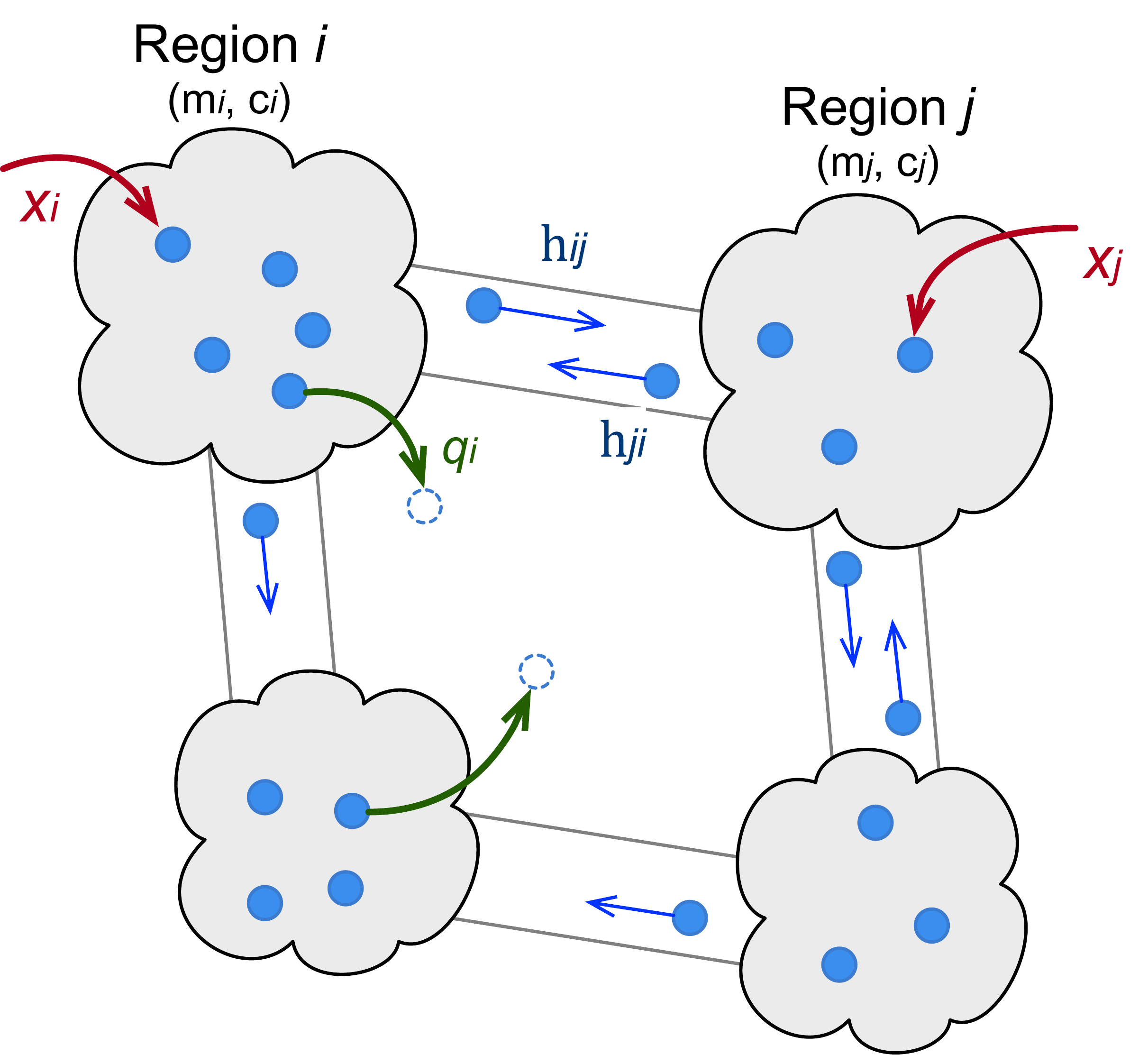}\label{fig:model_illus}} \hfill
	\subfloat[Diagram of the flow--density relation]
	{\includegraphics[width=0.3\textheight]{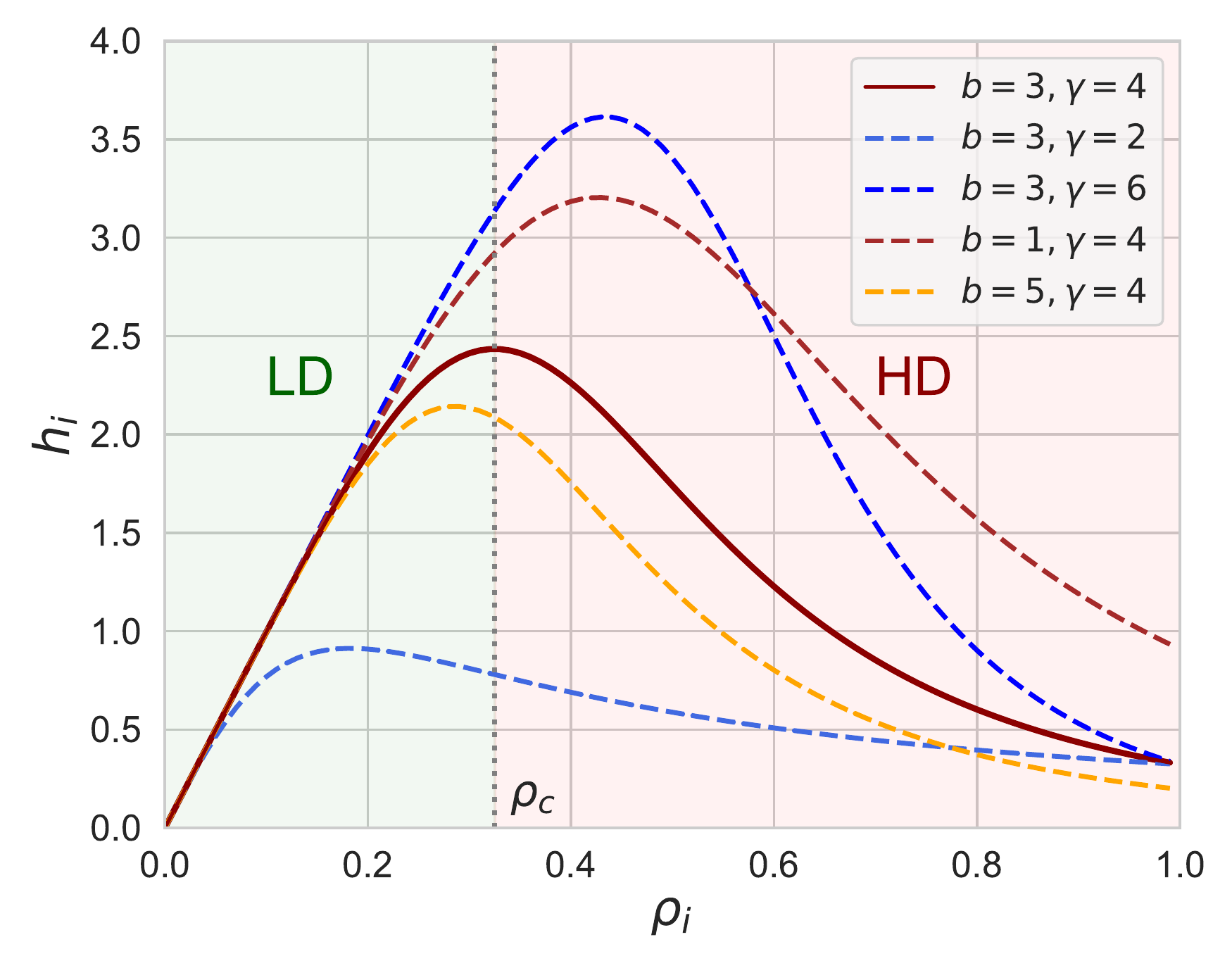}\label{fig:densityflow_fun}}
	\caption{(a) Illustration of the traffic model, where the blue dots represent the items, and the gray regions are the nodes in our model. (b) Diagram of the flow--density relation. The default setting of the following numerical results is: $a=0.1$, $b=3$, $\gamma=4$, if not specified.}
	\label{fig:model_introduction}
\end{figure}

The transport infrastructure formed by regions and paths can be abstracted by a flow network $G(\mathcal{N},\mathcal{L})$ consisting of set $\mathcal{N}$ with $N$ nodes and set $\mathcal{L}$ with $L$ directed links, where the items at each region can move between the nodes along the links. 
Figure \ref{fig:model_illus} illustrates the dynamic process on a network.
%For generalizing our model, we assume that the links in the underlying network are bi-directed, and the single-directed path is regarded as a special case.
The underlying flow network is predefined by a weighted adjacency matrix $F$, where entry $F_{ij}$ represents the demanded mass flow (i.e. number of items per unit time), rather than the actual flow, from node $i$ to node $j$, which is related to a specific route strategy. Since the items in each region either attempt to go downstream towards other regions or arrive at their destination region, instead of staying in the current region, the diagonal entry $F_{ii}$ for $i\in\mathcal{N}$ in the matrix $F$ equals $F_{ii}=0$. 
%represents the mass (i.e., number of items) flow rate attempting\footnote{Note that the flows in $F$ determine the predetermine exchanging flow $F_{ij}$ as demand, instead of the practical flow $h_{ij}$ between nodes. 
%The flow network $G$ is related to the specific route strategy.} to move from node $i$ to node $j$. 
The number of items per unit time injected at origin node $i$ is denoted by $x_i$ (also called the injection rate, which can be assimilated to an external reservoir). 
The item exits the system once its destination node is reached, and the mass flow departing from destination node $i$ is denoted by $E_i\geq0$.
We then obtain the flow transition matrix $P$, with entry $p_{ij} = \frac{F_{ij}}{\sum_{j=1}^N F_{ij}+E_i}$ expressing the mass fraction (interpreted as the transition probability\footnote{The matrix $P$ plays a similar role as the transition probability matrix in discrete-time Markov processes \cite{van2014performance}, except that the matrix $P$ has zero diagonal entries and obeys $\sum_{j=1}^N p_{ij} \leq 1$ due to the exiting probability $q_i$.} of items) at node $i$ moving downstream towards node $j$.
The mass fraction at node $i$ that exits the system is represented by $q_i = \frac{E_i}{\sum_{j=1}^N F_{ij}+E_i}$, so that $\sum_{j=1}^N p_{ij}+q_i=1$ for $i\in\mathcal{N}$.
%and $q_i = \frac{F_{ij}}{\sum_{j=1}^N F_{ij}+E_i}$ represents the mass fraction at node $i$ exits the system.
%and $q_i = \frac{d_i}{\sum_{j=1}^N F_{ij}+E_i}$, so that $\sum_{j=1}^N p_{ij}+q_i=1$. Thus, the entry $p_{ij}$ represents the mass fraction (interpreted as the transition probability of items) at node $i$ downstream to node $j$, and $q_i$ represents the mass fraction at node $i$ exits the system.
We assume the stationarity of the flow transition matrix $P$, i.e., the state of the transport system does not alter the matrix $P$. 
%We assume the dissimilarity of the items from different origins or towards different destinations, which implies the time-independent matrix $P$ in our model, i.e., the state of the transport system does not alter the constant flow transition matrix $P$. 

\subsection{Traffic dynamics processes:} 

We denote by $\rho_i(t)$ the time-dependent density of node $i$, which equals $\rho_i(t) = \frac{m_{i}}{{c_i}} \in[0,1]$, where $m_i$ is the number of items or the mass at node $i$ and $c_i$ is the total capacity (i.e. maximum volume to storage $c_i$ items) of node $i$. Thus, $\rho_i(t) = 0$ implies an empty node $i$ and $\rho_i(t) = 1$ implies that node $i$ is fully occupied. For simplicity, the capacity $c_i$ is the same for all nodes.
Further denoting by $v_i(t)$ the maximum volume velocity (corresponding to the movement of the items) outgoing from node $i$, the maximum allowed outflow $h_i(t)$ from node $i$ to other nodes is $h_i(t) = \rho_i(t)v_i(t)$, and then the maximum allowed flow from node $i$ to node $j$ is $p_{ij}h_i(t)$. 
Since a higher-density node practically degrades the mass inflow, we can assume that the actual flow $h_{ij}(t)$ from node $i$ downstream towards node $j$ follows 
\begin{align}\label{equ:flow_density}
	h_{ij}(t) = (1-\rho_j^\kappa(t))p_{ij}h_i(t) 
\end{align}
where the constant $\kappa\geq0$ tunes the items' penetrability that characterizes the decay effect of the flow $h_{ij}(t)$ due to the downstream density $\rho_j$ (as a dimensionless factor).
%todo harcore assumption
We mainly focus on the system with the constant $\kappa=1$, but we will exemplify that different $\kappa\geq1$ cannot substantially alter the behaviors of our proposed model \eqref{equ:model_vector_form} in Appendix \ref{sec:imapact_kappa}. %These hypotheses simplify the analysis but are not crucial to develop it.

%We impose an injection rate $ x_i$ at each node $i$, which can describe an injection rate for electronic migration, injected freight loads for transportation \cite{sole2016congestion}, or Langmuir kinetics for particle dynamics \cite{parmeggiani2004totally}.	
According to the flow conversation law, the density of node $i$	follows
\begin{align}\label{equ:flow_conversation}
	\rho_i(t) = \int_0^t \left( \sum_{j=1}^N h_{ji}(\tau) - \sum_{j=1}^N h_{ij}(\tau) +J_i^{in}(\tau)- J_i^{out}(\tau) \right) d\tau
\end{align}
where the term $J_i^{in}(t) = x_i(1-\rho_i^\kappa(t))$ is the external inflow at node $i$ by the constant injection rate $x_i$, and the exiting outflow is $J_i^{out}(t)=q_ih_i(t)$.
\begin{comment}
Reasonably assuming that the arrive processes and the driving processes are stochastic, the mean density, denoted by $x_i(t) = E[\rho_i(t)]$ translates to the probability pf node $i$ is full, i.e.,\\
 $x_i(t) = \Pr[\rho_j(t) = 1] $.

By the total probability law, the mean flow $E[h_{ij}(t)]$ from node $i$ to node $j$ is 
\begin{align}
	E[h_{ij}(t)]= 0\cdot\Pr[\rho_j(t) = 1] + \rho_i(t)v_i(t)\Pr[\rho_j(t) \neq 1] = (1-x_j(t))x_i(t)v_i(t)
\end{align}
By derivative on both side of \eqref{equ:flow_conversation}, we obtain the governing equation for the traffic dynamics on a network,
\begin{align}\label{equ:main_mf_model}
	\frac{d \rho_i(t)}{d t}  = \sum_{j=1}^N  p_{ji}(1-\rho_i)\rho_jv_j - \sum_{j=1}^N  p_{ij}(1-\rho_j)\rho_iv_i + f_i^{in} -f_i^{out}
\end{align}
where the term $f_i^{in} =  x_i(1-\rho_i)$ is the external injections as originals.
If node $i$ is a destination, the external outputs as destination $f_i^{out}$ is $f_i^{out}=q_i\rho_iv_i$.
%If node $i$ is a destination, i.e., $q_i = 1$, (otherwise $q_i = 0$ for non-destination), the maximum outflow of node $i$ is $q_ix_iv_i$, while the fraction of outflow to other nodes is $q_ix_iv_i\sum_{j=1}^N p_{ij}(1-x_j)$. 
%Thus, the external outputs as destination $f_i^{out}$ is the external outputs as destination can be computed by $f_i^{out}=q_ix_iv_i(1-\sum_{j=1}^N p_{ij}(1-x_j))$.
\end{comment}
Differentiating both side of \eqref{equ:flow_conversation} with respect to time $t$ yields
\begin{align}\label{equ:main_mf_model}
\frac{d \rho_i(t)}{d t}  = \sum_{j=1}^N p_{ji}(1-\rho_i)h_j - \sum_{j=1}^N p_{ij}(1-\rho_j)h_i + x_i(1-\rho_i) -q_ih_i
\end{align}
The governing equation in vector form reads 
\begin{align}\label{equ:model_vector_form}
	\frac{d \bm{\rho}(t)}{d t} = (I-diag(\bm{\rho}))P^T\bm{h}-diag(\bm{h})P(\bm{u}-\bm{\rho})+diag(\bm{ x})(\bm{u}-\bm{\rho})-diag(\bm{q})\bm{h}
\end{align}
for the density vector $\bm{\rho}:=(\rho_1,\rho_2,\dots,\rho_N)^T$, the flow vector $\bm{h}:=(h_1,h_2,\dots,h_N)^T$, the exiting fraction $\bm{q}:=(q_1,q_2,\dots,q_N)^T$ and the unit vector $\bm{u}$. 

\subsection{Nonlinear resistance:}

In this article, we investigate a generalized resistance $r_i(t)$ for outflows, acting as fluid resistance~\cite{white2006viscous} or impedance, of the form 
\begin{align}\label{equ:resistance_fun}
	r_i(t) = a+b\rho_i^\gamma(t)
\end{align}
which presents a nonlinear regression relationship in the density $\rho_i(t)$ and has three tunable parameters $a>0, b>0, \gamma\geq 0$. The parameters $a$ and $b$ tune the trade-off between the constant term and the nonlinear term, while the parameter $\gamma$ reflects the non-linearity of the resistance. The velocity $v_i(t)$ follows as $v_i(t) = r_i(t)^{-1}=\frac{1}{a+b\rho_i^\gamma(t)}$.
The resistance \eqref{equ:resistance_fun} determines the maximum outflow
\begin{align}\label{equ:flow_density_relation}
h_i(t) = \rho_i(t)v_i(t) = \frac{\rho_i(t)}{a+b\rho_i^\gamma(t)}
\end{align}
as a function of node density $\rho_i(t)$ and shown in Figure \ref{fig:densityflow_fun}. A larger $b$ implies a smaller flow $h_i|_{\rho_i=1} = \frac{1}{a+b}$ of a fully-congested node, and a large $\gamma$ increases the non-linearity of the resistance.
The flow $h_i(\rho_i)$ achieves a maximum at the critical density $\rho_c$ and increases almost linearly with the density $\rho<\rho_c$. This regime is defined as the low density (LD) or free regime.
The flow $h_i(\rho_i)$ decays with the increasing density in high density (HD) regime for the density above the critical density $\rho>\rho_c$ due to the nodal congestion.
The analogy between the proposed traffic model \eqref{equ:model_vector_form} and some related models are discussed in Appendix \ref{sec:analogy}.

\section{Phase coexistence of density}
\subsection{Analysis in the Bi-tank Model:}

\begin{figure}[htp]
	\captionsetup[subfloat]{farskip=2pt,captionskip=1pt}
	\begin{tabular}{*{2}{b{0.53\textwidth-2\tabcolsep}}}
	\subfloat[Phase diagram in the bi-tank model]{\includegraphics[width=0.98\hsize]{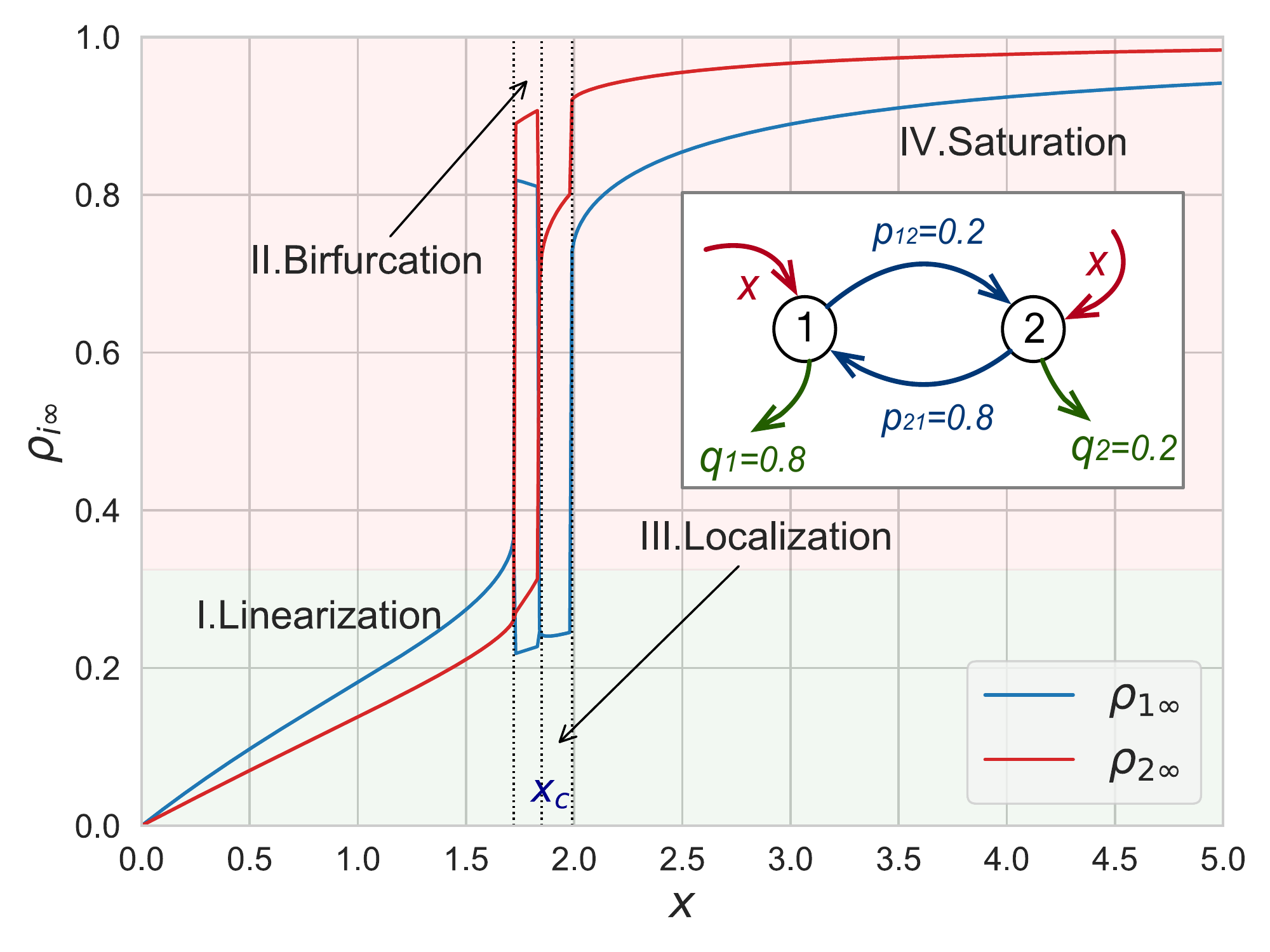}}&	
	\subfloat[$x=1.2$ (Linear)]{\includegraphics[width=0.17\textheight]{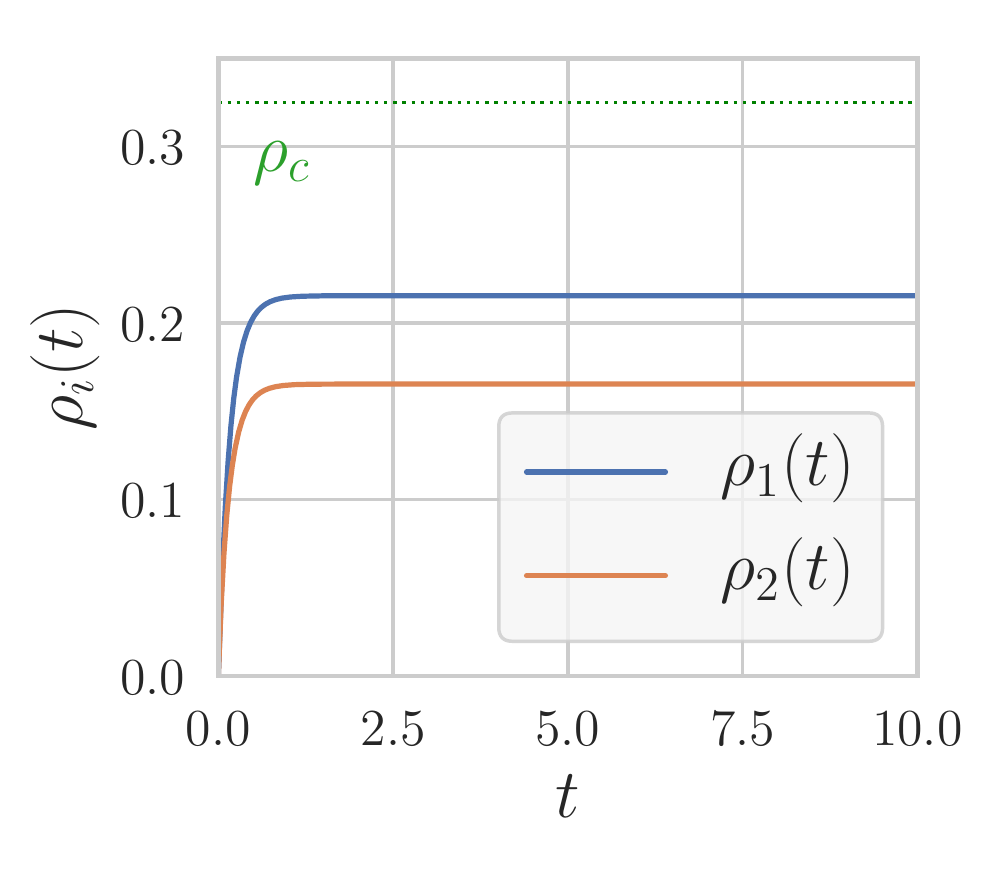}}
	\subfloat[$x=1.8$ (Birfurcation)]{\includegraphics[width=0.17\textheight]{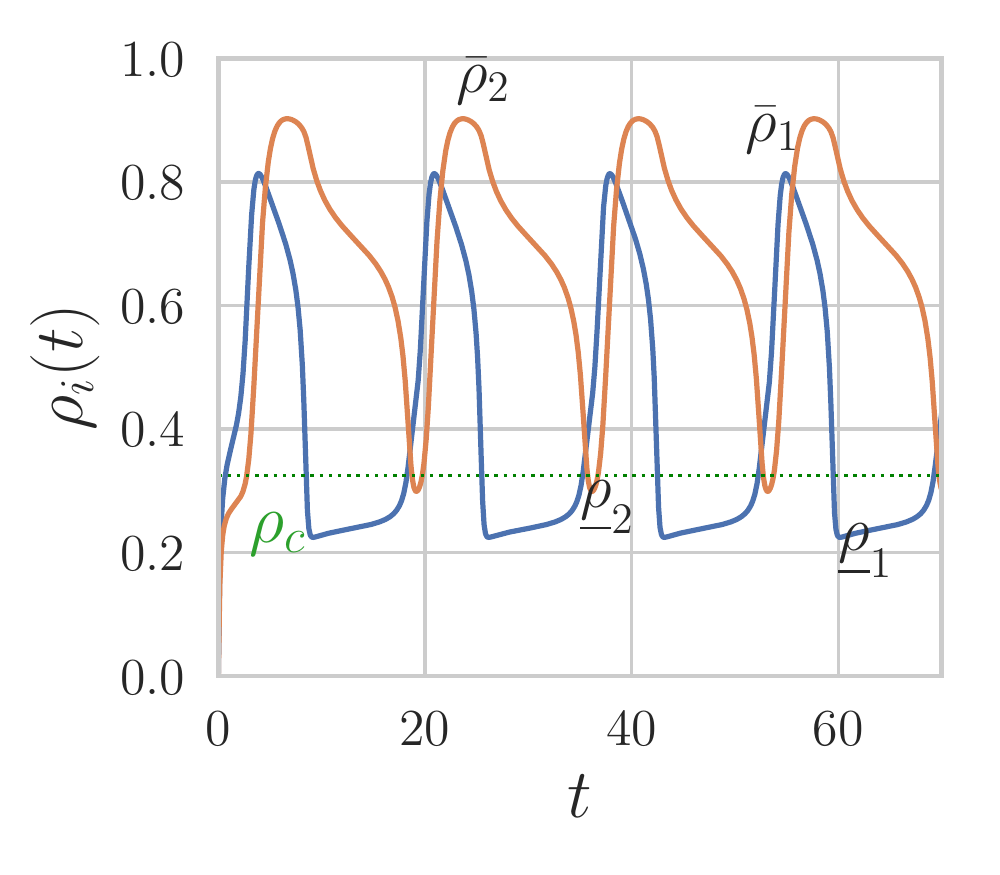}} 
	
	\subfloat[$x=1.9$ (Localization)]{\includegraphics[width=0.17\textheight]{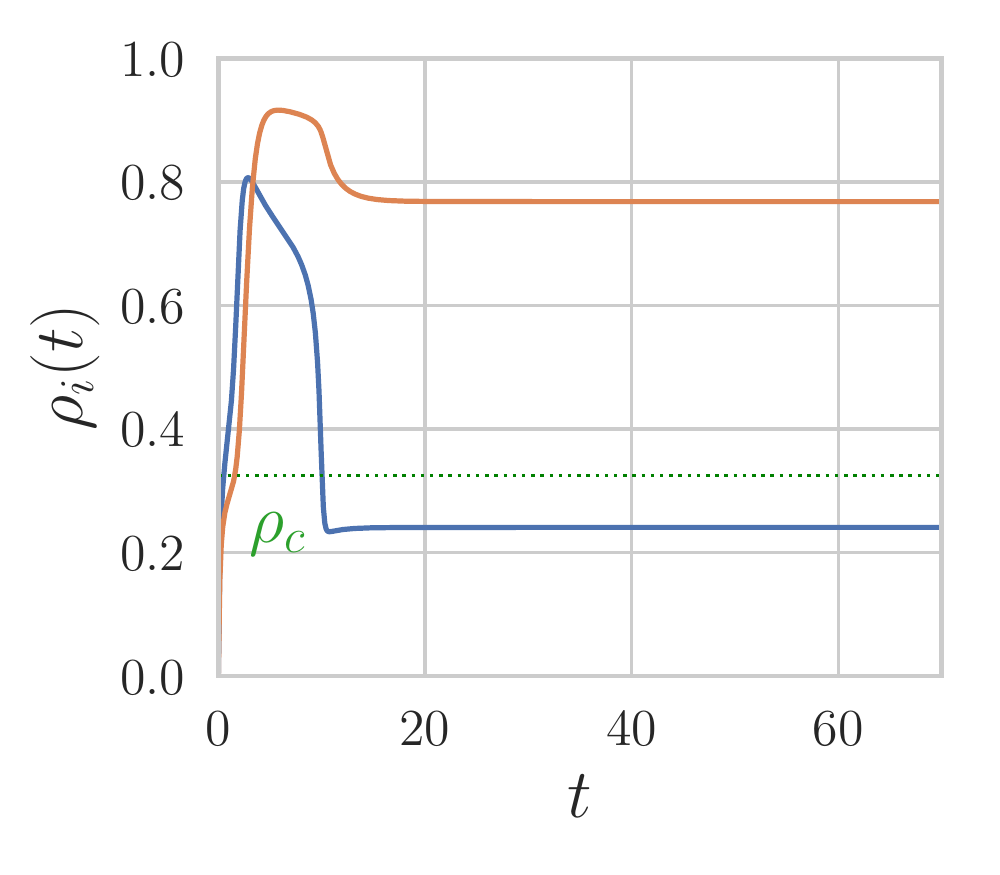}} 
	\subfloat[$x=2.4$ (Saturation)]{\includegraphics[width=0.17\textheight]{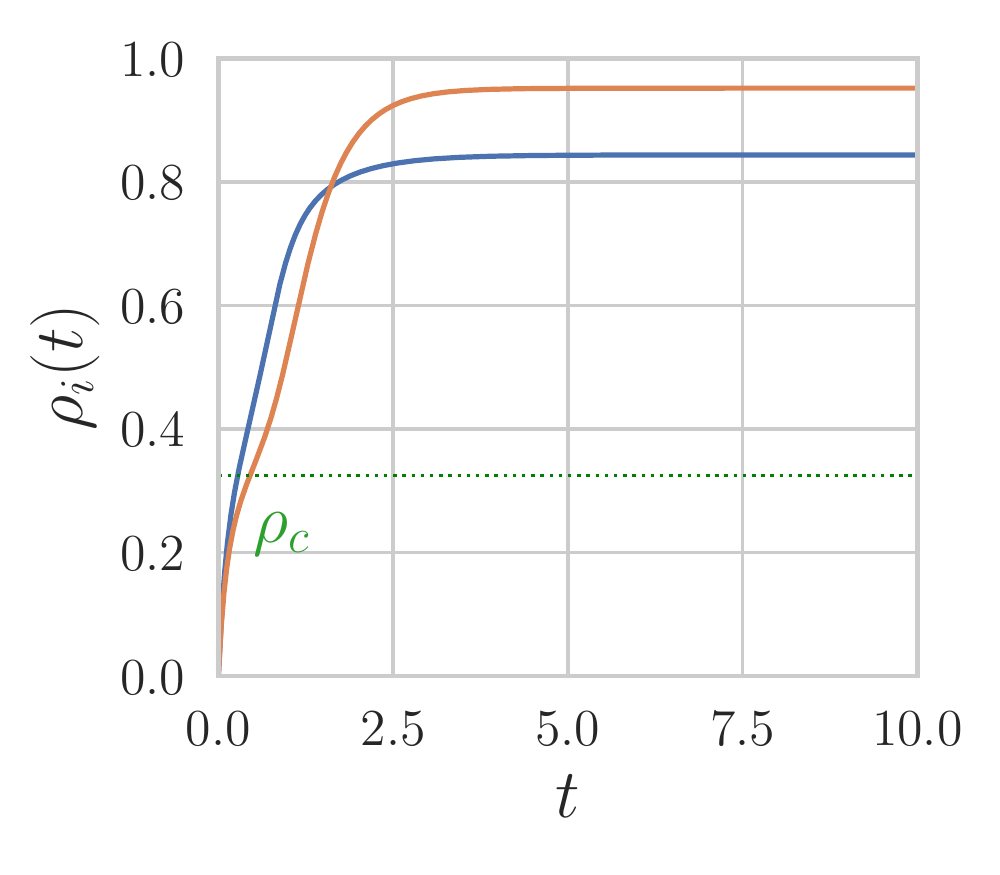}} 
	\end{tabular}
	\caption{(a) Node density $\bm{\rho}_\infty$ with respect to the injection rate $ x$ in the bi-tank model. The embedded plot illustrates the reconfiguration of the traffic model with the same injection rates $ x_1= x_2= x$ and different exiting flows $q_1\neq q_2$. The green part indicates the LD (free) regime and the red indicates the HD (congestion) regime. In the bifurcation phase, we show the upper and the lower bounds, $\bar{\rho}_i$ and $\underline{\rho}_i$, of the oscillating densities in Figure 2(c). (b)--(e) Time-dependent node density $\bm{\rho}(t)$ with respect to different injection rate $ x$ in the bi-tank model.}
	\label{fig:2binphase}
\end{figure}

We start with a toy model with two nodes, which we call the bi-tank model, to explain the detailed behavior of the traffic processes. 
We confine ourselves to investigate the steady state of the processes all with empty initial state ($\bm{\rho}_0=\bm{0}$) and consider the external injections as a constant homogeneous field, i.e., $x_i=x$ for $i=\{1,\dots,N\}$.
Figure \ref{fig:2binphase} illustrates the phase diagram of the steady-state density $\bm{\rho}_\infty:= \bm{\rho}(t)|_{t\rightarrow\infty}$ in the bi-tank model, which exhibits four phases: the linear phase, the bifurcation phase, the localization phase and the saturation phase.

In the \textit{linear} phase, the steady-state density of each node $\bm{\rho}_\infty$ grows proportionally with the external injection rate $x$ in a constant rate that depends upon the nodal topological properties. The linear phase holds until the density of a node slightly exceeds the critical density $\rho_c$, i.e., reaches the HD regime from the LD regime. 
Then, the system arrives at the second phase, the (Hopf) \textit{bifurcation} phase, for an injection rate above the critical threshold $ x> x_c$, where the time-dependent densities of both nodes oscillate between the LD regime and the HD regime with a constant frequency. 
One can numerically verify the possible inexistence of the bifurcation phase in large networks (e.g., the tri-tank model in Appendix \ref{sec:existence} and lattices in the following section).
For a larger injection rate $x>x_c$, the traffic system reaches the \textit{localization} phase. 
In the localization phase, the LD node and the HD node coexist, i.e., the density of node 1 is stable in the LD regime and node 2 stays in the HD regime. 
By a perturbation method (see Appendix \ref{sec:bitank_detail}), we can approach the density of two nodes in the steady state as
\begin{equation}\label{equ:localization_density}
\rho_{\text{LD}}\approx 1-\frac{q_1 / a}{\frac{ p_{21}}{a+b}+ x+\frac{q_1}{a}}, \quad \rho_{\text{HD}}\approx 1-\frac{1}{a+b} x^{-1}
\end{equation}
Finally, if the injection rate is large enough $x>x_s$, the system reaches the \textit{saturation} phase where the density of nodes increases relatively slowly with the injection rate $x$ in the HD regime. 
More details about the phase transitions in the bi-tank model are deferred to Appendix \ref{sec:bitank_detail}.

Understanding the onset of congestion requires an investigation of the localization phase. Recalling the governing equation \eqref{equ:main_mf_model}, each of the steady-state node density $\rho_{i\infty}$ is influenced by flows from two sources: the inflow $y_{i\infty}^{in}=\sum_{j=1}^N h_{ji}$ and outflow $y_{i\infty}^{out}=\sum_{j=1}^N h_{ij}$ between neighbors and the external in/out-flow, $J_{i\infty}^{in}$ and $J_{i\infty}^{out}$, which follows the balance condition for mass conversation $y_{i\infty}^{in}+J_{i\infty}^{in} = y_{i\infty}^{out}+J_{i\infty}^{out}$.
Intuitively, the existence of the localization phase arises from the rebalance between the inflows and the outflows of a node around the critical injection rate $ x_c$. The outflow $y_{i\infty}^{out}$ decreases more sharply than the inflows $y_{i\infty}^{in}$ and $J_{i\infty}^{in}$ with the increasing node density $\rho_{i\infty}$. Meanwhile, the outflow $J_{i\infty}^{out}$ is limited by the maximal outflow $h(\rho_c)$ specified by \eqref{equ:flow_density_relation} and also decreases for $\rho_{i\infty}\geq\rho_c$, failing to reach the flow balance as in the linear phase. In this case, the equilibrium in the linear phase shifts to another equilibrium where node $i$ is congested (i.e. $\rho_{i\infty}\rightarrow 1$), while the density of its neighbor nodes decreases a little due to the decreased outflow $y_{i\infty}^{out}$ from the congested node $i$ to these neighbors.
The interplay between the flow difference incurred by the network heterogeneity and the existence of the equilibrium guaranteed by the resistance non-linearity \eqref{equ:resistance_fun} essentially leads to the localization phase (see Appendix \ref{sec:existence} and Appendix \ref{sec:homo_stability} for more illustrations). 

\subsection{Processes on general networks} 

\begin{figure}[htb]
	\captionsetup[subfloat]{farskip=2pt,captionskip=1pt}
	\begin{tabular}{*{2}{b{0.5\textwidth-2\tabcolsep}}}
		\subfloat[$x=0.5$]{\includegraphics[width=0.13\textheight]{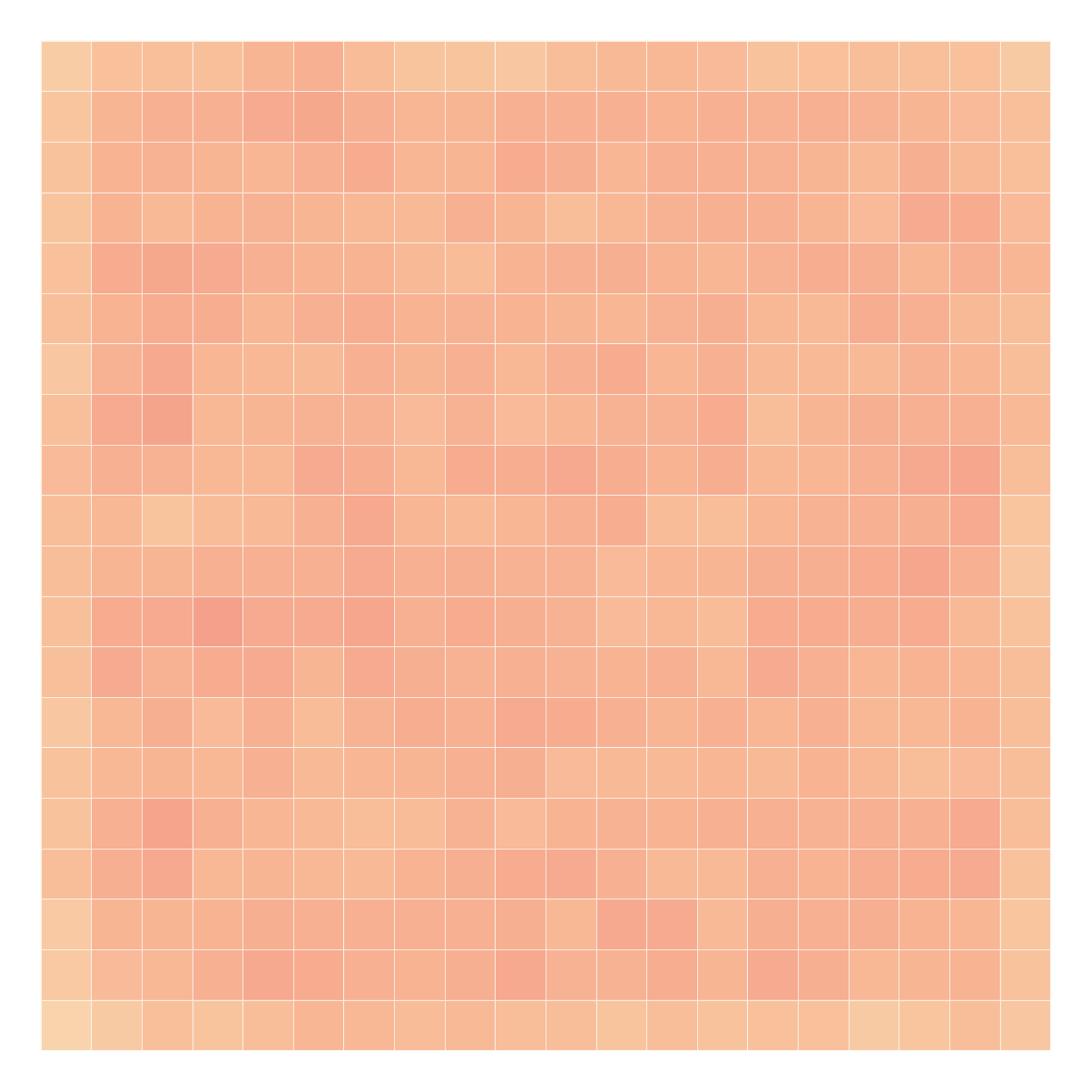}}
		\subfloat[$x=0.7$]{\includegraphics[width=0.13\textheight]{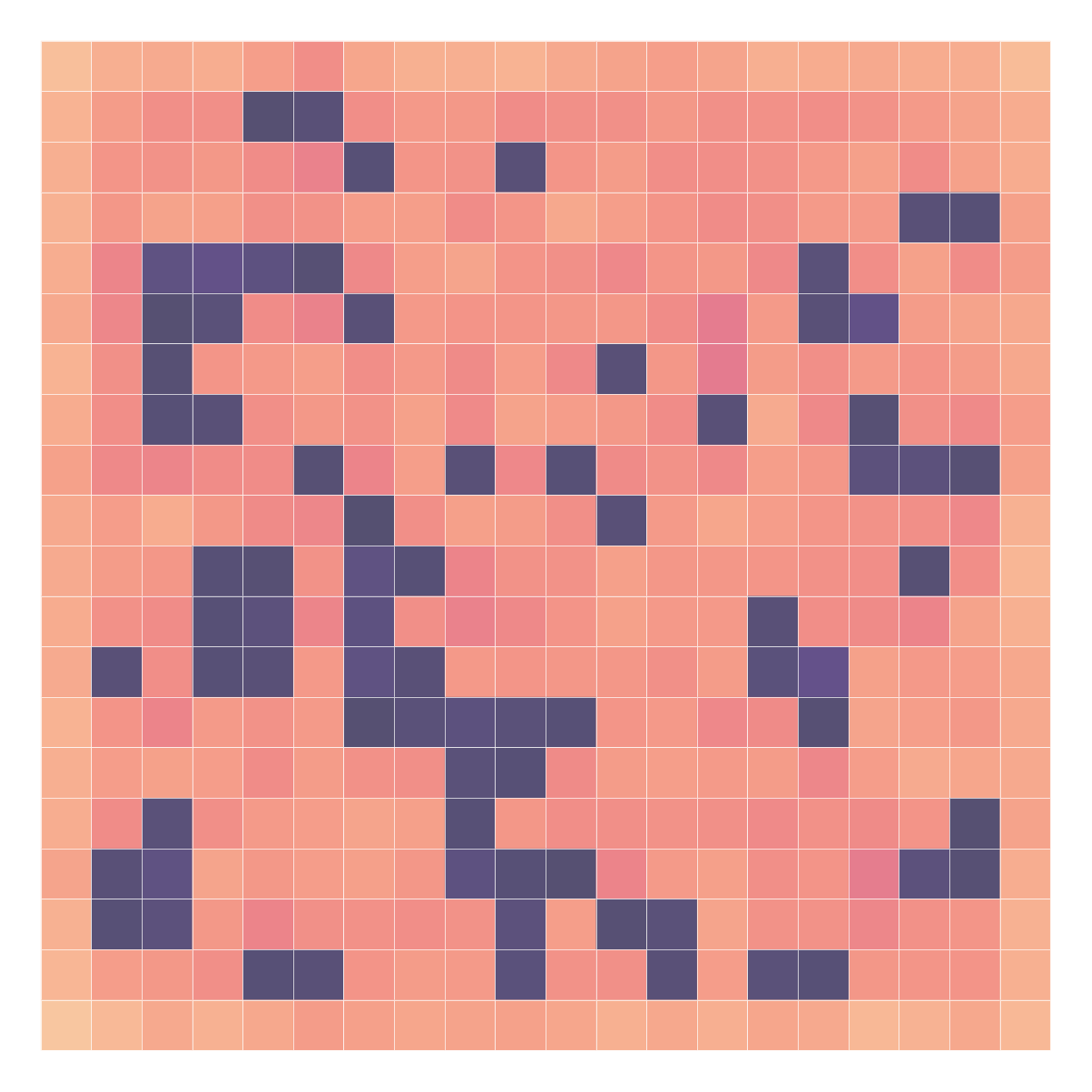}} 
	
		\subfloat[$x=0.75$]{\includegraphics[width=0.13\textheight]{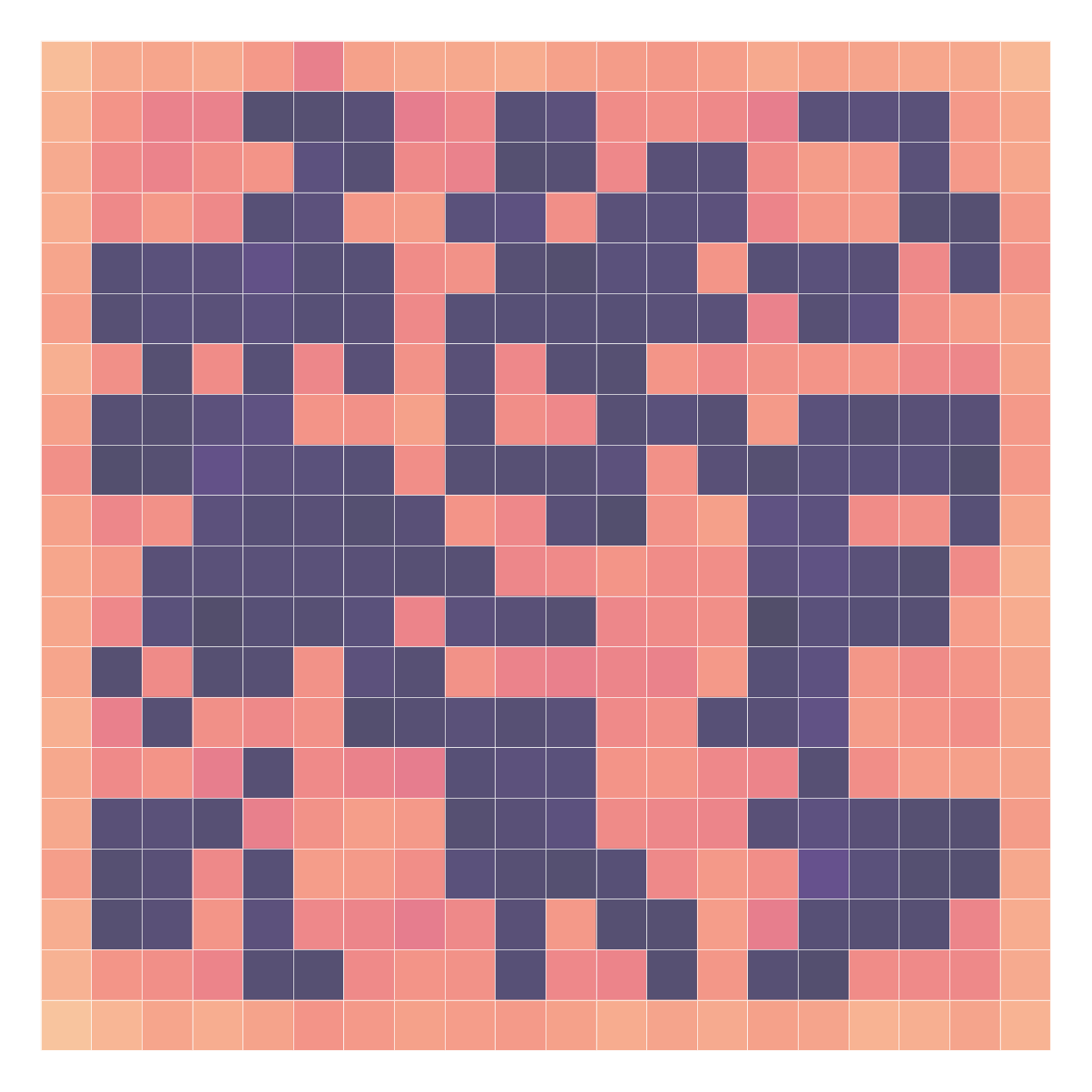}} 
		\subfloat[$x=0.9$]{\includegraphics[width=0.155\textheight]{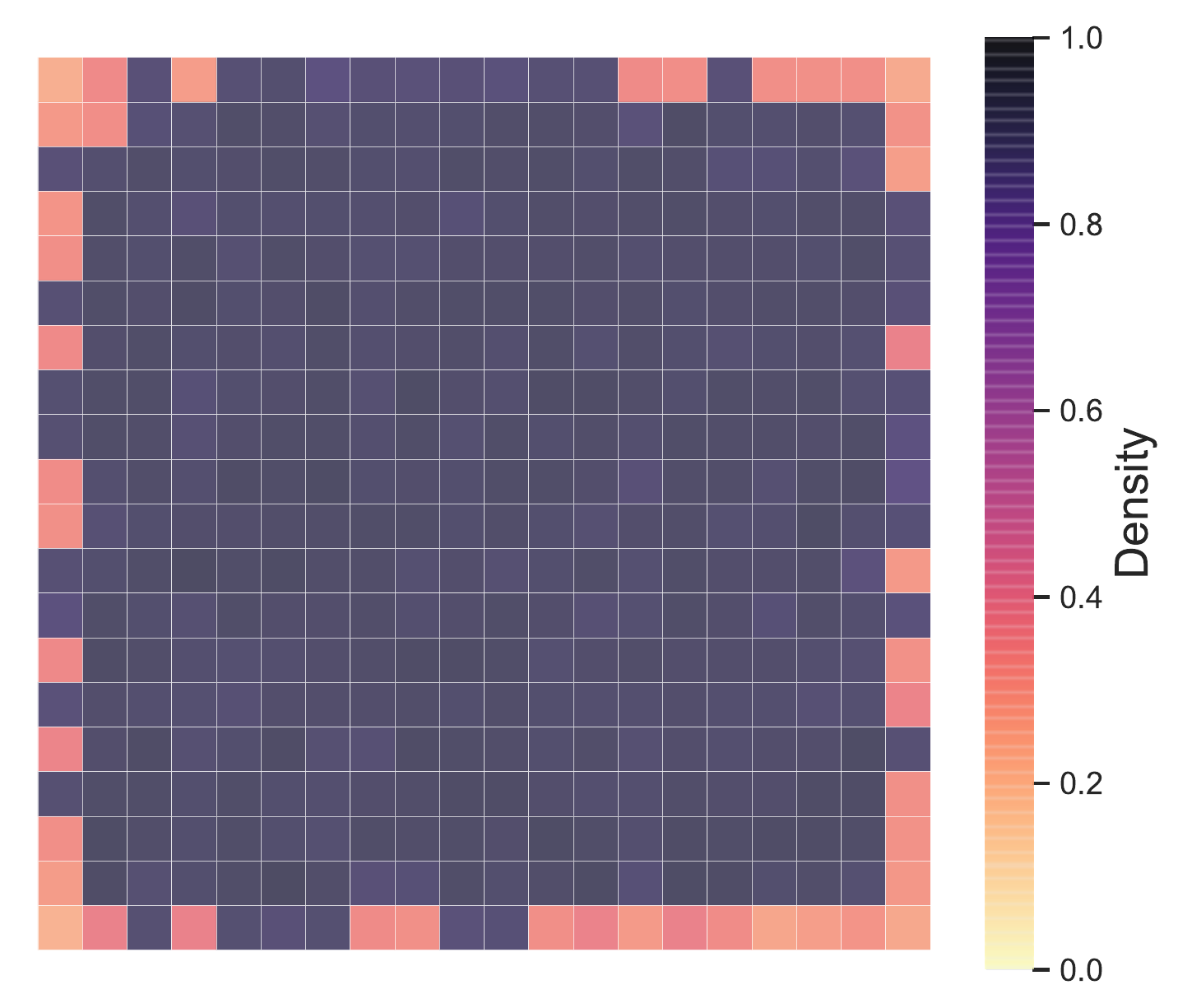}} 
		&
		\subfloat[]{\includegraphics[width=0.96\hsize]{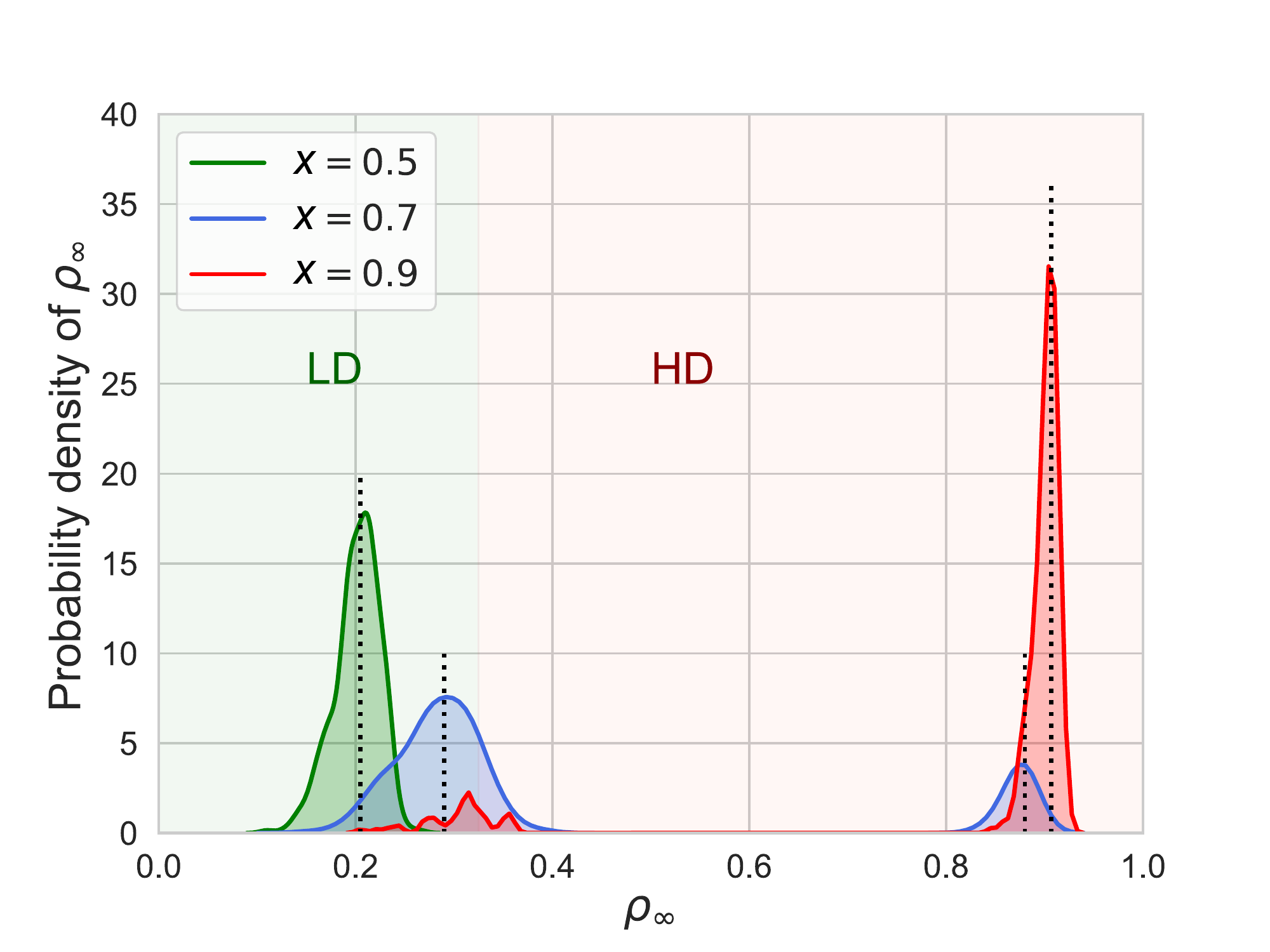}}
	\end{tabular}
	\caption{Localization of traffic congestion in a Lattice with $N=20\times 20$ nodes. The flows $F_{ij}$ and the exiting flows $E_i$ for $i\in\mathcal{N}$ and $j\in\mathcal{N}$ are random variables with a Gaussian distribution $N(\mu, \sigma^2)$ with mean $\mu=1$ and standard deviation $\sigma=0.2$. (a)--(d) Evolution of the onset and the propagation of congested nodes under different homogeneous injection rate $ x$. The darkness of the spot indicates the node density. (e) Histogram of the node density $\bm{\rho}_\infty$ for different injection rate $ x$.}
	\label{fig:lattice_coexistence}
\end{figure}

In a general network, the phase coexistence of free and congested nodes occurs for a wider range of homogeneous injection rates $ x$ and exhibits a spatial localization of congestion. Figure \ref{fig:lattice_coexistence} illustrates the evolution of the density $\bm{\rho}_\infty$ in the steady state with respect to the injection rate $ x$ in a lattice with $N=400$ nodes. 
We denote by $(I-P^T)^{\dag}$ the pseudoinverse of matrix $I-P^T$, and $(I-P^T)^\dag=(I-P^T)^{-1}$ if matrix $I-P^T$ is invertible.
Specifically, in the linear phase, we can approach the node density (see Appendix \ref{sec:free_saturation}) as
\begin{align}\label{equ:density_phase1}
	\bm{\rho}_\infty = a(I-P^T)^{\dag}\bm{ x}+\mathcal{O}(\bm{ x}^2)
\end{align}
for the injection rates $\bm{ x}\rightarrow \bm{0}$, where the vector $\bm{ x}^{m} := ( x_1^m,  x_2^m,\dots,  x_N^m)$. Some separated congested nodes emerge in the network in the localization phase for a specific injection rate $\bm{ x}$. More emerging congested nodes gradually aggregate into several components with the increasing injection rate $\bm{ x}$. Finally, almost all nodes are congested in the saturation phase, and the density of nodes follows
\begin{align}\label{equ:density_phase4}
	\bm{\rho}_\infty=\bm{u}-\frac{1}{a+b}diag(\bm{q})\bm{ x}^{-1}-\mathcal{O}(\bm{ x}^{-2})
\end{align}
with respect to the large injection rates $\bm{ x}\rightarrow \bm{\infty}$ (see Appendix \ref{sec:free_saturation}). 
Figure \ref{fig:lattice_coexistence} and Figure \ref{fig:general_histogram} shows the bimodal distribution of node density in a lattice and some random networks, where the node densities $\bm{\rho}_\infty$ tend to be centered around two constants in the LD and the HD regime, respectively.

The heterogeneity of the underlying flow network $G$ is crucial for the phase coexistence. 
In the homogeneous cases, such as k-regular networks and extremely large lattices, (where the fraction $ p_{ij}$ for each node $i$ are the same, and the injection rates $ x_i$ and the exiting probability $q_i$ are also the same for all nodes), the density of all nodes can shift from a LD density $\rho_{i\infty}\approx \frac{a}{q_i} x_i$ to a HD density $\rho_{i\infty}\approx 1-\frac{q_i}{(a+b) x_i}$ for an above-critical field $ x> x_c$, but the congestion localization does not exist (see Appendix \ref{sec:homo_stability}). 

\section{Properties of traffic congestion}
\subsection{Onset of congestion:}

The onset of congestion occurs if some nodes leave the linear phase due to an above-critical injection rate $\bm{ x}$. In this critical situation, the density of some nodes approaches the critical density $\rho_c$ and the flow $h$ of these nodes approximates the maximum $h(\rho_c)$.
Invoking the node density in the linear phase \eqref{equ:density_phase1}, we define the \textit{congestion centrality} $\omega_i$ of node $i$ as the corresponding $i$-th entry in the vector $(I-P^T)^{\dag}\bm{ x}$, that is
\begin{align}\label{equ:congestion_centrality}
 \omega_i = e_i(I-P^T)^{\dag}\bm{ x}
\end{align}
where $e_i$ is a basis vector.
The phase diagrams in Figure \ref{fig:2binphase} and Figure \ref{fig:3binphase} illustrate that the first congested node (e.g., node $i$) is not the node with the largest $\omega_i$ in the network, but the node $j$ with the largest congestion centrality $\omega_j$ among the neighbors of node $i$. The density of node $i$ with the largest congestion, on the contrary, decreases a little when its neighbor $j$ stays in the HD regime. Thus, the onset of the congestion occurs if 
\begin{align}
a \cdot \max_{j\in{\mathcal{N}(i)}}\omega_j \geq \rho_c \quad \text{for} \quad i=\arg \max_{i\in\mathcal{N}}\omega_i
\end{align}
where $\mathcal{N}(i)$ is the set of all neighbors of node $i$.
Specially, the congestion threshold is $ x_c \approx \frac{\rho_c}{a\cdot e_j(I-P^T)^{\dag}{\bm{u}}}$ for the homogeneous injection rates.

\subsection{Impact of non-linearity and heterogeneity:}

\begin{figure}[htp]
	\centering
	\subfloat[$b$]
	{\includegraphics[width=4.8cm]{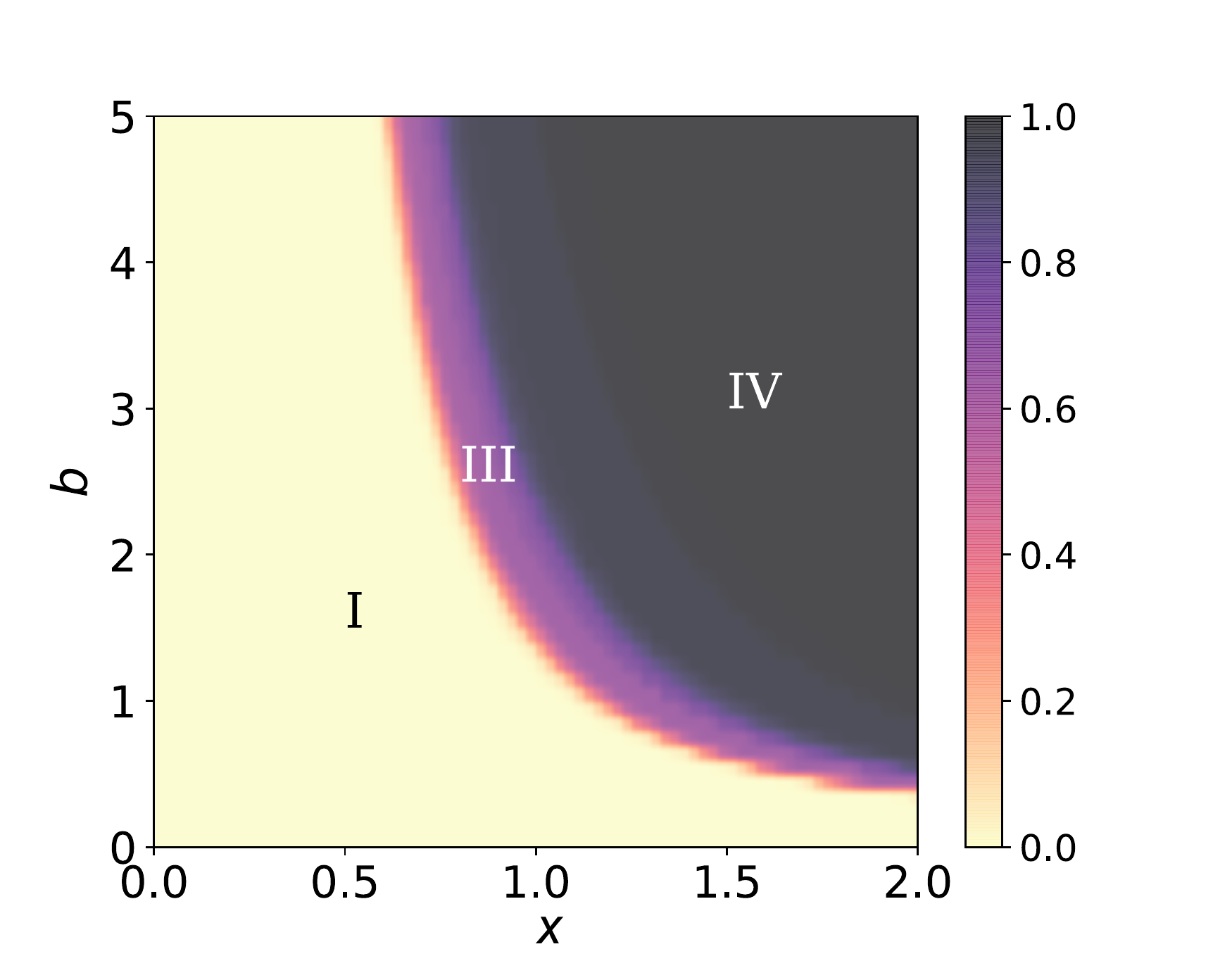} \label{fig:pd_b}}\hspace*{-2.0em}
	\subfloat[$\gamma$]
	{\includegraphics[width=4.8cm]{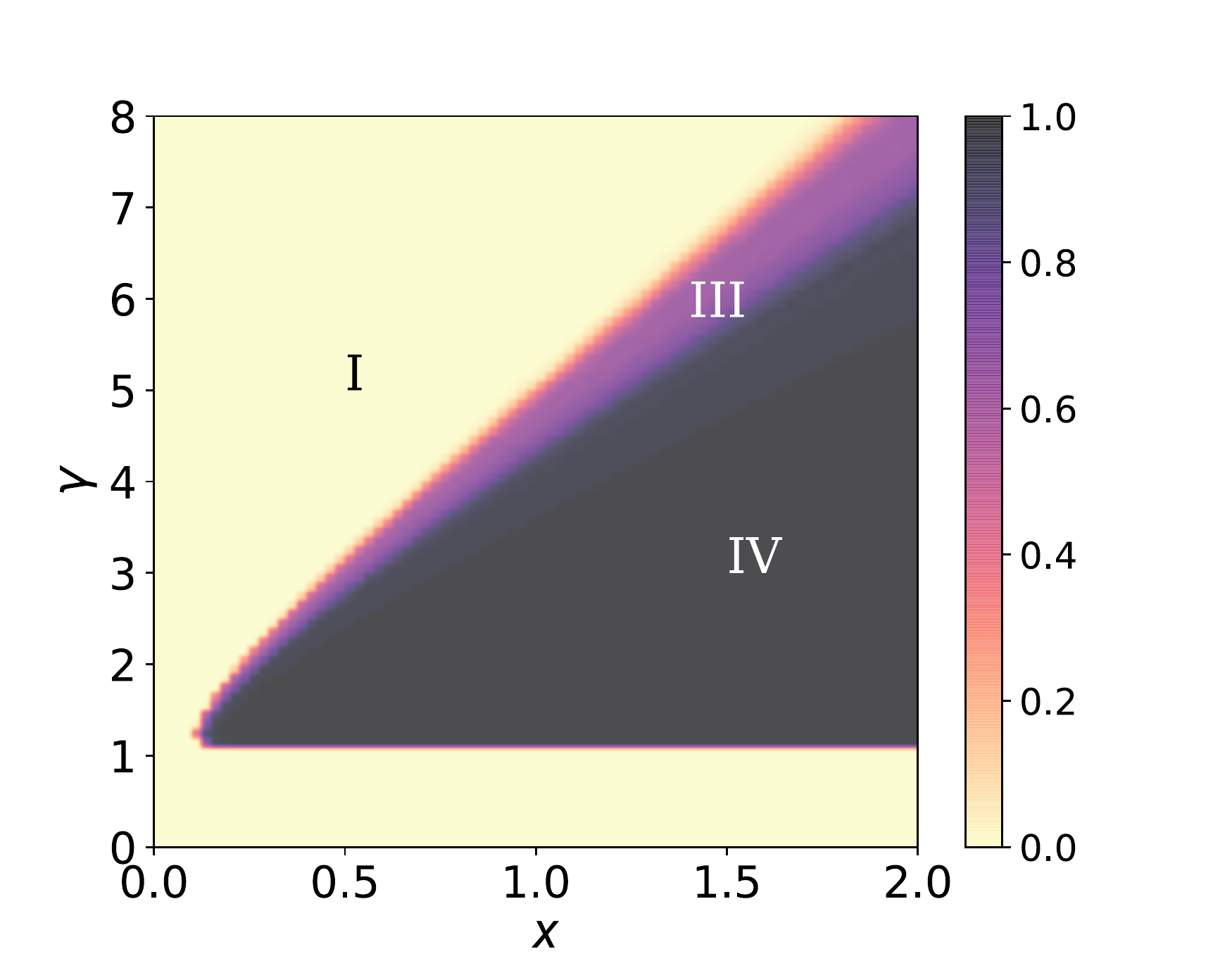}\label{fig:pd_g}}\hspace*{-1.3em}
	\subfloat[$\sigma$]
	{\includegraphics[width=4.8cm]{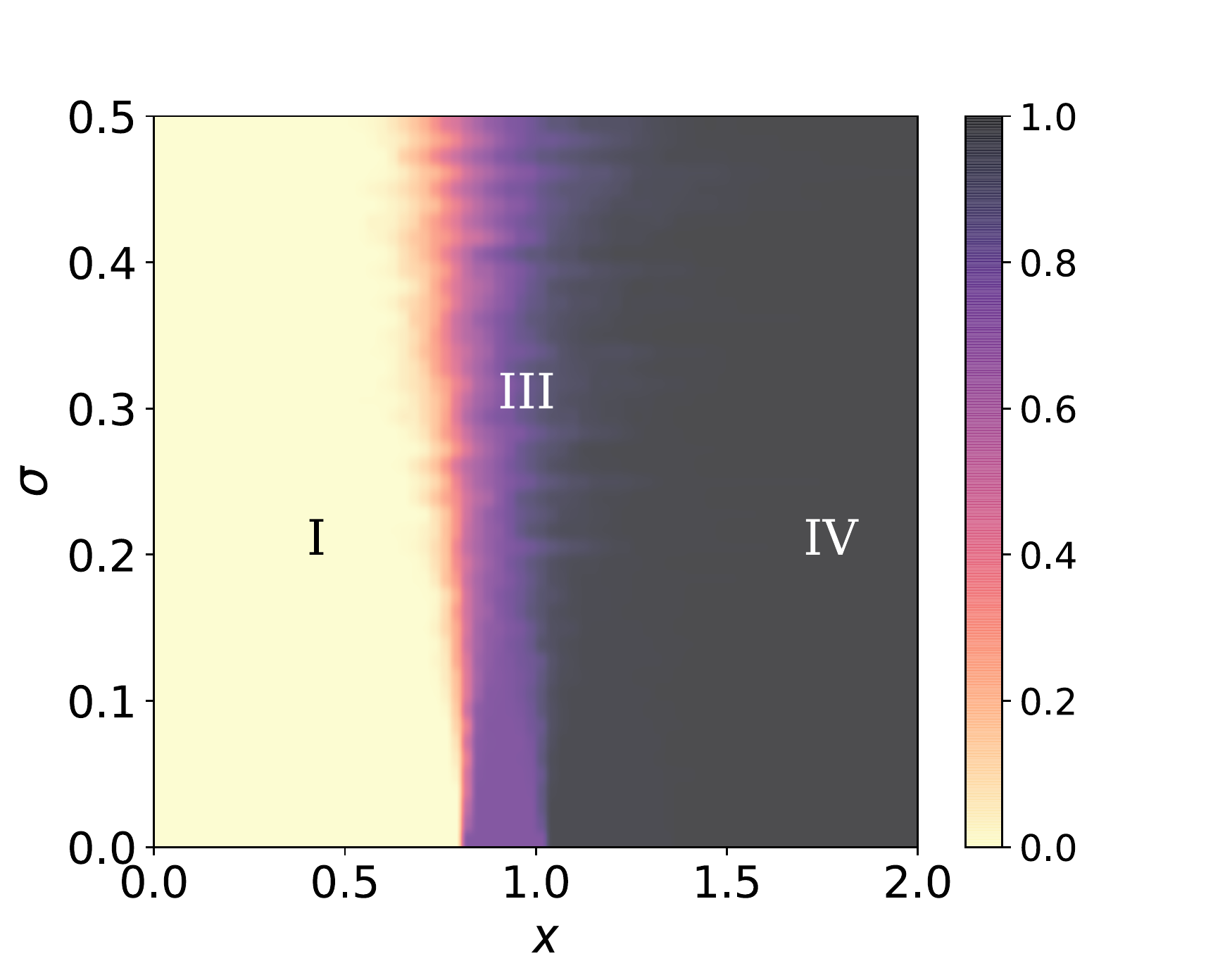}\label{fig:pd_s}} \hspace*{-1.3em}
	\subfloat[$\sigma_p$]
	{\includegraphics[width=4.8cm]{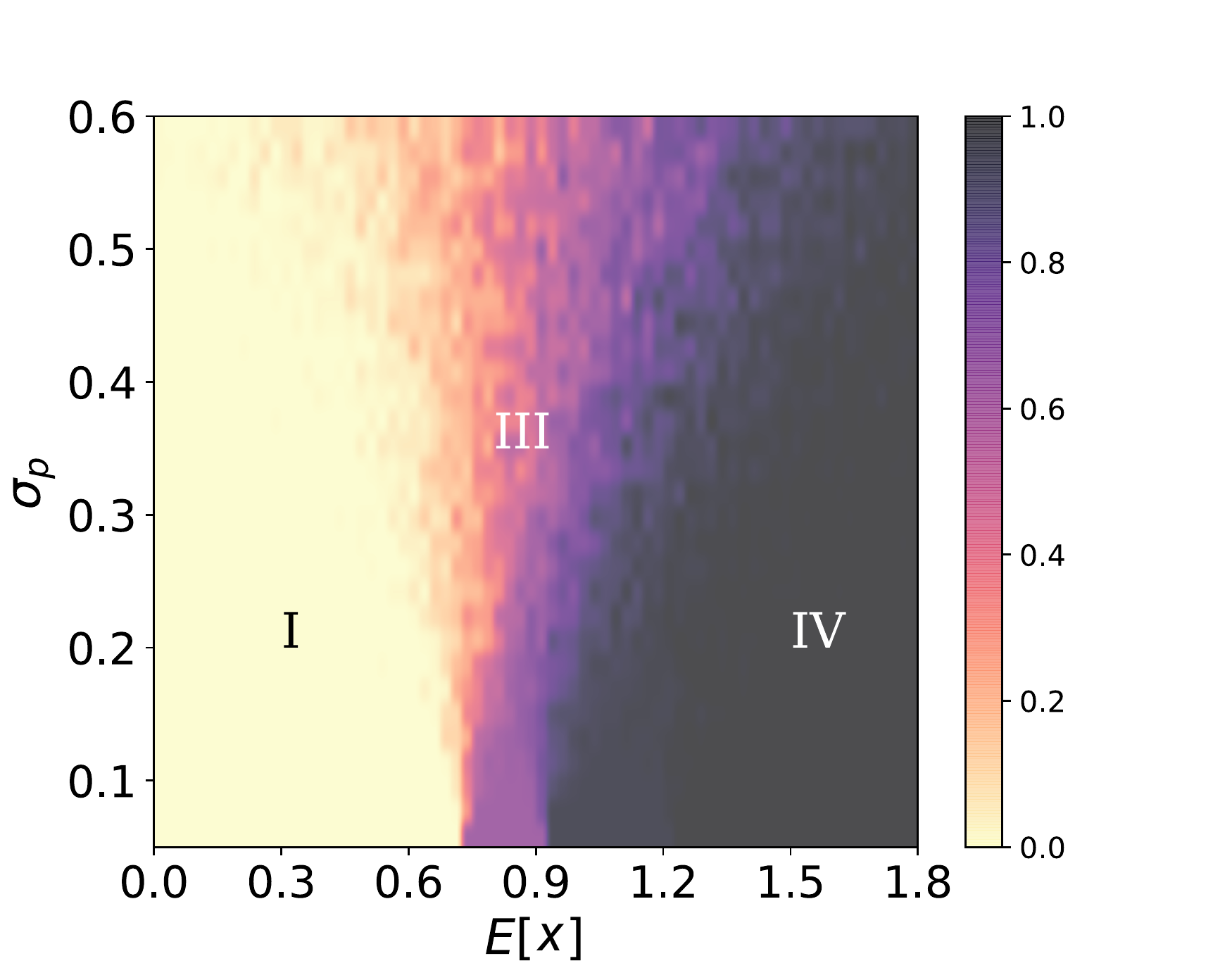}\label{fig:pd_p}} 
	\caption{Phase diagrams of fraction of congested nodes $\chi$ in a lattice with $10\times10$ nodes under a homogeneous injection rate $ x$, with respect to different factors. 
	The flows $F_{ij}$ and the exiting flows $E_i$ for $i\in\mathcal{N}$ and $j\in\mathcal{N}$ are positive random variables with a Gaussian distribution $N(\mu, \sigma^2)$ with mean $\mu=1$ and standard deviation $\sigma=0.2$ in (a), (b) and (d). The injections $\bm{ x}$ in (d) are positive random variables with a Gaussian distribution $N_p(\mu_p, \sigma_p^2)$ with mean $\mu_p=\mathbb{E}[ x]$ and standard deviation $\sigma_p$.
	The factors include: (a) the tunable parameter $b$ for the the non-linear term in the resistances specified in \eqref{equ:resistance_fun}; (b) the non-linearity indicator $\gamma$ specified in \eqref{equ:resistance_fun}; (c) the heterogeneity of topology indicated by the standard deviation $\sigma$ of the flow rate $F$; (d) the standard deviation of the injections $\sigma_p$ indicating the heterogeneity of the injection rate. Different phases are marked by \RNum{1}. the linear phase, \RNum{3}. the localization phase, and \RNum{4}. the saturation phase.}
	\label{fig:phaseDiag_congest}
\end{figure}

Since the congested nodes, as the bottlenecks, diminish the performance of transport on the network, we focus on the order parameter $\chi = \frac{1}{N}\sum_{i=1}^N 1_{\rho_{i\infty}>\rho_c}$ as the fraction of congested nodes in the steady state.
Both the flow network $G$ and the resistance specified in \eqref{equ:resistance_fun} could affect the behaviour of congestion on networks.
Figure \ref{fig:phaseDiag_congest} shows the phase diagrams in a lattice with respect to various factors including (a) the parameter $b$ in the resistance \eqref{equ:resistance_fun}; (b) the parameter $\gamma$ in the resistance \eqref{equ:resistance_fun}; (c) the standard deviation $\sigma$ of the flows $F_{ij}$ and the exiting flows $E_i$; (d) the standard deviation $\sigma_p$ of the injection rates $x_i$. The critical density $\rho_c$ can be obtained according to the maximum flow as the flow-density relation \eqref{equ:flow_density_relation} by $\frac{d h(\rho_i)}{\rho_i}|_{\rho_i=\rho_c} =0$, which yields $\rho_c=\left(\frac{a}{b(\gamma-1)}\right)^\frac{1}{\gamma}$ for $\gamma>1$.

The congestion threshold $ x_c$ that divides the free regime ($\chi=0$) and the localization regime ($0<\chi<1$) in a determined network $P$ follows $ x_c \sim \rho_c \sim b^{-\frac{1}{\gamma}}$, as shown in Figure \ref{fig:pd_b}, which demonstrates a larger nonlinear term in the resistances \eqref{equ:resistance_fun} degrades the congestion threshold $ x_c$.
The congestion threshold $ x_c$ follows $ x_c \sim \left(\frac{a}{b}\right)^\frac{1}{\gamma} (\gamma-1)^{-\frac{1}{\gamma}}$ for $\gamma>1$, as shown in Figure \ref{fig:pd_g}, which increases with the non-linearity constant $\gamma$ and tends to a constant $\frac{a}{b}$ for a very large $\gamma$. 
Figure \ref{fig:pd_g}, \ref{fig:pd_s} and \ref{fig:pd_p} illustrate that a large non-linearity of resistance (indicated by $\gamma$) as well as a large heterogeneity of the flow network $F$ and the injection rate $x$ (indicated by $\sigma$ and $\sigma_p$) generally extend the localization phase\footnote{The localization phase presented in Figure \ref{fig:pd_s} for $\sigma=0$ is due to the dissimilarity of the congestion centrality of nodes on the periphery of the lattice, which prevents the lattice to be strictly homogeneous.} and incur the phase coexistence for a wider range of the injection rates $x$. 
The phase diagrams respecting the average density of nodes $\langle\rho_\infty\rangle = \frac{1}{N}\sum_{i=1}^N\rho_{i\infty}$ in the network exhibit similar behavior as the order parameter $\chi$ (see Appendix \ref{sec:phasediag_avgdensity}).

\begin{figure}[htp]
	\centering
	\subfloat[Histogram of node density in some general networks]
	{\includegraphics[width=0.3\textheight]{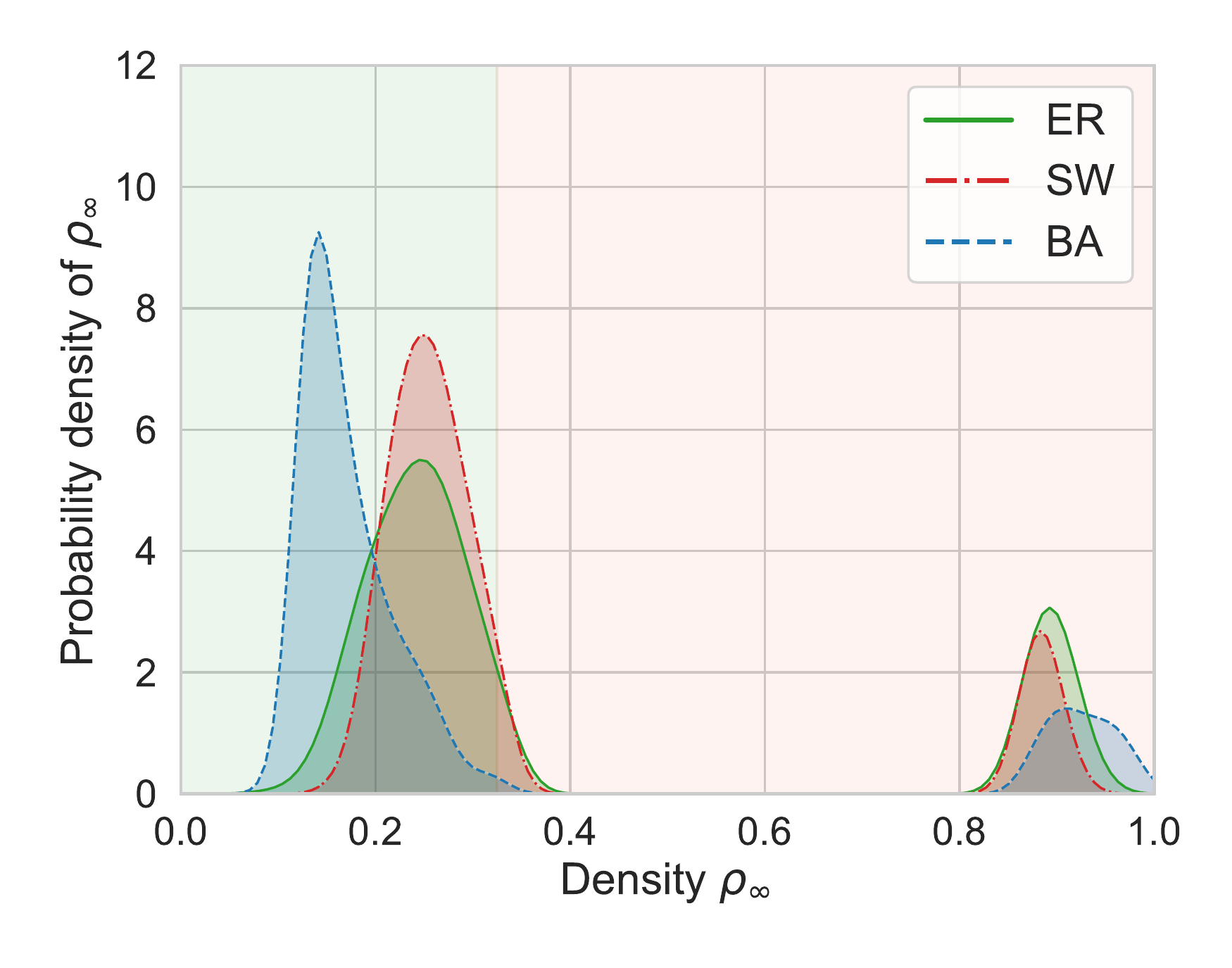}\label{fig:general_histogram}} \hfill
	\subfloat[Correlation between the node density and the congestion centrality]
	{\includegraphics[width=0.3\textheight]{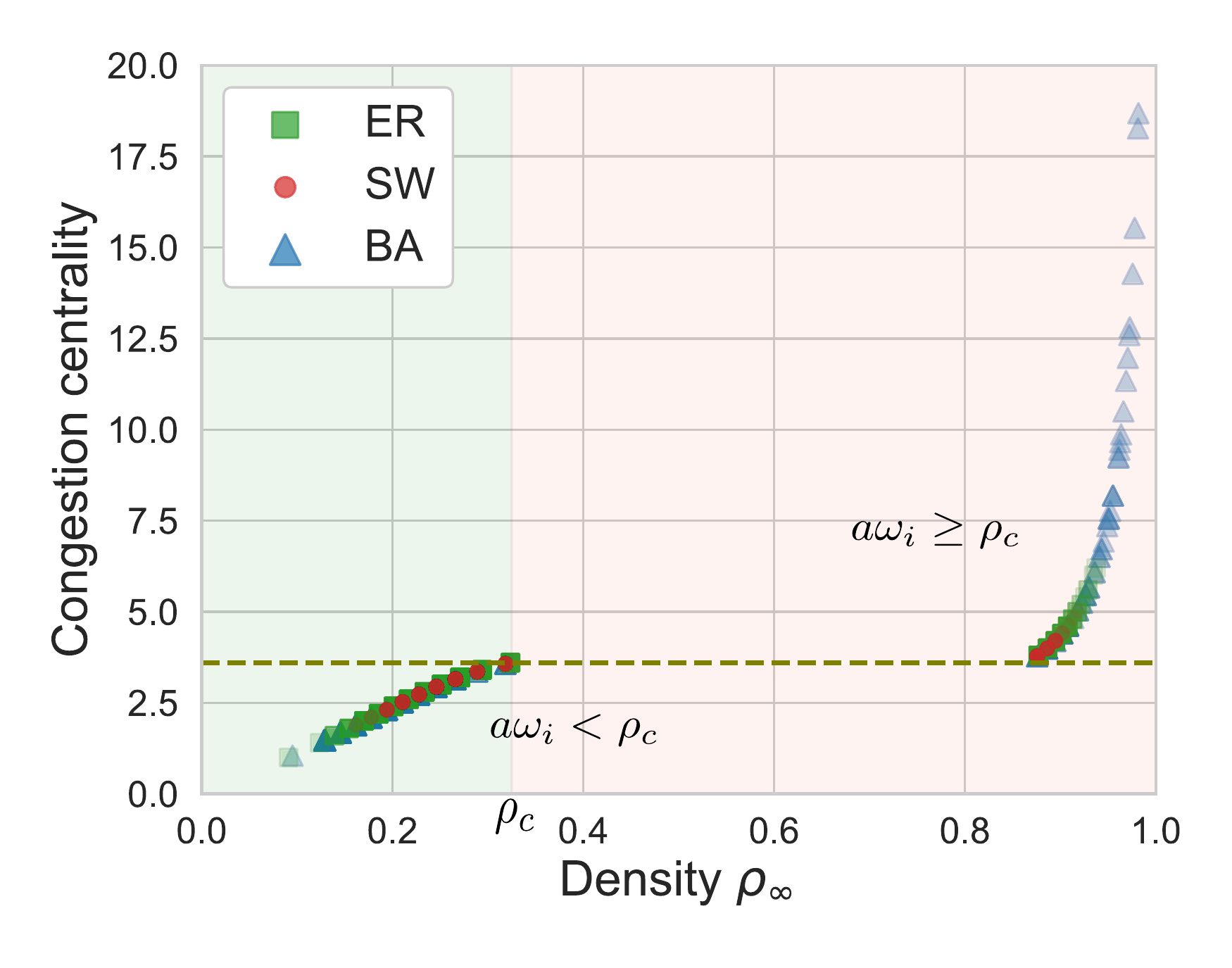}\label{fig:density_coorelation}}\\
	\subfloat[Correlation between the congestion time and the congestion centrality]
	{\includegraphics[width=0.3\textheight]{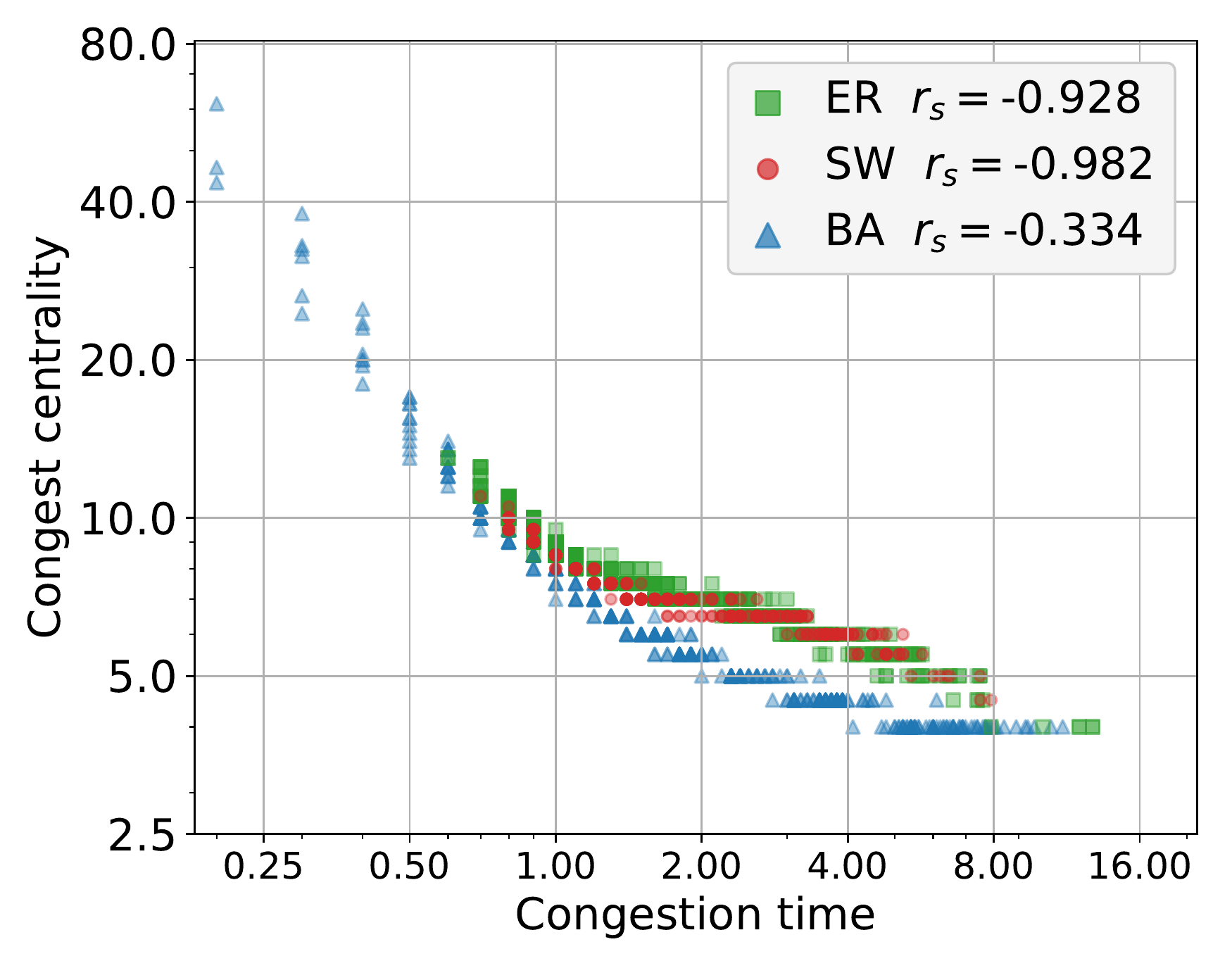}\label{fig:time_corelation}} \hfill
	\subfloat[Correlation between the decongestion time and the congestion centrality]
	{\includegraphics[width=0.3\textheight]{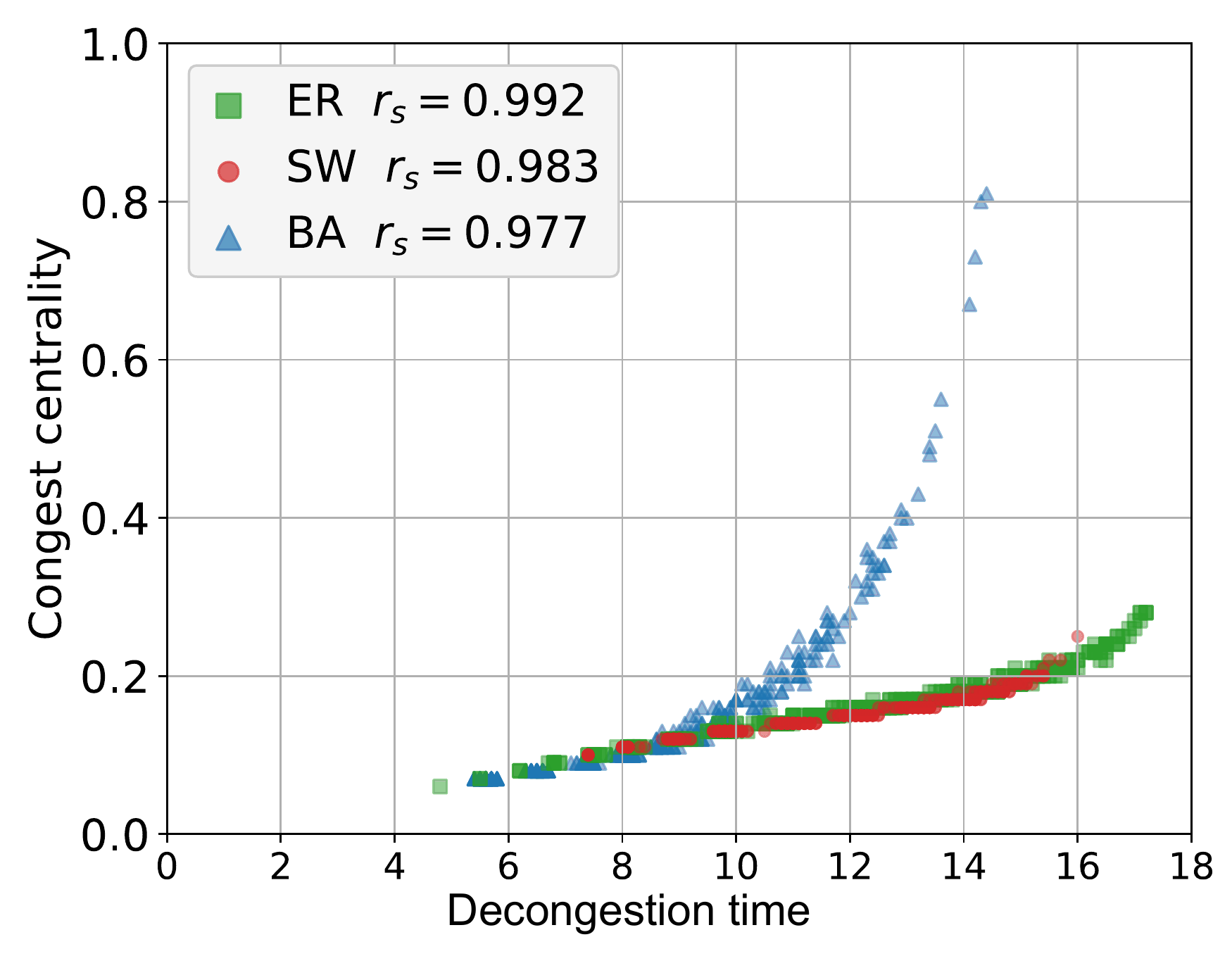}\label{fig:time_corelation_full}}
	\caption{Traffic congestion in general networks, e.g., (i) an Erd{\H{o}}s--R{\'e}nyi (ER) network with $N=500$ nodes and link density $0.03$, (ii) a small-world (SW) network with $N=500$ nodes, link density $0.03$ and rewiring probability $0.5$, (iii) a Barab{\'a}si–-Albert (BA) network with $N=500$ nodes and 6 attached links. 
	The flows $F_{ij}$ and the exits $E_i$ for $i\in\mathcal{N}$ and $j\in\mathcal{N}$ in the above three networks are equal to $F_{ij}=E_i=1$.
	(a) Bi-modal distribution of node density for the homogeneous injection rate $ x=0.2$. (b) Correlation between the congestion centrality and the node density. (c) Correlation between the congestion centrality $\omega$ and the time $t_{HD}$ when the node enters the HD regime for the process started with the all-empty state with a large homogeneous external injection $ x=0.5$. The plot is on a log-log scale. (d) Correlation between the congestion centrality $\omega$ and the time $t_{LD}$ when the node leaves the HD regime for the process started with the all-full state with a small homogeneous external injection $ x=0.01$. The Spearman's rank correlation coefficients $r_s(\omega,t_{HD})$ and $r_s(\omega,t_{LD})$ are presented in (c) and (d). }
	\label{fig:general_network}
\end{figure}

\subsection{Identifying congested nodes:}

We identify the congested nodes based on the defined congestion centrality \eqref{equ:congestion_centrality}.
%Figure \ref{fig:density_coorelation} illustrates the relation between the congestion centrality and the node density. 
Although the node with the largest congestion centrality is usually not the first congested node (as discussed before), the congestion centrality strongly correlates to with the steady-state node density $\bm{\rho}_{\infty}$ for a determined injection $ x$, as illustrated by Figure \ref{fig:density_coorelation}.
In general, the congested nodes can be identified by the congestion centrality obeying $a\omega_i\geq \rho_c$, while another analytic estimation of the lower bound of the fraction of congested nodes $\chi$ is presented in Appendix \ref{sec:est_chi_lb}.
The defined congestion centrality is related to the resolvent of matrix $P^T$ and resembles the alpha-centrality\footnote{The alpha-centrality is defined as $(I-\alpha A^T)^{-1}$, where $A$ is the adjacency matrix and $\alpha$ is a constant that trades off the importance of external influence against the importance of connection. The exiting fraction $\bm{q}$ in our model plays a similar role as the constant $\alpha$, but in a heterogeneous way.} referring to \cite{bonacich2001eigenvector}, which has been applied to identify the nodal importance for asymmetric social relations \cite{ghosh2012rethinking}. Broadly, the term $(I-P^T)^{\dag}$ in the congestion centrality \eqref{equ:congestion_centrality} also resembles the pseudoinverse of the Laplacian matrix as a diffusion operator in linear systems \cite{van2017pseudoinverse}, e.g., reflecting the relation between the voltage and the injected current in impedance networks.

\subsection{Congestion propagation}

We are concerned about the temporal evolution of node density $\bm{\rho}(t)$ regarding congestion propagation.
%The mass flow obeys  $h_i \approx \rho_i\frac{d h(\rho_i)}{d\rho_i}\big|_{\rho_i=0} \leq \frac{\rho_i}{a}$ due to the Lipschitz constant $\sup_{\rho_i}\frac{d h(\rho_i)}{d\rho_i} = \frac{d h(\rho_i)}{d\rho_i}\big|_{\rho_i=0} = \frac{1}{a}$ for $\rho_0\rightarrow 0$. 
Invoking the governing equation \eqref{equ:model_vector_form}, the density vector $\bm{\rho}(t)$ in the initial stage ($t\rightarrow0$ and $\bm{\rho}(t)\rightarrow\bm{0}$) follows
\begin{align}\label{equ:density_increase}
	\frac{d\bm{\rho}(t)}{dt}\approx \bm{ x}-\frac{1}{a}\left(I-P^T+a\cdot diag(\bm{ x})\right) \bm{\rho}(t)
\end{align}
for the flow that obeys $h_i(t) \approx \frac{\rho_i(t)}{a}$.
%which yields $\bm{\rho}(t)\approx\bm{ x}-e^{-\frac{1}{a}M_1 t}\bm{\rho}_0$
Thus, the density $\bm{\rho}(t)\approx\bm{\rho}_\infty-e^{-\frac{1}{a}(I-P^T+a\cdot diag(\bm{ x}))t}\bm{\rho}_\infty$ exponentially converges to a steady state $\bm{\rho}_\infty$ in the initial stage, where the upper-bound of the convergence rate is determined by both the free flow constant $\frac{1}{a}$, and the spectral radius $\varrho(M_1)$ of the matrix $M_1 := I-P^T+a\cdot diag(\bm{ x})$. 
Similarly, for a decongestion process started from the fully-occupied state ($\bm{\rho}_0=\bm{u}$) with a small injection vector $\bm{ x}\approx\bm{0}$, the decay rate of the density $\bm{\rho}(t)$ to a empty state in the final stage is determined by the spectral radius $\varrho(M_2)$, where the matrix $M_2 := I-P^T$. 
Generally, a more homogeneous network $G$, leading to a smaller $\varrho(M_1)$ for the same injection rates (i.e. $x_i=x$ for $i\in\mathcal{N}$), decreases both the growth speed of the density $\bm{\rho}(t)$ and the steady-state density\footnote{The sum of entries $\sum_{i=1}^N\sum_{j=1}^N p_{ij}$ in the matrix $P$ is a constant for a fixed exiting flow $\bm{q}$. The spectral radius is $\varrho(I-P^T)\leq ||I-P^T||_F = \left(N+\sum_{i=1}^N\sum_{j=1}^N p_{ij}^2\right)^{1 \over 2}$. The upper bound of the spectral radius $\varrho(M_2)$ nearly achieves the minimum for the same entries $p_{ij}$ and become larger for the entries $p_{ij}$ with a large standard deviation in large networks.}, which also benefits the congestion alleviation via a smaller $\varrho(M_2)$.
The flow-based network emphasizes the importance of a better routing schedule and a load balancing strategy. 

If we consider the binary-state of nodes, i.e., the free (or LD) and the congested (or HD), the congestion propagation resembles a percolation process. The nodes that satisfy the condition $a\omega_i\geq \rho_c$ shift to the congestion state one by one during congestion propagation, while the nodes obeying the condition $a\omega_i< \rho_c$ return to the free state during decongestion.
Figure \ref{fig:time_corelation} and \ref{fig:time_corelation_full} show both the time sequence of the node becoming congested from the all-empty state and the decongestion sequence from the all-full state have a strong correlation with the defined congestion centrality.
%The flow network and the external injections essentially determine the congestion propagation according to the distribution of the nodal congestion centrality, %which explain the congestion spreading started from high-degree nodes in a road network.

\subsection{Hysteresis of congestion}

\begin{figure}[htp]
	\centering
	\includegraphics[width=10cm]{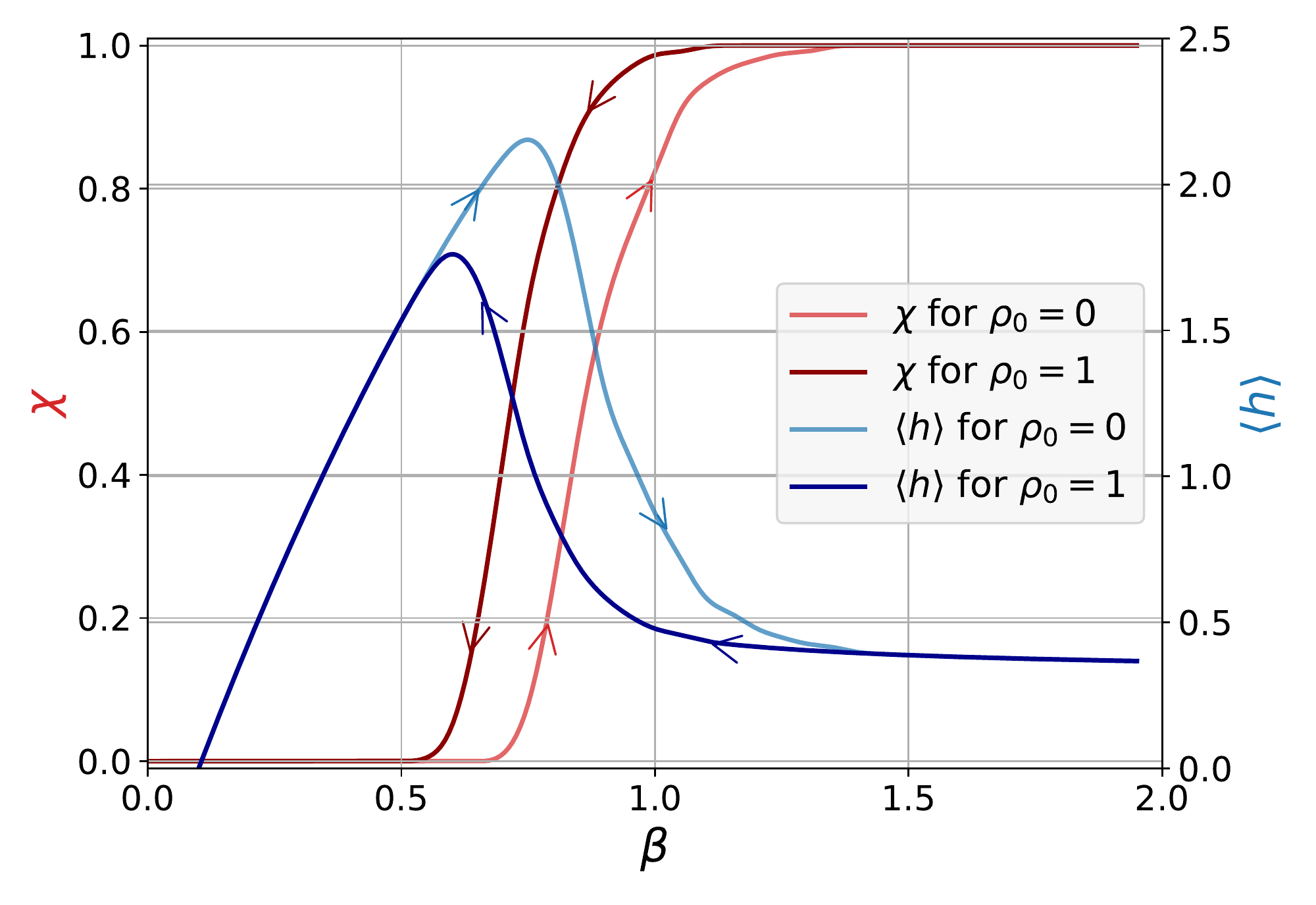}
	\caption{The fraction of congested nodes $\chi$ and the average current $\langle h\rangle$ with respect to the homogeneous injection rate $x$ in a Lattice with $N=10\times 10$ nodes for different initial states $\bm{\rho}_0$.  The flow $F_{ij}$ and the exiting flow $E_i$ for $i\in\mathcal{N}$ and $j\in\mathcal{N}$ are absolute random variables with a Gaussian distribution $N(\mu, \sigma^2)$ with mean $\mu=1$ and standard deviation $\sigma=0.2$.}
	\label{fig:hysteresis}
\end{figure}

%The previous investigation mainly focuses on the congestion onset processes started from the all-empty state $\bm{\rho}(0)=\bm{0}$. 
Due to the multiple equilibrium states for the dynamics, the processes started with different initial states could be dissimilar, with the different fraction of congested nodes $\chi$ and the different average current $\langle h\rangle = \frac{1}{N}\sum_{i=1}^N h_i$ in the steady state, as shown in Figure \ref{fig:hysteresis}. Under the homogeneous injection, the congestion threshold $ x_{c,\bm{\rho}_0=0}$ for the all-free initial state is smaller than the threshold $ x_{c,\bm{\rho}_0=\bm{u}}$ from the all fully-occupied initial condition, which demonstrates that decongestion is more difficult than promoting traffic congestion. 
The hysteresis effect for the average current $\langle h\rangle$ also demonstrates that the congestion localization could have richer behaviors for different initial states, which may corresponds to different transition behavior at different time (e.g. rush hour and off-peak hour) observed in real traffic cases \cite{li2015percolation}\cite{gayah2011clockwise}.
%The hysteresis effect can be qualified by the indicators $ p x_c =  x_{c,\rho_0=1}- x_{c,\rho_0=0}$ and $ p H = \max_ x()$.

\section{Applications}
\subsection{Vehicle transport}

The proposed traffic model in Section \ref{sec:traffic_model} describes the dynamic process of vehicle transportation.
In a microscope view of road systems, the node in the model represents a lane as a region, and the links represent the junctions as downstream transitions. 
Then, the proposed model \eqref{equ:model_vector_form} has a similar form as the Lighthill-–Whitham-–Richards (LWR) model \cite{lighthill1955kinematic} for shock wave dynamics of spatial discretization. In addition, the nonlinear resistance \eqref{equ:resistance_fun} is consistent with the formula of travel time \cite{dafermos1969traffic} proposed by the Bureau of Public Roads (BPR). Thus the flow-density relation in our model is in line with the traffic fundamental diagram \cite{geroliminis2008existence}. 
The traffic model captures most features of vehicle transportation, though the flow network $F$ is assumed to be fixed instead of practically changing due to possible congestion, which is enough to describe the traffic behaviors in a mild congestion state. 

We have shown that the heterogeneity of the flow transitions and the injections on nodes highlight the generality of the localization phase. The one-way lanes in empirical road systems can be regarded as the zero flow transition, while drivers usually enter the system only from a small part of nodes. Both the above factors boost the heterogeneity and possibly lead the transport system to the congestion localization. Following the analysis in Section 4, the most vulnerable nodes with a relatively high probability to be congested can be identified by the proposed congestion centrality. The testable prediction about the congestion propagation helps working out contingency plans by limiting the number of entering drivers that are abstracted by the injection rates at some specific nodes. On the other aspect, a better routing schedule for distributing traffic loads more balanced, which is manifested by a more homogeneous flow network, and can promote a more efficient transport system.
%We expect to investigate the correlation between the proposed congestion centrality and the real-world congested regions in the Dutch traffic system in the future work.

%todo The impact of the resistance properties on the congestion threshold implies the upgradation of road conditions can ease the traffic congestion. On the other aspect, a better routing schedule for distributing traffic loads more evenly, which is manifested by a more homogeneous flow network, can promote a more efficient transport system.

\begin{comment}
distribution of lane density in empirical data

congestion propagation in empirical data

switch of giant component for PNAS

In directed network, tends to be a propagation

effect of directed network

lieaer congestion propagation in directed network?
The constant provides the intuitive to alter the threshold.
\end{comment}

\subsection{Targeted diffusion}

Target diffusion is a general task inspired by many real applications in nature, which aims to induce the matter to some specific regions instead of distributing evenly over the network. For example, in the chemotherapy application, targeted diffusion can be applied for inducing biochemical cascades to treat cancer \cite{roberts2007targeting}, which prefers to guide the drug to aggregate in affected areas in a long period but with a minimum dose to reduce side effects for other organs. In the financing field, the start-ups aim to attract capital from the financial market as a flow model \cite{portes2001information} by building relations with 
investment institutions.

Despite the targeted spreading in epidemic models \cite{lokhov2017optimal}\cite{he2018optimal}, the localization phase in the proposed traffic model provides a promising approach for targeted diffusion on networks. The targeted diffusion can be reduced to an optimization problem of different forms. Promoting the system to a localization state, the specific nodes can spontaneously reach a high-density state $\rho_{HD}$ with a minimum sum of injection rates $\bm{u}^T\bm{x}$ as budgets. Further, the high-density nodes can be designated by adjusting the underlying topology or allocating the injections on different nodes in an optimized way. 
%Appendix \ref{sec:application_target} provides an approach of injection allocation for targeted diffusion.

\subsection{Neural networks}

The cortical-network dynamics in the brain have been investigated based on the Kuramoto model and the epidemics models. The Kuramoto model helps to explore the functional topology of the brain by correlating the brain waves at different regions \cite{cabral2011role}, while the epidemics model establishes a strong connection between topological disorder and rare-region effects \cite{moretti2013griffiths}\cite{liu2018network}.

More intuitively, the nerve conduction induced by neuropotentials among cortex yields the similarity between electrical networks and neural networks. The proposed traffic model with the nonlinear time-dependent resistance is in accord with the electrical network with memorizes modeled for brain systems \cite{de2006self}. Interestingly, we verify the localization of the high-density (implying frequent neuronal activity) in some specific nodes, as a similar behaviour as the Griffiths phases in epidemic models \cite{moretti2013griffiths}, both stemming from the structural heterogeneity.
This investigation helps controlling brain dynamics among states characterized by the activation of various cognitive systems \cite{gu2017optimal}, e.g., alleviating the symptoms of epilepsy by decreasing the congestion centrality of the seizure focus. 
%Appendix \ref{sec:application_congestion} provides a heuristic greedy algorithm for minimizing the density of some specific nodes via link removals.

\section{Discussion and Summary}
In this work, we proposed and studied a traffic model on networks considering a generalized flow-density relation and the heterogeneity of the flow network. 
Various phase transitions, i.e., the linear phase, the localization phase and the saturation phase have been observed in general networks with respect to the external injections, which exhibit distinct behaviors.
Both the flow network and the injections in real-work transport systems are of a high heterogeneity, which promote the existence of the localization phase.
Thus, the coexistence of high-density nodes and low-density nodes, manifested by the spatial localization of congested nodes, could be a generic phenomenon and merits a deeper investigation.

Quenched and topological disorder, stemming from topological heterogeneity, has been shown to induce novel behavior on non-equilibrium dynamics on networks, e.g., the Griffiths phases in contact process and epidemics on networks \cite{munoz2010griffiths}.
%which usually occurs around the spreading threshold for the transcritical bifurcation in a mean-field \cite{goltsev2012localization}. 
The localization incurred by the interplay between nonlinear conductivity and disorder has also been witnessed in fracture processes \cite{andrade2009fracturing}.
%Similar with the previous localization due to topological rare regions effect, the phase transition in our model also occurs accompanying with a bifurcation state.
Specifically, the localization in our traffic model is induced by the trade-off in interaction between neighbors against the external influences.
The non-linearity of the resistance allows the rebalance between these two effects on nodes due to the heterogeneity of topology and injections, which introduces a special equilibrium, and the congested localization emerges. %The localization incurred by the interplay between nonlinear conductivity and disorder has also been witnessed in a fracture \cite{andrade2009fracturing}.
%The dominance between these two effects on nodes switches non-simultaneously due to the heterogeneity of topology and injections, which introduces a special equilibrium, and the congested localization emerges. \cite{andrade2009fracturing}
%todo
%The congested nodes in the localization phase then interplay little with its neighbour and present an ``Meissner effect" \cite{parmeggiani2004totally}, i.e. change little with respect to the increasing external injections.  

Another important contribution of our work is the proposed congestion centrality that depends upon the flow network topology and the external injections. 
We observed that the onset of nodal congestion occurs if the inflows of some nodes are large enough so as fail to be balanced by the limited maximum outflow $h_i$ determined by the flow-density relation. The congestion centrality plays an important role in the congestion threshold, identification of congested nodes and estimating congestion propagation velocity. The tunable attributes in the resistance, describing the local movement behaviour, allow traffic optimization by the improvement of regional traffic capability. 
Previous research on transport networks usually assumes the all-or-nothing traffic assignment with the shortest path routes, which emphasis the nodes with a high betweenness as the congested nodes \cite{zhao2005onset}\cite{sole2016congestion}. Practical traffic under stochastic routing choices \cite{ccolak2016understanding} is more akin to driven diffusion processes \cite{schmittmann1998driven}, which implies a promising significance of our proposed model for identifying congested nodes.
%具有一定的独立性，并非cascading failure
\begin{comment}
We discussed some practical applications of our further work. The problem of targeted diffusion and congestion dissipation in large networks can be reduced to tractable programmings in terms of the congestion centrality.
Some questions merit further study. For example, what is the performance of the traffic with routing choice? The drivers usually choose less congested nodes as the downstream, which essentially corresponds to a density-dependent flow network. Also, a deeper investigation on the temporal behaviour of traffic processes could trigger the interest of dynamic optimization for a better traffic system.
\end{comment}

\begin{comment}
Impact of routing strategy

Non-linearity efforts memory effect. Non-linearity and heterogeneous leads to localization.

We focus on the process started from intial states, which empass the generation of congesess in rush hours. The method can easily applies to the recovery proceess in

After a birfurcation (sis transcritical birfurcation)	

interdependent and collective dynamic

Griffith phases and quenched disorder

Quenched disorder is well known to induce noval behavior in phase transitions both in equilibrium and away from it \cite{munoz2010griffiths}.

path-routing choosing

localization \cite{burda2009localization} \cite{marquie1995observation}
\end{comment}

\begin{appendices}
\section{Analogy with other models}\label{sec:analogy}
\textbf{Hydromechanics:} 
According to the Hagen–-Poiseuille law \cite{sutera1993history} which describes an incompressible and Newtonian fluid in laminar flow passing through a tube, the volumetric flow rate is equal to the pressure drop $ \Delta p_i$ along the tube $i$ divided by the resistance $r_i$ to flow.  
The dynamic viscosity $\eta_i$ with respect to the density $\rho_i$ of liquid, e.g. LST leavy liquid \cite{caffrey2013use}, can be generalized by a non-linear function as
\begin{align}
\eta_i = a +b\rho_i^\gamma
\end{align} 
where $a>0$, $b>0$, $\gamma>0$ are fitting constants.
The resistance $r_i$ to flow in turn is directly proportional to the dynamic viscosity $\eta_i$, i.e., $r_i\sim\eta_i$, which can be regarded as $r_i=\eta_i=a +b\rho_i^\gamma$ for simplicity.
Assuming the flow approximates to be incompressible and Newtonian in a small time interval, the mass flow follows
\begin{align}\label{equ:hydromechanics}
h_{ij} = \rho_i\frac{ \Delta p_i}{r_i} = \rho_i\frac{ \Delta p_i}{a +b\rho_i^\gamma} 
\end{align}
Considering that the demanded transition probability defined by $P$ in Section 2 plays a similar role as the external driven force for the pressure drop $ \Delta p_i$ along the tube, the mass flow in \eqref{equ:hydromechanics} has a similar form as the proposed flow $h_{ij}$ specified by \eqref{equ:flow_density} and \eqref{equ:flow_density_relation}.
\begin{comment}
The external pressure in our model is determined by the transit fraction $ p_{ij}$ in the flow transition matrix $P$, which is countered by the pressure from the downstream (depending on the downstream density $\rho_j$), i.e. the pressure drop $ \Delta p_{ij} =  p_{ij}(1-\rho_j^\kappa)$, where the constant $\kappa$ tunes the effect of the downstream density $\rho_j$.
The resistance to flow in turn is directly proportional to the dynamic viscosity $\eta$. The dynamic viscosity $\eta$ with respect to the density $\rho$ of liquid, e.g. LST leavy liquid \cite{caffrey2013use}, can be generalized by a non-linear function as
\begin{align}
 r_{ij}\propto\eta = a +b\rho_i^\gamma
\end{align} 
where $a>0$, $b>0$, $\gamma>0$ are fitting constants.
Assuming the flow approximates to be incompressible and Newtonian in a small time interval, the mass flow obeys the relation
\begin{align}
h_{ij} \propto \rho_i\frac{ \Delta p_{ij}}{r_{ij}} = \rho_i\frac{ p_{ij}(1-\rho_j^\kappa)}{a +b\rho_i^\gamma} 
\end{align}
which has a similar form with the proposed flow-density relation \eqref{equ:flow_density}.
\end{comment}

\textbf{Exclusion processes:} 
Our model has a similar
 implication as the totally asymmetric simple exclusion process (TASEP) with Langmuir kinetics \cite{parmeggiani2004totally} \cite{tsuzuki2018effect}.
The TASEP is defined on a one-dimensional lattice of size $L$. Each site can be occupied by at most one particle. The particle at site $i$ can hop to the site $i+1$ with a rate, provided the target site $i+1$ is empty. Particles can enter the lattice from the external reservoir at the empty site 1 with a rate, leave the lattice from site $L$ with another rate. Langmuir kinetics defines an independent attachment and detachment process for each site.
The TASEP characterizes the occurrence of boundary-induced phases transitions between nonequilibrium stationary states. A domain wall is a discontinuity over a microscopic regime in the lattice connecting a low-density state to a high-density state \cite{popkov2003localization}, which can be explained by a phenomenological domain--wall theory \cite{kolomeisky1998phase}

Although the phase diagram and steady states of the TASEP in one-dimensional lattice have been investigated intensively, the behaviour of exclusion processes, though confined on some simplified networks, have been considered only in a few works \cite{neri2011totally}\cite{denisov2015totally}\cite{raguin2013role}.
The generalization of our proposed model for traffics encompasses two aspects: (a) The transportation between two nodes can be asymmetrically bi-directed, (i.e., the single-directed link, corresponding to the definition of TASEP, can be regarded as a special case.); (b) The flow network can be of a general heterogeneous topology, and the external injection and exiting rates on nodes can also be different, which is coincident with the real-world situations that sources and destinations are only located in some specific regions.

\textbf{Epidemic spreading:} 

If the velocity $v_i(t)$ reduces to a constant instead of a function of the node density, i.e., $\gamma=0$, the term $\sum_{j=1}^N  p_{ji}(1-x_i)x_jv_j$ describes an spreading process \cite{pastor2015epidemic} with the infection rate $ p_{ji}v_j$. The model \eqref{equ:model_vector_form} can be regarded as two competing Susceptible-Infected (SI) processes combining with spontaneous birth (as the term $J_{in}$) and death (as the term $J_{out}$) processes. This similarity is related to the reason why some epidemics model can roughly describe the process of congestion propagation verified by real data \cite{saberi2019simple}.

\section{Analysis of the bi-tank model}
\subsection{Phase transition in the bi-tank model}\label{sec:bitank_detail}
\begin{figure}[ht]
	\centering
	\subfloat[$ x=1.2$]
	{\includegraphics[width=0.2\textheight]{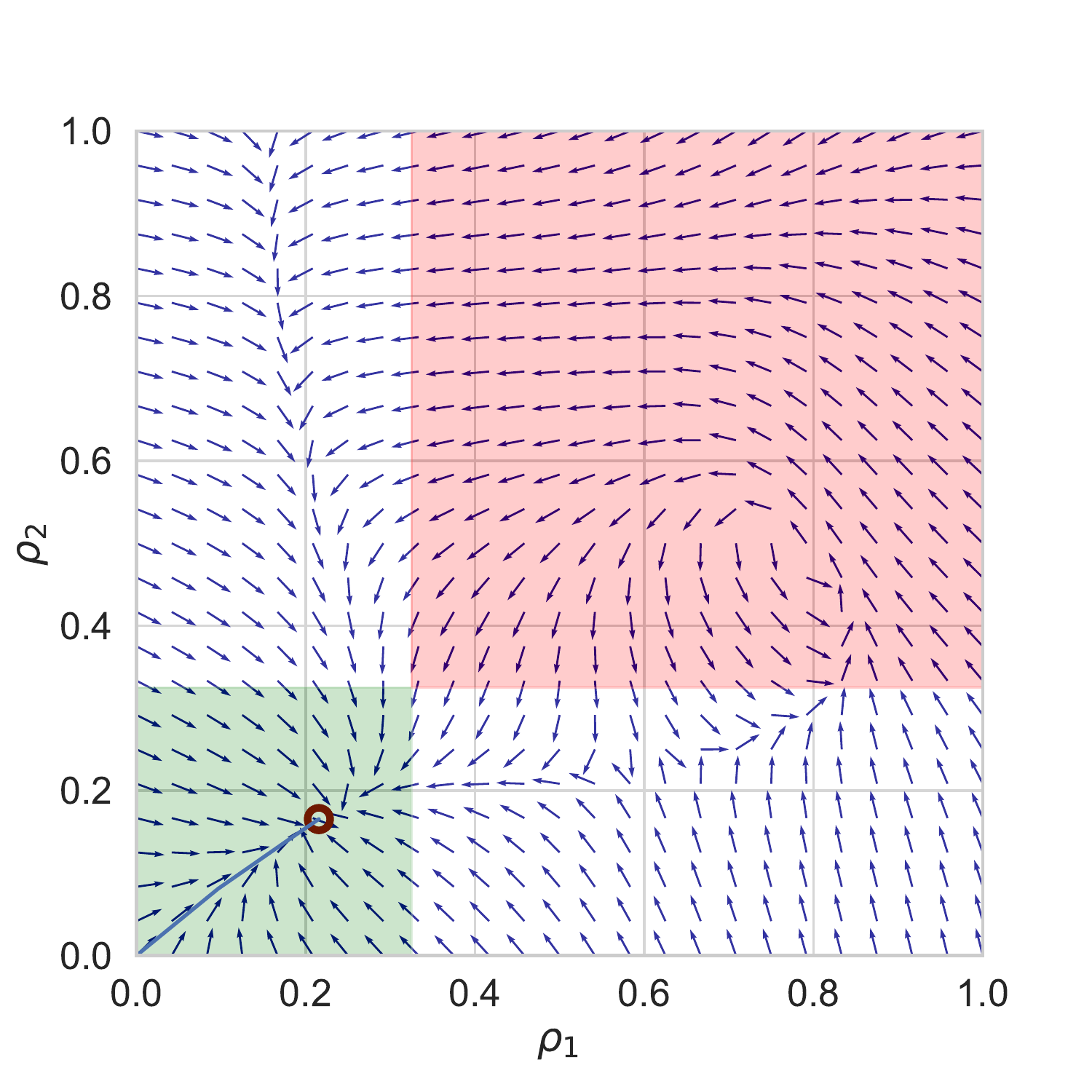}} 
	\subfloat[$ x=1.7$]
	{\includegraphics[width=0.2\textheight]{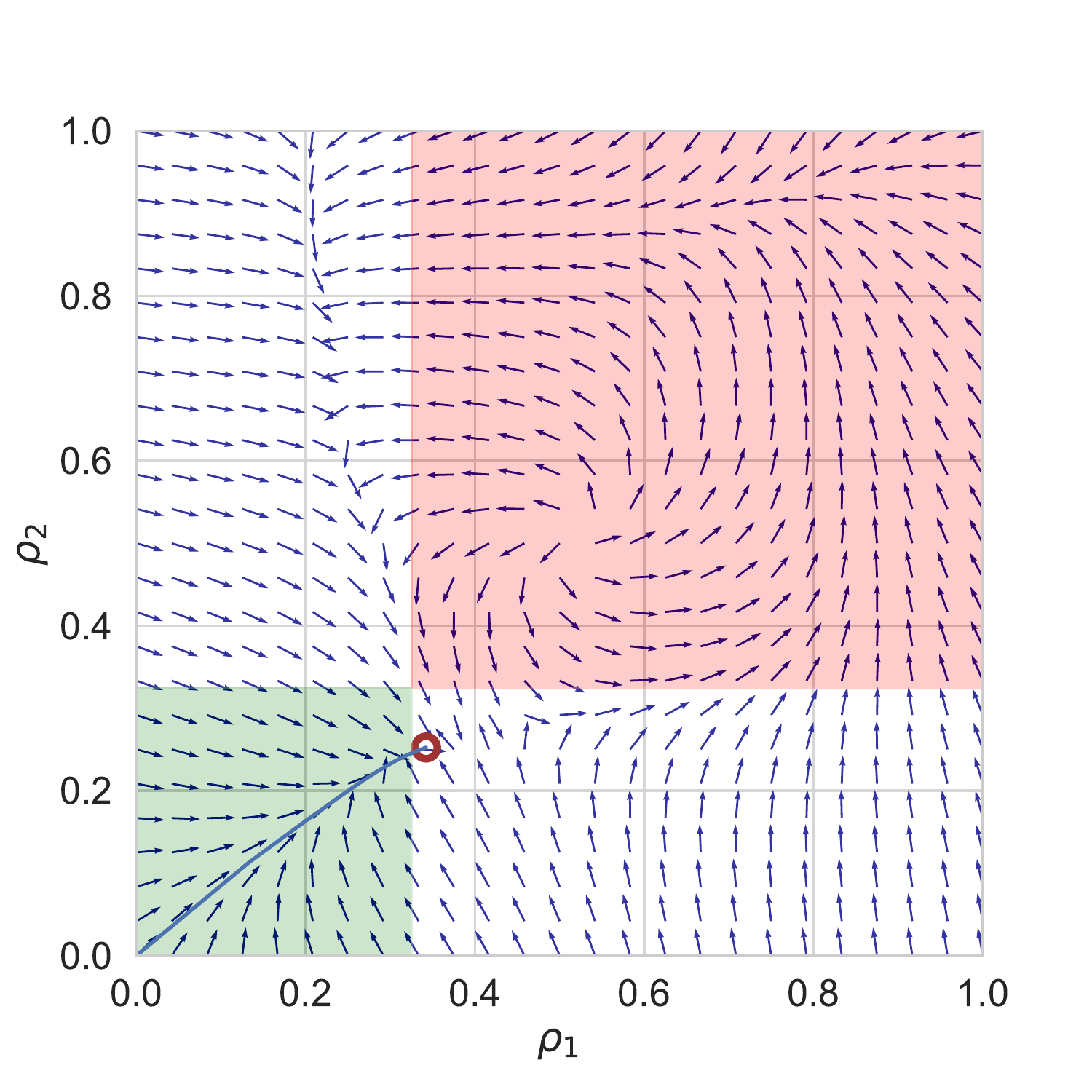}} 
	\subfloat[$ x=1.8$]
	{\includegraphics[width=0.2\textheight]{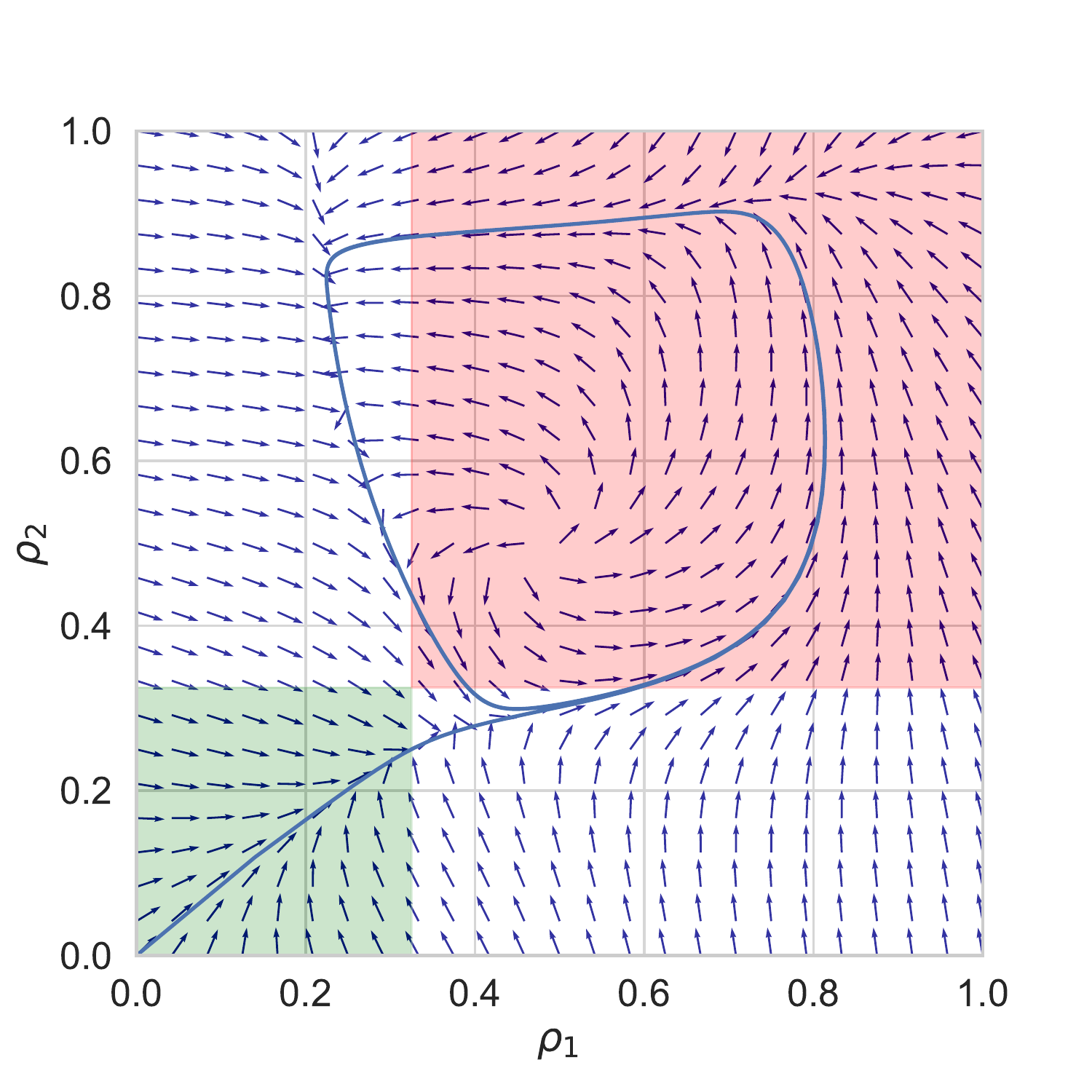}} 
	\\
	\subfloat[$ x=1.9$]
	{\includegraphics[width=0.2\textheight]{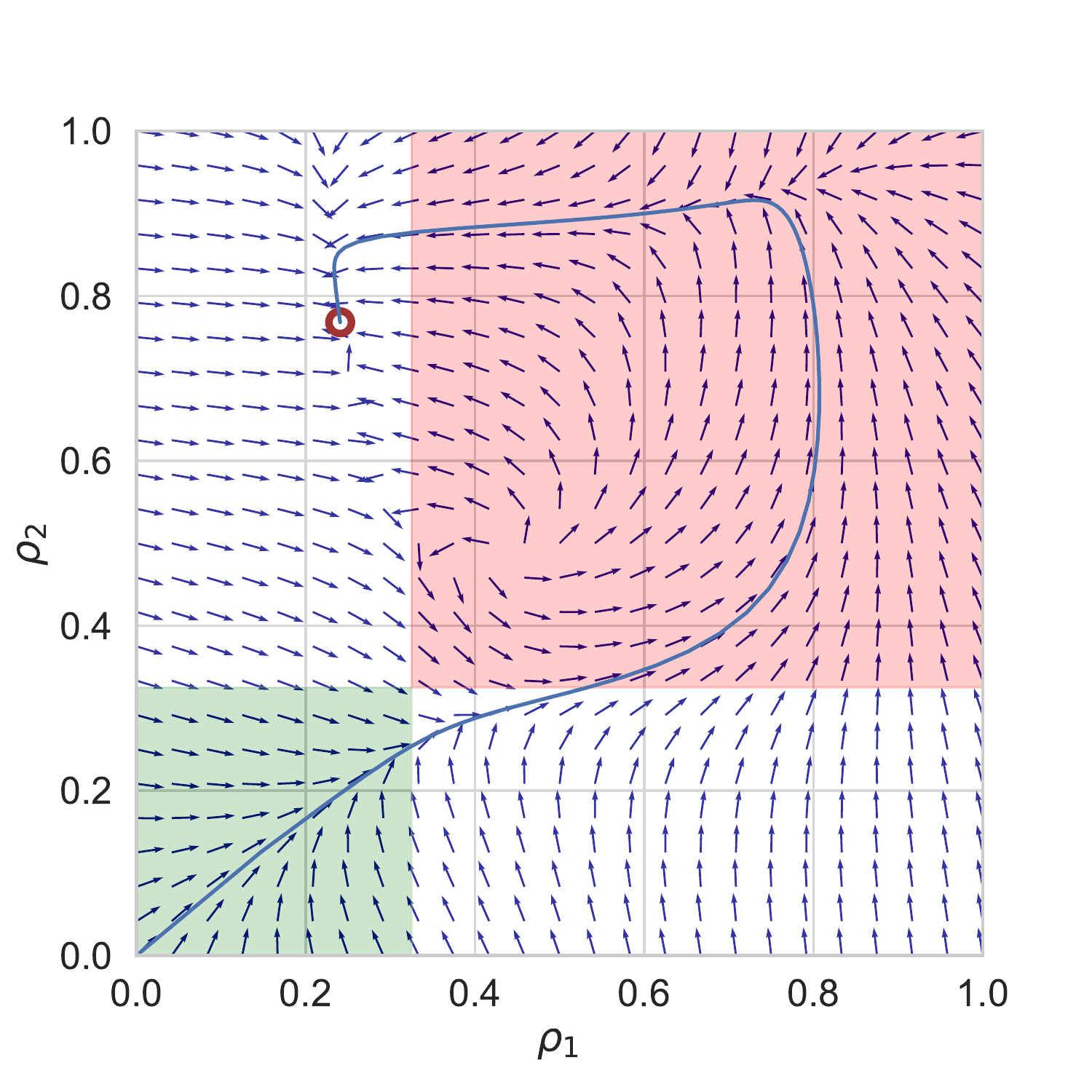}} 
	\subfloat[$ x=2.0$]
	{\includegraphics[width=0.2\textheight]{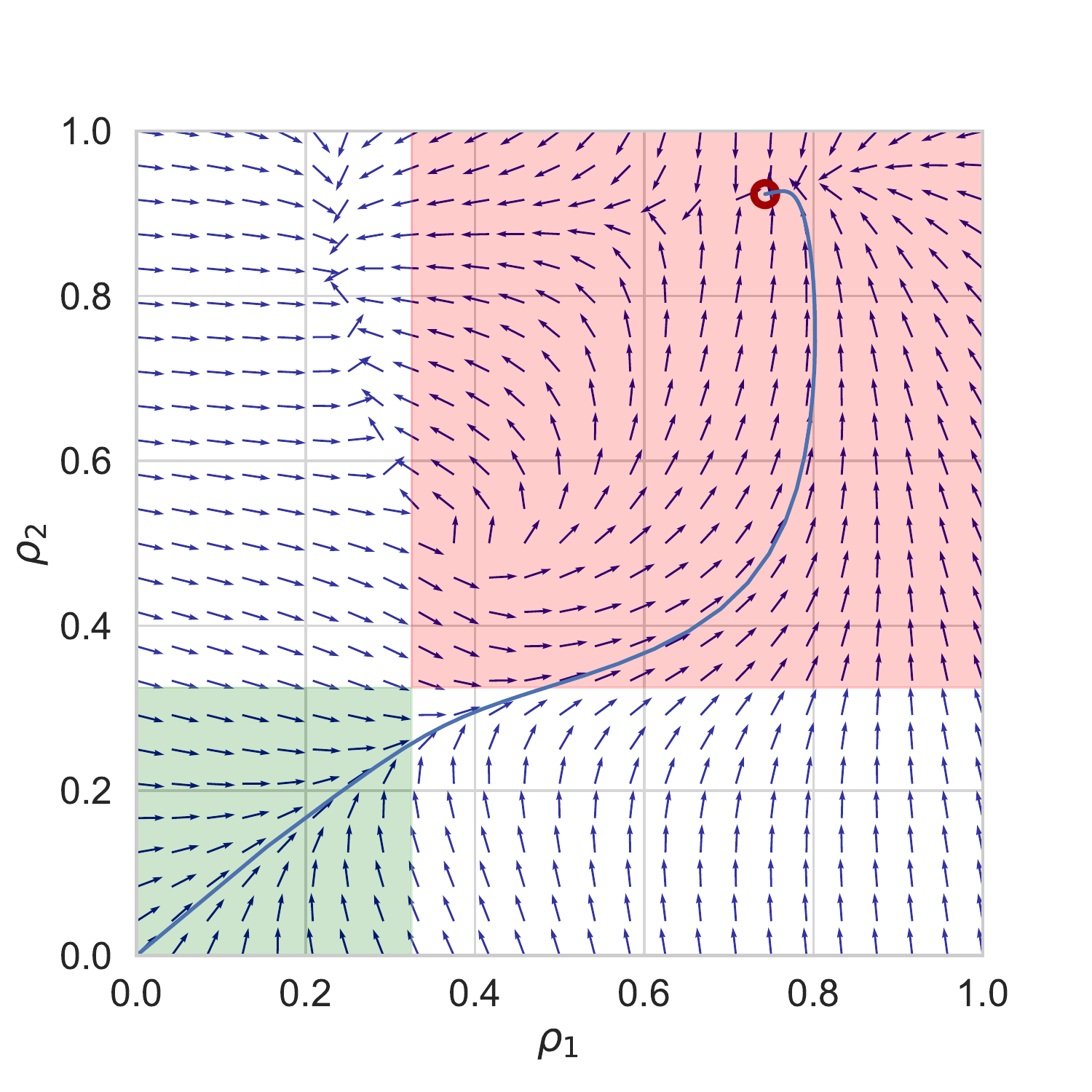}} 
	\subfloat[$ x=2.4$]
	{\includegraphics[width=0.2\textheight]{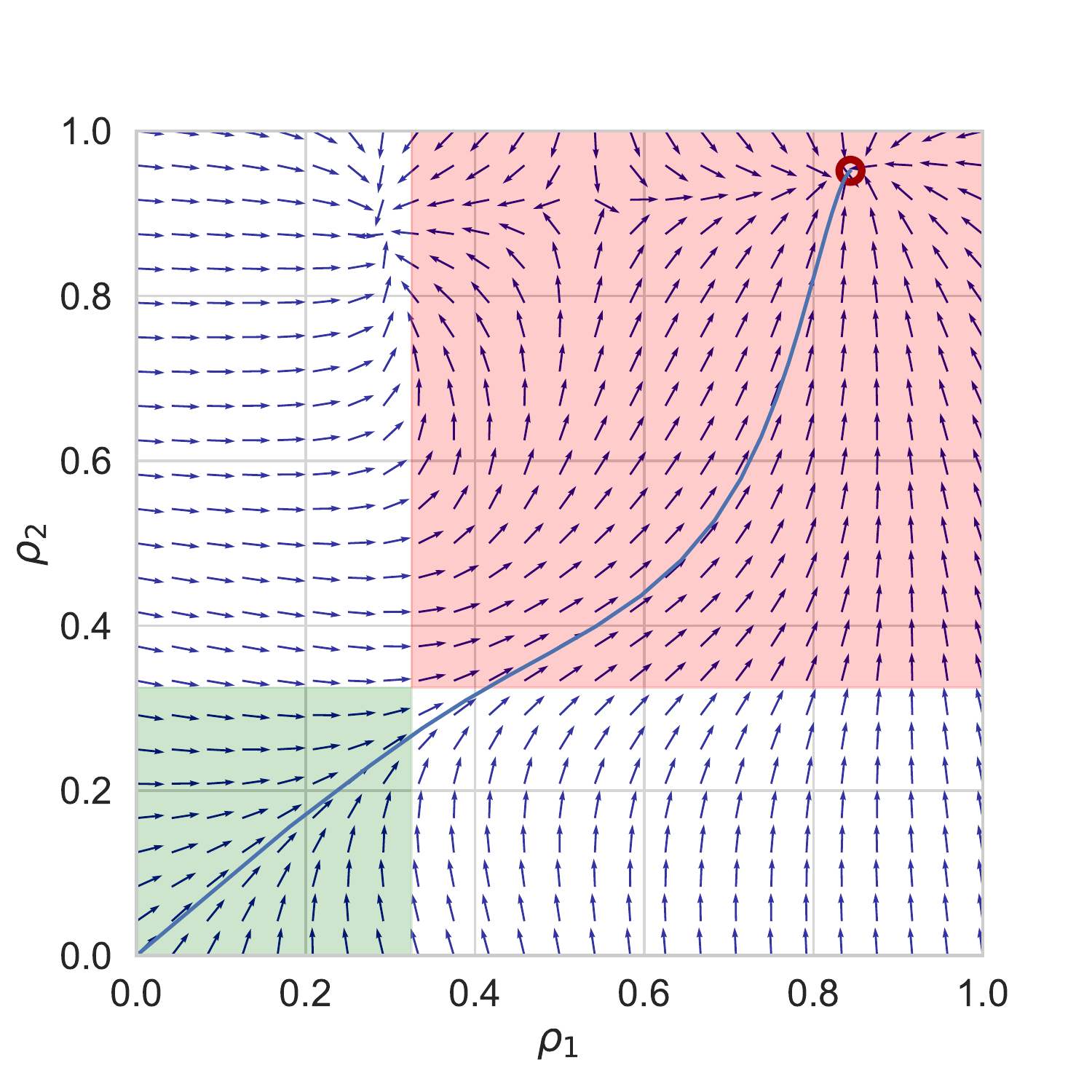}}
	\caption{Phase portraits with respect to the increasing homogeneous injection rate $ x$ in the bi-tank model. The red marker represents the equilibrium point, and the blue line represents the trajectory starting from the all-empty state. The equilibrium transits with the increasing injection rate $ x$: (a) and (b) illustrate the linear phase, (c) illustrates the bifurcation phase, (d) illustrates the localization phase, (e) and (f) illustrate the saturation phase.  }
	\label{fig:2binmodel_phaseplane}
\end{figure}

Figure \ref{fig:2binmodel_phaseplane} illustrates the phase portraits with respect to the increasing injection rate $ x$ in the bi-tank model. We hereby approximate the equilibrium point in the localization phase.
The equilibrium of the density $\rho_{1\infty}$ is reached when the state of node 2 is almost saturated, i.e., $\rho_{2\infty}\rightarrow 1$ and $h_{2\infty}\rightarrow \frac{1}{a+b}$. Invoking $h_{1\infty}\approx \frac{\rho_{1\infty}}{a}$ for a small density $\rho_{1\infty}$, we obtain from \eqref{equ:main_mf_model} that 
\begin{align}
\frac{d\rho_{1}}{dt}\bigg|_{\rho_{2\infty}=1} \approx  p_{21}(1-\rho_1)+ x(1-\rho_1)- \frac{q_1}{a}\rho_1 =0
\end{align}
which yields the density
\begin{align} \rho_{1\infty}\approx1-\frac{\frac{q_1}{a}}{\frac{ p_{21}}{a+b}+ x+\frac{q_1}{a}}
\end{align}

On the other side, the equilibrium of the density $\rho_{2\infty}$ is reached when the density of node 1 is very small (in the linear phase), i.e. $\rho_{1\infty}\rightarrow 0$ and $h_{1\infty}\rightarrow 0$. Then, the density $\rho_2$ follows from \eqref{equ:main_mf_model} as
\begin{align}\label{equ:local_rho2_fun}
\frac{d\rho_{2}}{dt}\bigg|_{\rho_{1\infty}=0} \approx - p_{21}h_2+ x(1-\rho_2)-q_2h_2 = 0
\end{align}
Further, we assume that $\rho_{2\infty}=1-c x^{-1}$ and $h_{2\infty}=\frac{1}{a+b}+kc x^{-1}$ where $c$ is a coefficient and $k$ is defined in \eqref{equ:horder_largebeta}. Equating the coefficient of the (constant) term $ x^0$ for the equation $ x(1-\rho_{2\infty})-h_{2\infty}=0$ by reducing \eqref{equ:local_rho2_fun}, we arrive the coefficient $c=\frac{1}{a+b}$. Thus, we can obtain the density $\rho_{2\infty}$ approximates $\rho_{2\infty}\approx1-\frac{1}{a+b} x^{-1}$.

\subsection{Existence of the localization phase in the bi-tank model}\label{sec:existence}
\begin{figure}[htp]
	\centering
	\subfloat[$\gamma=1$]
	{\includegraphics[width=0.22\textheight]{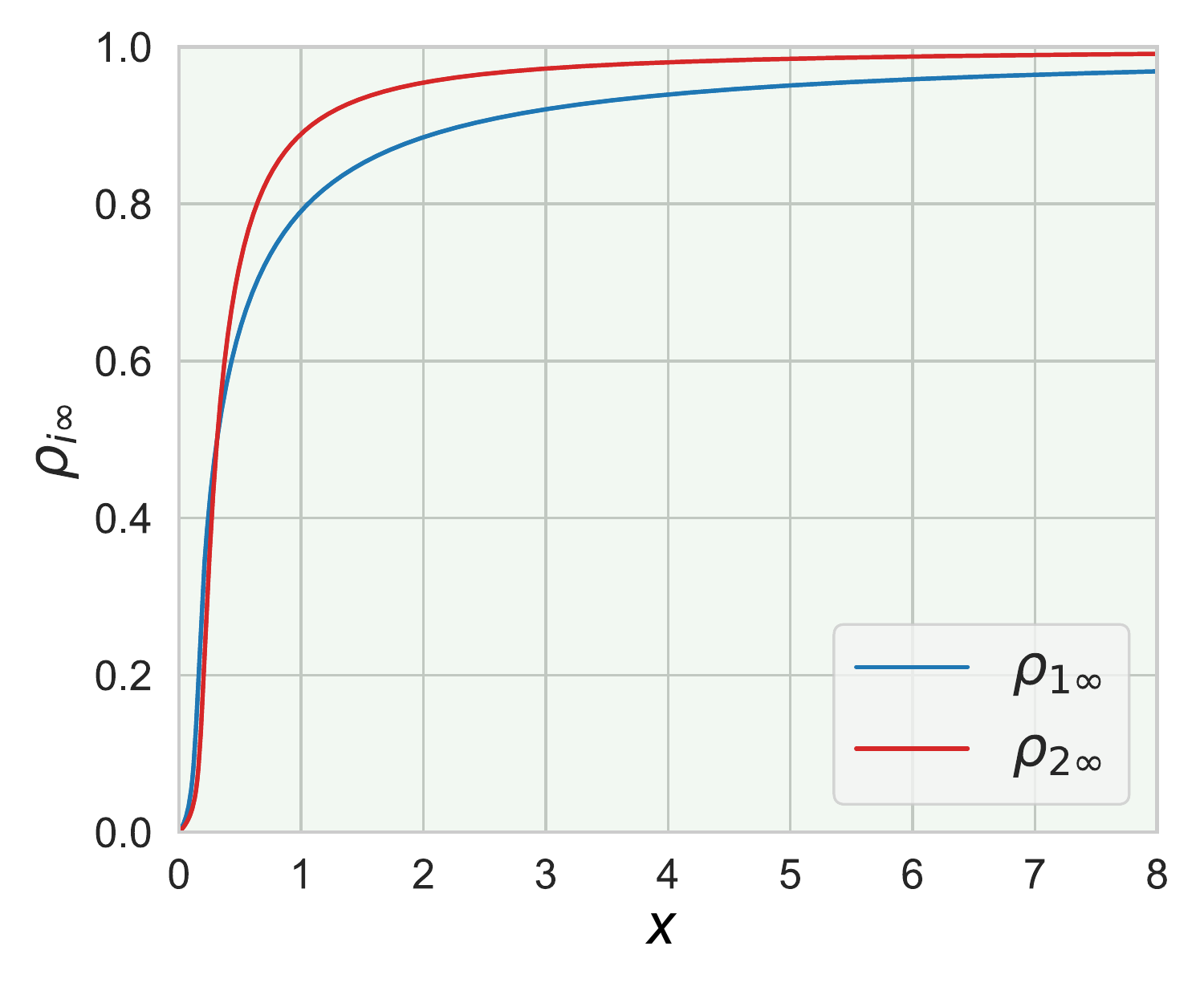}}
	\subfloat[$\gamma=2$]
	{\includegraphics[width=0.22\textheight]{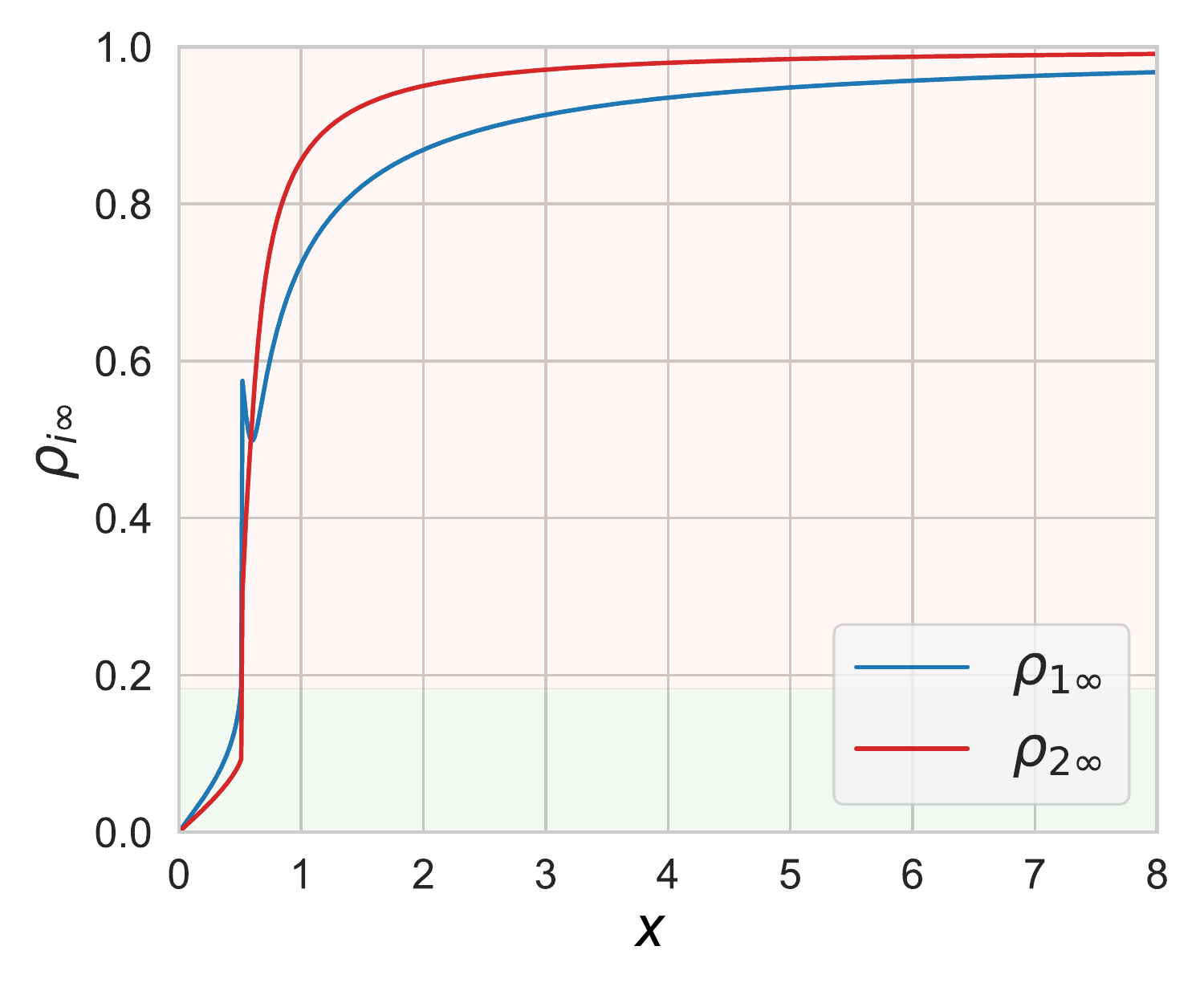}} 
	\\
	\subfloat[$\gamma=6$]
	{\includegraphics[width=0.22\textheight]{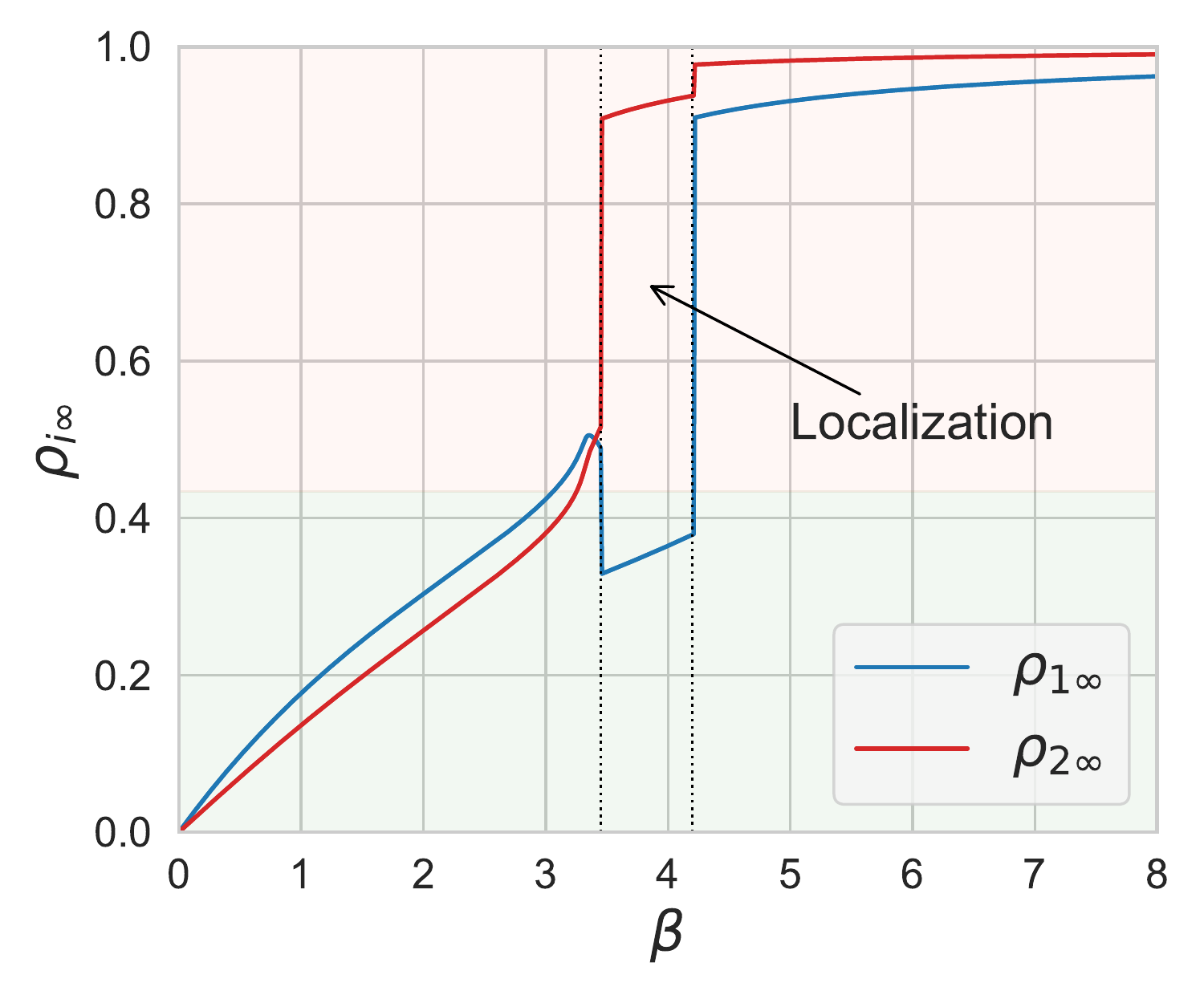}} 
	\subfloat[$\gamma=8$]
	{\includegraphics[width=0.22\textheight]{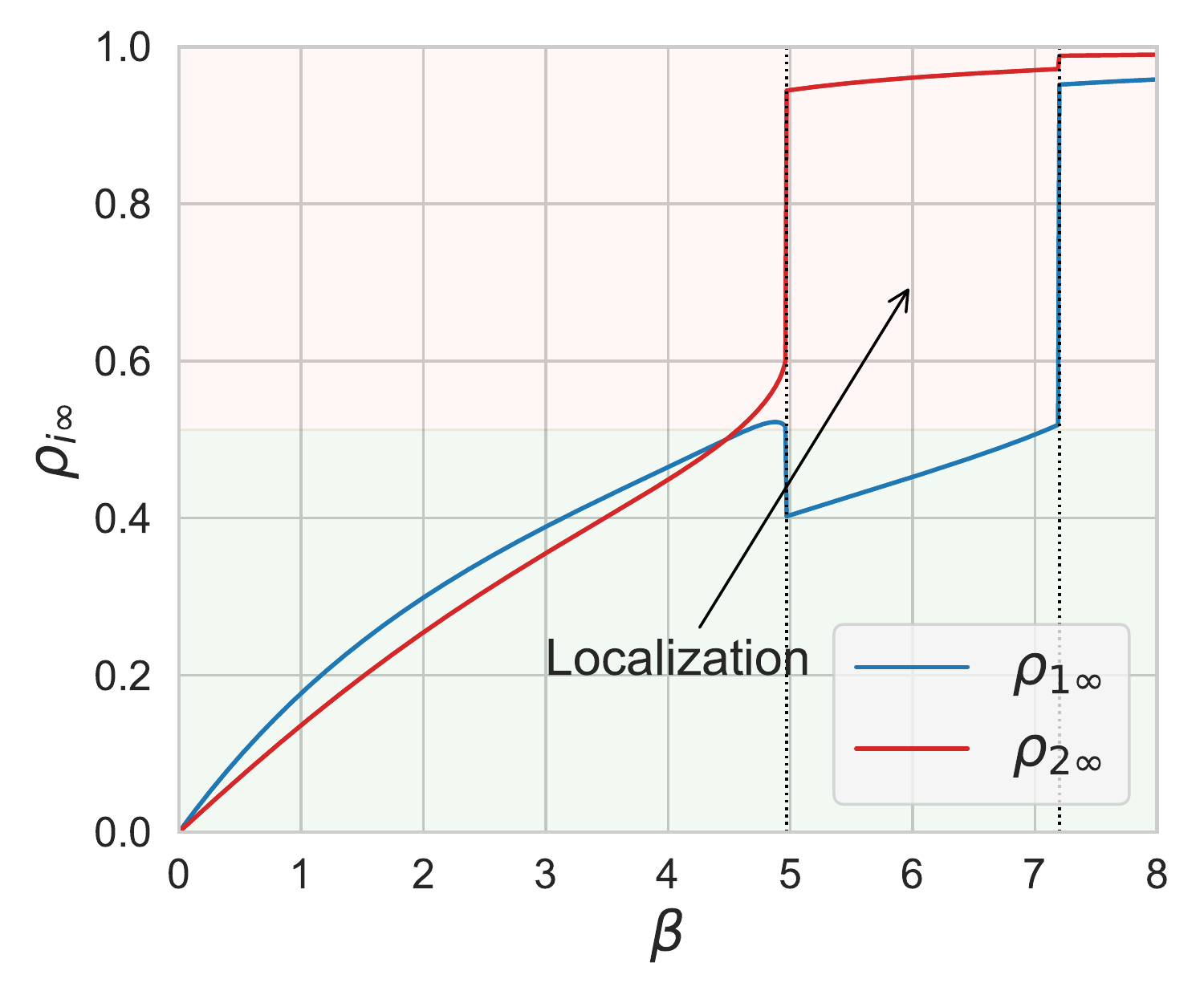}} 
	\caption{The node densities $\bm{\rho}_{\infty}$ in the bi-tank model as a function of the homogeneous injection rate $ x$ for different parameter $\gamma$ in the resistance \eqref{equ:resistance_fun}.}
	\label{fig:bitank_gamma}
\end{figure}

\begin{figure}[!htp]
	\centering
	\subfloat[Flow-density relations]
	{\includegraphics[width=0.23\textheight]{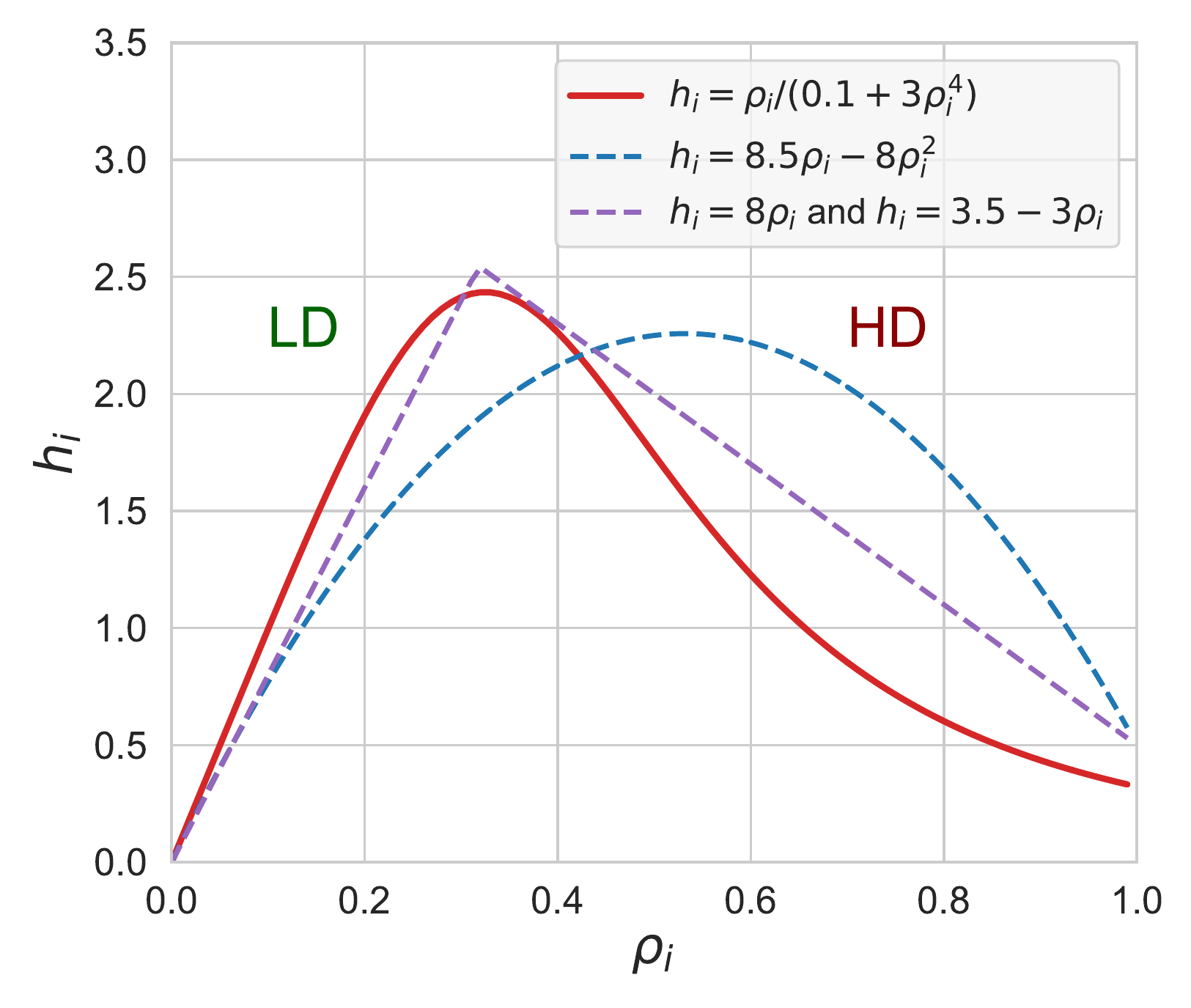}}
	\subfloat[Quadratic relation]
	{\includegraphics[width=0.23\textheight]{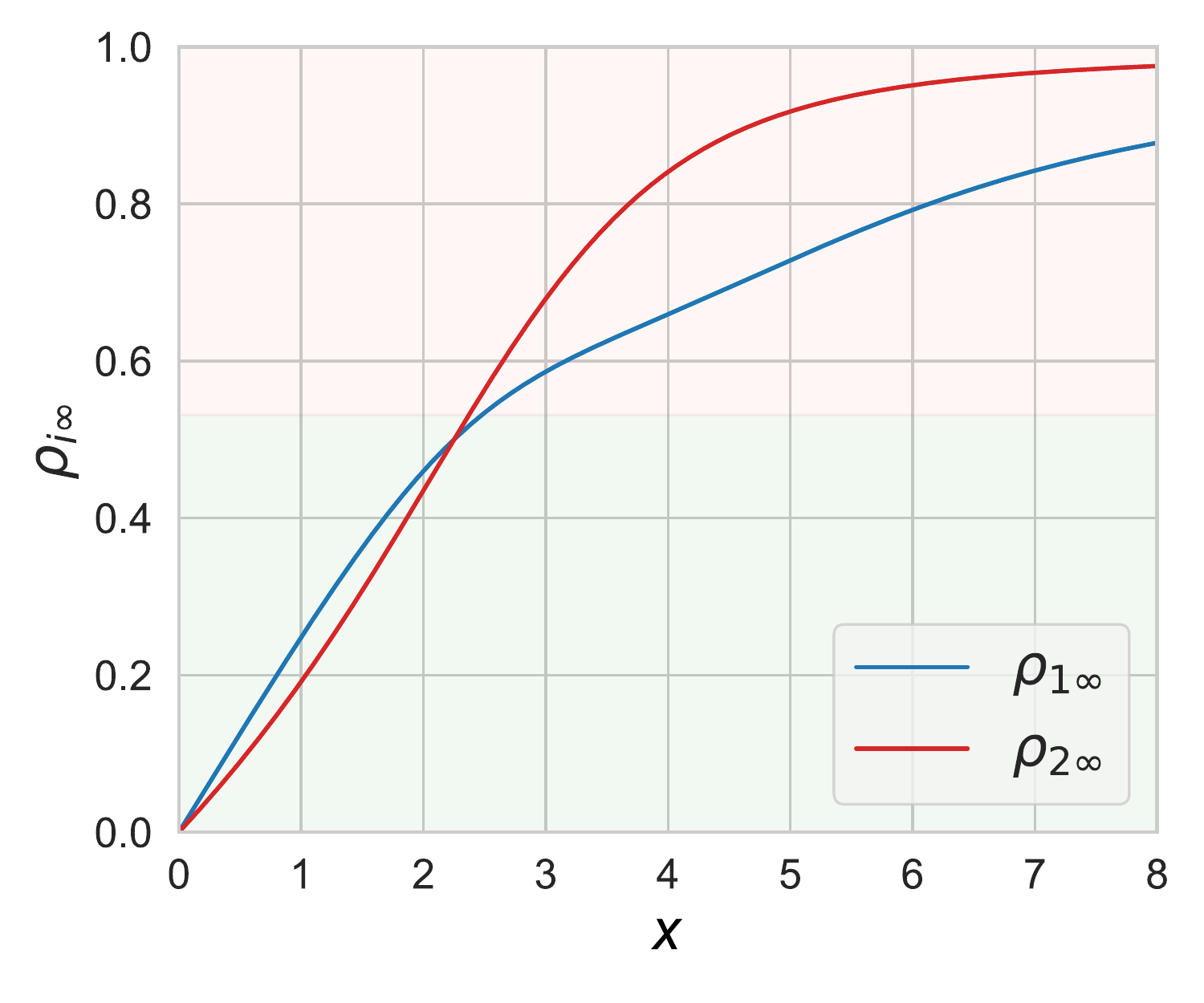}}
	\subfloat[Piecewise relation]
	{\includegraphics[width=0.23\textheight]{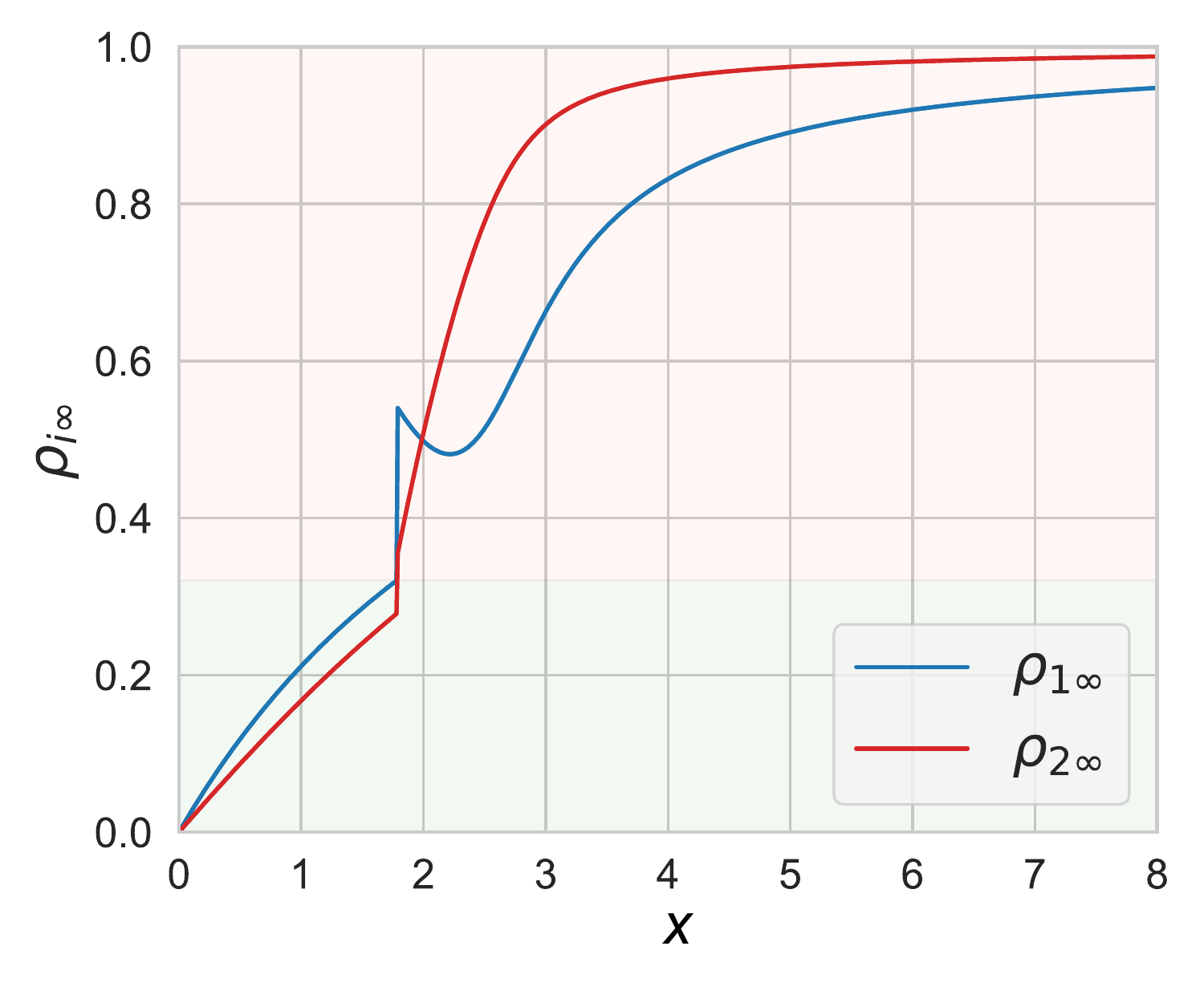}} 
	\caption{The steady-state node densities $\bm{\rho}_{\infty}$ in the bi-tank model as a function of the homogeneous injection rate $ x$. (a) Illustration of different flow-density relation.  (b) Phase diagram for a quadratic flow-density relation $h_i = 8.5\rho_i-8\rho_i^2$ with $\rho_c=0.53$. (c) T Phase diagram for a piece-wise flow-density relation $h_i = 8\rho_i$ for $\rho_i\leq0.32$ and $h_i = 3.5-3\rho_i$ for $\rho_i\geq0.32$, with $\rho_c=0.32$.}
	\label{fig:2bin_quadratic}
\end{figure}

We explore further the existence conditions of the localization phase in the bi-tank model.
Figure \ref{fig:bitank_gamma} shows the density in the bi-tank model as a function of the homogeneous injection rate $ x$ for different non-linearity of the resistance indicated by the constant $\gamma$. The localization phase exists only if the non-linearity indicated by $\gamma$ is large enough in a heterogeneous network. A larger non-linearity parameter $\gamma$ also enlarges the existence of the localization phase for a wider range of the external injection $ x$.
Figure \ref{fig:2bin_quadratic} shows that other forms of flow-density relation, e.g. the quadratic and the piecewise, cannot incur the localization phase, though the flow $h_i$ as a function of the density $\rho_i$ has a similar shape as the generalized flow $h_i(\rho_i)$ in \eqref{equ:flow_density_relation} and with a maximal $h_i$ at the critical density $\rho_c$. 

\begin{figure}[!htp]
	\centering
	\includegraphics[width=8cm]{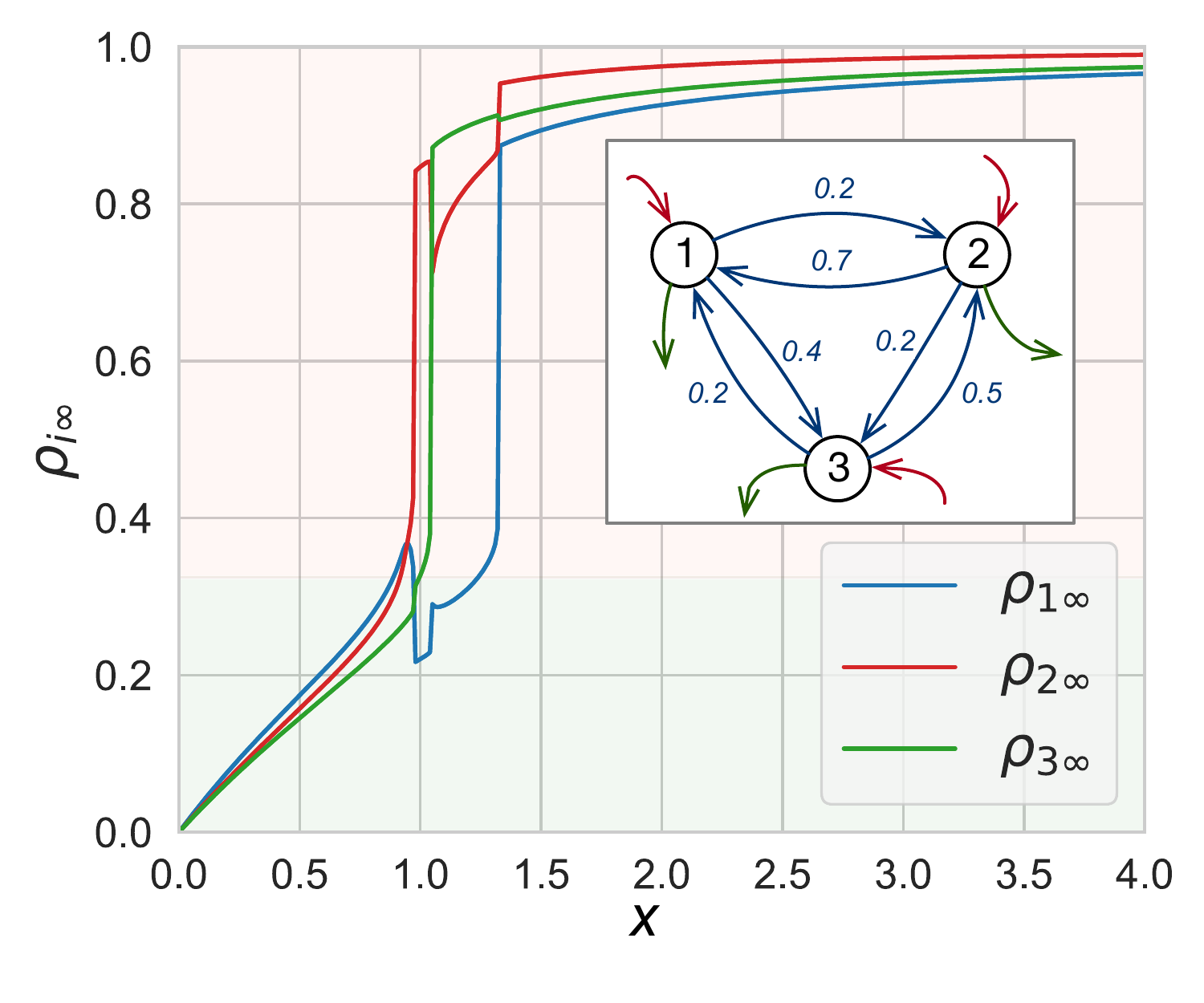}
	\caption{The node densities $\bm{\rho}_{\infty}$ in the bi-tank model as a function of the homogeneous injection rate $ x$. The embedded plot shows the flow transition network $P$.}
	\label{fig:3binphase}
\end{figure}

Figure \ref{fig:3binphase} illustrates the evolution of the node density with the increasing injection rate $ x$ in a tri-tank model with 3 nodes. The bifurcation phase does not exist. The localization starts when the node with the second-largest congestion centrality (node 2) enters the HD regime and the density of the node with the largest congestion centrality (node 1) inversely decreases a little. With the increasing injection rate $ x$, the third-largest congestion centrality (node 3) then shifts to the HD regime. Finally, all nodes enter the state of saturation if the injection rate $ x$ is large enough and the localization phase ends.

%todo 
\begin{comment}
one by on shift from the LD to the HD

correlation of congested nodes with topological properties

With the increasing of injection, the congestion can be spatially localized in some specific nodes.
$\rho_i>\rho_j$ $h_{ji}>h_{ij}$

Decreasing the injection or increasing the capacity of roads may increase the number of congested nodes. Block some path may be better, which can be regarded as Braess's Paradox.

Reverse flow - trap effect
\end{comment}

\section{Analysis for general networks}
\subsection{Analysis for the linear phase and the saturation phases}\label{sec:free_saturation}
Inspiring by two different equilibrium points close to $\bm{\rho}_{\infty} \rightarrow 0$ and $\bm{\rho}_{\infty} \rightarrow 1$, we apply a perturbation method to approach the densities in two steady states, i.e,  the linear phase and the saturation phase, respectively.
The node  density in the steady state $$\bm{\rho}_{\infty}= \bm{0}+C_1\bm{ x}+\mathcal{O}(\bm{ x}^2)$$ 
for a small external injections $\bm{ x}\rightarrow \bm{0}$, where $C_1$ is a coefficient matrix. Substituting the steady-state density $\rho_{\infty}$ and the corresponding steady-state flow $\bm{h}_{\infty} = \frac{1}{a}C_1\bm{ x}+\mathcal{O}(\bm{ x}^2)$ into the governing equation \eqref{equ:model_vector_form} reaches
\begin{align}\label{equ:free_purturbation}
P^T C_1\bm{ x} - diag(P\bm{u})C_1\bm{ x} +a\bm{ x}-diag(\bm{q})C_1\bm{ x} = \bm{0}
\end{align}
for the term $\bm{ x}$.
Following \eqref{equ:free_purturbation} and invoking the relation $\bm{q}+P\bm{u} = \bm{u}$, the condition $(diag(\bm{q}+P\bm{u})-P^T)C_1\bm{ x} = a\bm{ x}$ yields  the coefficient matrix $C_1 = a(I-P^T)^{\dag}$.

The density vector in the steady state is $\bm{\rho}_{\infty}= \bm{u}-C_2\bm{ x}^{-1}-\mathcal{O}(\bm{ x}^{-2})$ with the coefficient matrix $C_2$ for a large external injection $\bm{ x}\rightarrow\infty$. The corresponding flow vector $\bm{h}_{\infty}$ follows
\begin{align}\label{equ:horder_largebeta}
\bm{h}_{\infty} = \bm{h}|_{\bm{\rho}=\bm{1}}-\frac{d h_i(\rho_i)}{d\rho_i}\bigg|_{\rho_i=1}C_2{\bm{ x}_i^{-1}} -\mathcal{O}(\bm{ x}^{-2})  = \frac{1}{a+b}-kC_2\bm{ x}^{-1}-\mathcal{O}(\bm{ x}^{-2})
\end{align}
where $k:=\frac{d h_i(\rho_i)}{d\rho_i}\big|_{\rho_i=1}=-\frac{a+b(1-\gamma)}{(a+b)^2}$.
Substituting into the governing equation \eqref{equ:model_vector_form} and equating the coefficients of the term $\bm{ x}^0$ yields the coefficient matrix $C_2=\frac{1}{a+b}diag(\bm{q})$.
\begin{comment}
For a higher-order perturbation, we define 
\begin{align}
\bm{\rho}_{\infty} &= \bm{u}-C_2\bm{ x}^{-1}-C_3\bm{ x}^{-2}-\mathcal{O}(\bm{ x}^{-3}) \notag\\
\bm{h}_{\infty} &=\frac{1}{a+b}-kC_2\bm{ x}^{-1}--kC_3\bm{ x}^{-2} -\mathcal{O}(\bm{ x}^{-3}) \notag
\end{align}
and arrive the terms $\bm{ x}^1$ as
\end{comment}
%\begin{align}
%\frac{1}{a+b}A^T\bm{u}-\frac{1}{a+b}A\bm{u}-C_2\bm{u}-\frac{1}{a+b}\bm{q}=0
%\end{align}
%Invoking $A\bm{u}+\bm{q}=\bm{u}$, we arrive the coefficient matrix $C_2=\frac{1}{a+b}(A^T-I)$.

We can obtain that 
\begin{equation}\label{equ:density_bothphase}
\bm{\rho}_\infty=\left\{
\begin{aligned}
&a(I-P^T)^{\dag}\bm{ x}+\mathcal{O}(\bm{ x}^2), &\bm{ x}\rightarrow \bm{0}\\
&\bm{u}-\frac{1}{a+b}diag(\bm{q})\bm{ x}^{-1}-\mathcal{O}(\bm{ x}^{-2}) , & \bm{ x}\rightarrow \bm{\infty}
\end{aligned}
\right.
\end{equation}

\subsection{Steady state in homogeneous cases}\label{sec:homo_stability}
We consider the homogeneous cases, where the link weights $ p_{ij}$ incident to each node $i$ are the same, the external injection rate $ x_i$ and outputs $q_i$ are also the same for all nodes. The homogeneous cases are characterized by the processes on regular networks and infinity large lattices. 
The condition for the equilibrium reduces to $g(\rho_i) =  x_i(1-\rho_i)-q_i\rho_iv_i = 0$.
The derivative of $g(x_i)$ is
\begin{align}\label{equ:homo_case_model}
	\frac{d g(\rho_i)}{d\rho_i} = - x_i-q_i\frac{a+b(1-\gamma)\rho^\gamma}{(a+b\rho^\gamma)^2}
\end{align}
which is always negative if $\gamma \leq 1$.
Invoking $g(0) =  x_i>0$ and $g(1) = -\frac{q_i}{a+b} <0$, the model exists only one fix point, i.e., the solution of $g(\rho_i)=0$.

\begin{figure}[htp]
	\centering
	\includegraphics[width=9cm]{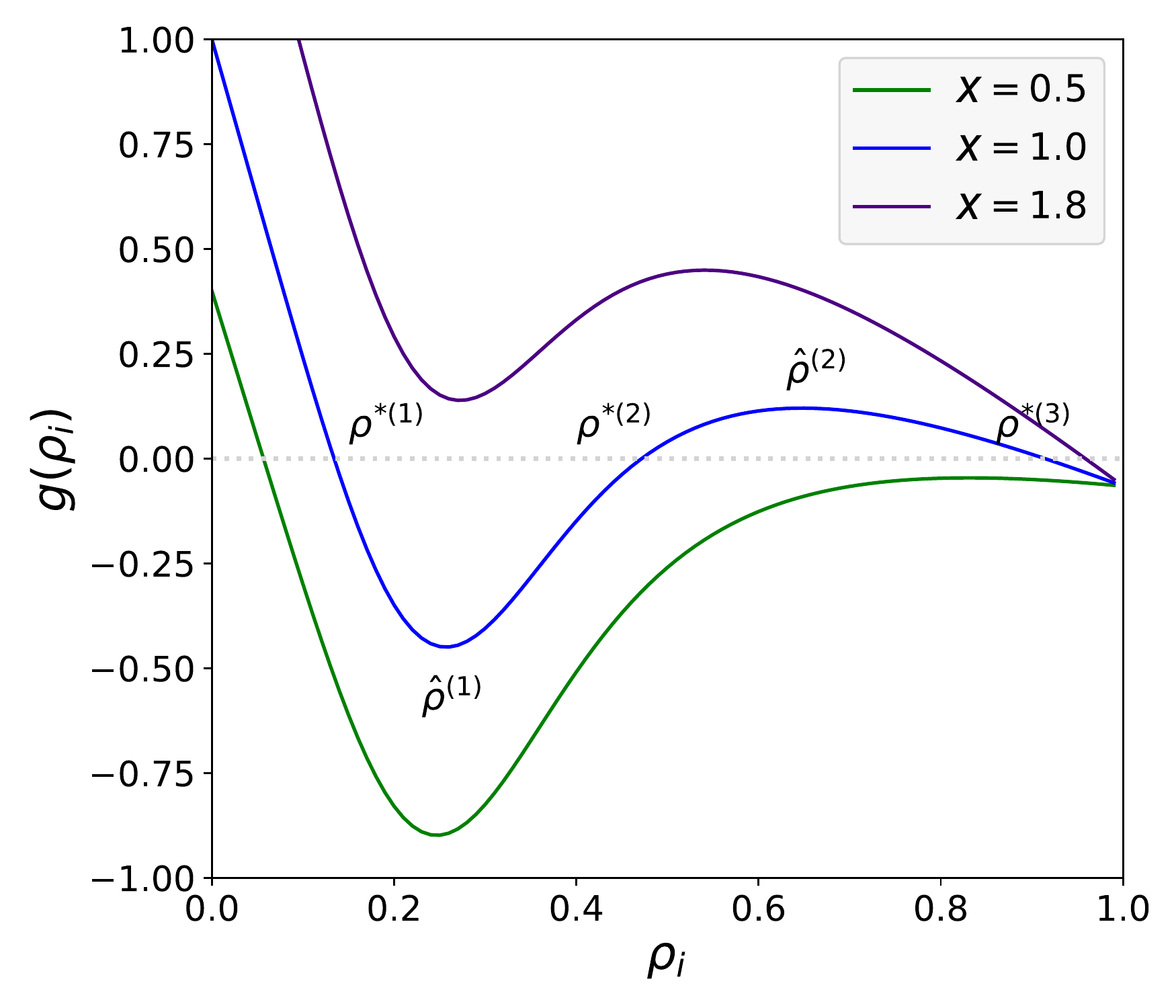}
	\caption{The function g($\rho_i$) in terms of the density $\rho_i$ in homogeneous cases.}
	\label{fig:homogenous}
\end{figure}

For $\gamma>1$, we can obtain at most two solutions for the equation $\frac{d g(\rho_i)}{d\rho_i}=0$, denoted by $0<\hat{\rho}^{(1)}<\hat{\rho}^{(2)}<1$. The condition that the value $g(\rho^{(1)})\leq0$ and $g(\rho^{(1)})\geq0$ implies at most three solutions for the equation $g(\rho_i)=0$, i.e., three fixed points, denoted by $\rho^{*(1)}\leq \rho^{*(2)}\leq \rho^{*(3)}$, as shown in Figure \ref{fig:homogenous}.
\begin{comment}
\begin{figure*}[t]
	\centering 
%	\subfloat[Karate]
%	{\includegraphics[width=5.8cm]{fig/unkarate.eps}}
%	\subfloat[Les Mis\'{e}rables]
%	{\includegraphics[width=5.8cm]{fig/lesmis.eps}}	
%	\subfloat[Netscience]
%	{\includegraphics[width=5.8cm]{fig/netscience.eps}}
	\includegraphics[width=5cm]{3sol.png}
	\caption{Function $g(\rho_i)$}
	\label{fig:undir_performance}
\end{figure*}
\end{comment}

Denoting $\rho^*$ a fixed point and substitute $\rho_i=\rho^*+u_i$ into the reduced model \eqref{equ:homo_case_model} yields the linearization of the system
\begin{align}
	\frac{d\rho_i}{dt} = u_i\frac{d g(\rho_i)}{d\rho_i}\bigg|_{\rho_i=\rho^*}
\end{align}
One can verify that $\frac{d g(\rho_i)}{d\rho_i}\big|_{\rho_i=\rho^{*(1)}}<0$ and $\frac{d g(\rho_i)}{d\rho_i}\big|_{\rho_i=\rho^{*(3)}}<0$, which implies that the system has two non-trivial equilibrium points of local stability, i.e., $\rho^{*(1)}$ and $\rho^{*(3)}$. The equilibrium points $\bm{\rho}^{*(1)}$ and $\bm{\rho}^{*(3)}$ are consistent with the density in \eqref{equ:density_bothphase}, but the phase coexistence cannot happen.

\begin{comment}
We further apply perturbation method to approach the density in the steady state. Substituting the expansion of the density in steady state $\rho_{i\infty}^{*(1)}= 0+c_1 x_i+\mathcal{O}( x^2)$ yields that $c_1=a/q_i$ and $\rho_{i\infty}^{*(1)}\approx \frac{a}{q_i} x_i$ for a small $ x_i\rightarrow0$. On the other side, Substituting the expansion of the density in steady state $\rho_{i\infty}^{*(3)}= 1-c_1 x_i^{-1}-\mathcal{O}( x^{-2})$ yields that $c_1=\frac{q_i}{a+b}$ and $\rho_{i\infty}^{*(3)}\approx 1-\frac{q_i}{(a+b) x_i}$ for a large $ x_i^{-1}\rightarrow0$.

The difference of two equilibrium states for the same $ x_i$, which indicates the hysteresis level can be approximated by 
\begin{align}
	 p\mathcal{H} = 1-\frac{q_i}{(a+b) x_i}-\frac{a}{q_i} x_i
\end{align}
which reaches a maximum at $ x_i=\frac{q_i}{\sqrt{a(a+b)}}$.
\end{comment}

\subsection{Estimation of the fraction of congestion nodes}\label{sec:est_chi_lb}
We recall that the node density is $\bm{\rho}_{-}\approx a(I-A^T)^{-1}\bm{ x}$ in the linear phase \eqref{equ:density_phase1}, and the congested nodes in the localization phase follows $\bm{\rho}_{+} \approx \bm{u}-\frac{1}{a+b}\bm{ x}^{-1}$ as \eqref{equ:localization_density}.
Assuming that most nodes are still in the free state and a small fraction of nodes are congested, the density vector can be represented by $\bm{\rho} = (I- \mathcal{E})\bm{\rho}_{-} + \mathcal{E}\bm{\rho}_{+}$, where $\mathcal{E}$ is a diagonal matrix with the entry $\mathcal{E}_{ii}\in\{0,1\}$ for $i\in\mathcal{N}$.
Correspondingly, the flow vector is $\bm{h} = (I- \mathcal{E})\bm{h}_{-} + \mathcal{E}\bm{h}_{+}$, where the flow $\bm{h}_{-} = \bm{h}(\bm{\rho}_{-})$ and $\bm{h}_{+} = \bm{h}(\bm{\rho}_{+})$ that can be approximated by \eqref{equ:density_bothphase}.

Following the governing equation \eqref{equ:model_vector_form} and summing up the derivatives of the density $\frac{d\rho_i(t)}{dt}$ of all nodes $i\in\mathcal{N}$, we obtain the necessary condition for the matrix $\mathcal{E}$ as
\begin{align}
\bm{ x}^T(\bm{u}-\bm{\rho}_{-})-\bm{q}^T\bm{h}_{-} 
&=\bm{ x}^T\mathcal{E}(\bm{\rho}_{+}-\bm{\rho}_{-})+\bm{q}^T\mathcal{E}(\bm{h}_{+}-\bm{h}_{-}) \notag\\
&\leq ||\mathcal{E}||_F\cdot|\bm{ x}^T(\bm{\rho}_{+}-\bm{\rho}_{-})+\bm{q}^T(\bm{h}_{+}-\bm{h}_{-})|
\end{align}
Thus, the fraction of congested nodes follows
\begin{align}
\chi \approx \frac{1}{N}||\mathcal{E}||_F^2\geq \frac{1}{N} \bigg|\frac{\bm{ x}^T(\bm{u}-\bm{\rho}_{-})-\bm{q}^T\bm{h}_{-} }{\bm{ x}^T(\bm{\rho}_{+}-\bm{\rho}_{-})+\bm{q}^T(\bm{h}_{+}-\bm{h}_{-})}\bigg|^2
\end{align}

\section{More numerical results in Lattice}
\subsection{Phase diagram for the average density in Lattice} \label{sec:phasediag_avgdensity}
We here present the phase diagrams of the average density $\langle\rho_\infty\rangle$ in a Lattice, which presents similar behaviors as the fraction of congested nodes $\chi$.
\begin{figure}[!htp]
	\centering
	\subfloat[$b$]
	{\includegraphics[width=0.18\textheight]{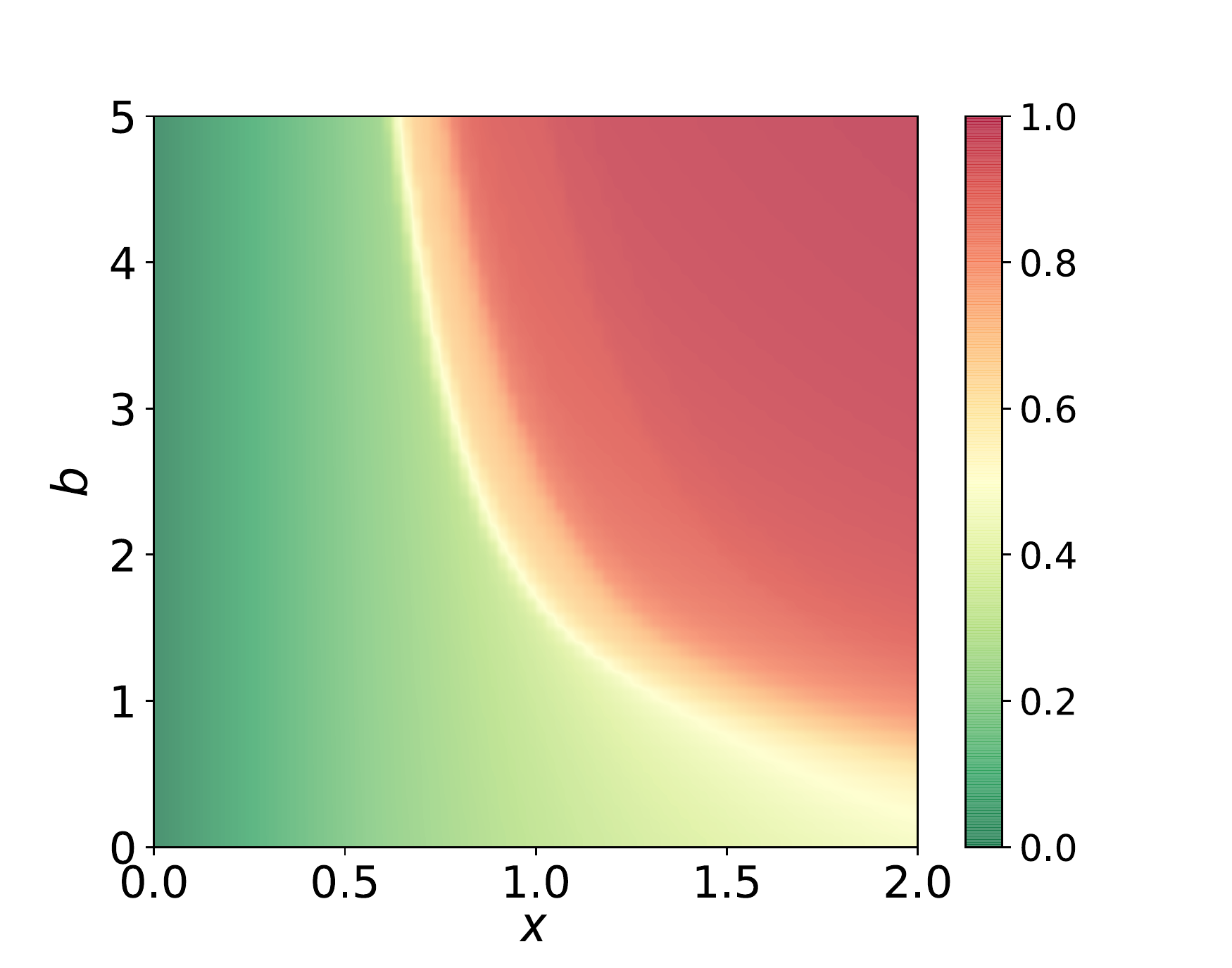}}
	\subfloat[$\gamma$]
	{\includegraphics[width=0.18\textheight]{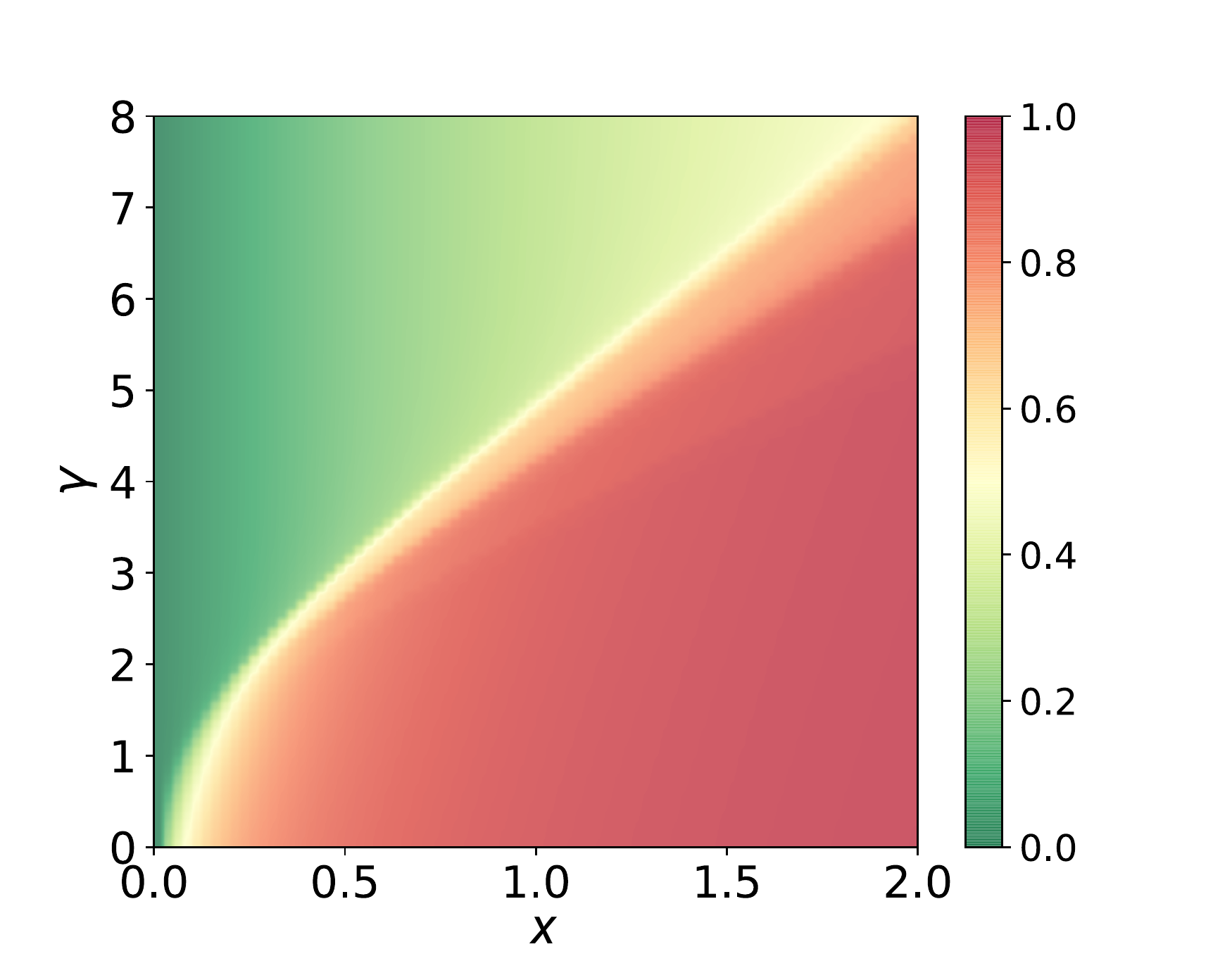}}
	\subfloat[$\sigma$]
	{\includegraphics[width=0.18\textheight]{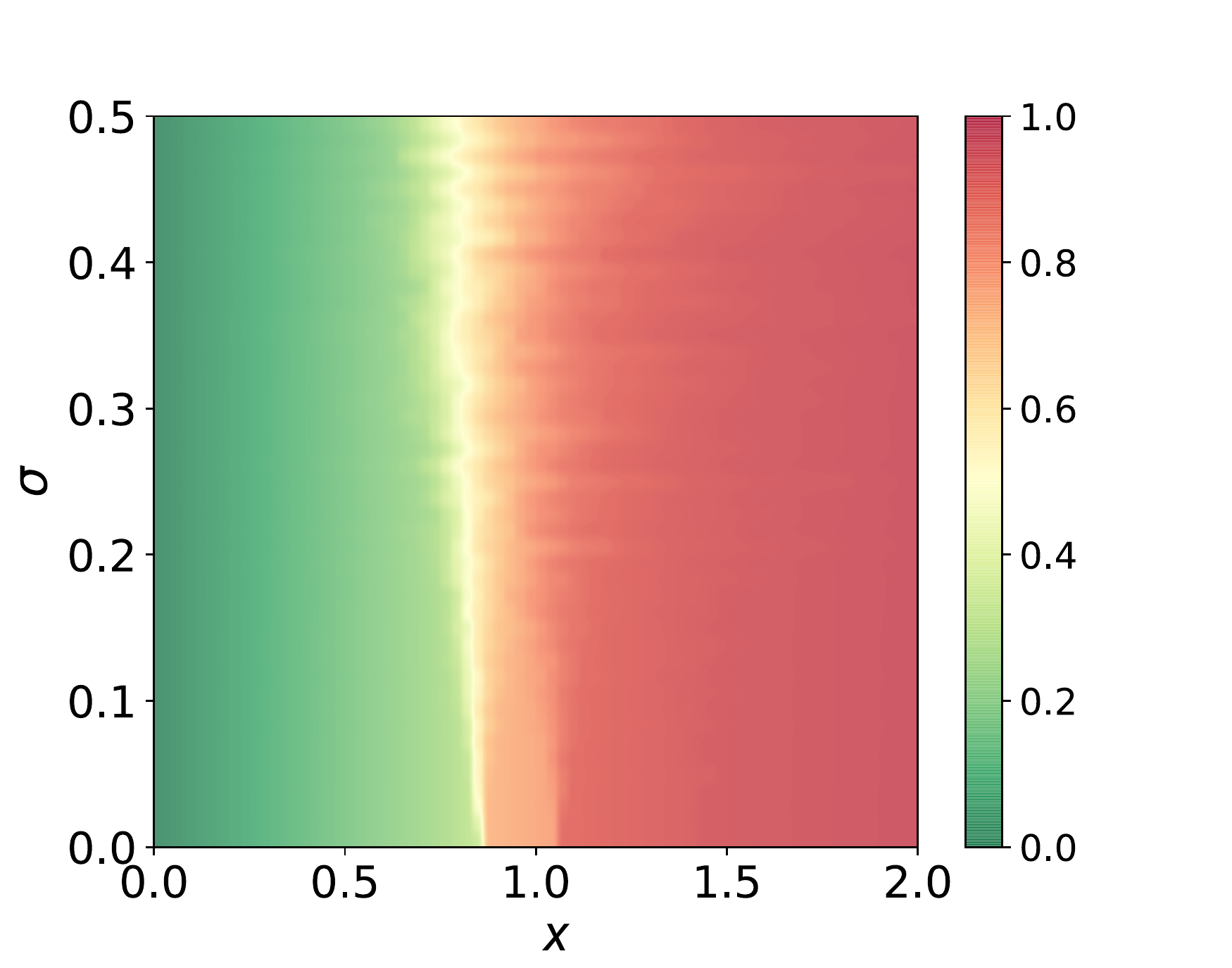}} 
	\subfloat[$\sigma_p$]
	{\includegraphics[width=0.18\textheight]{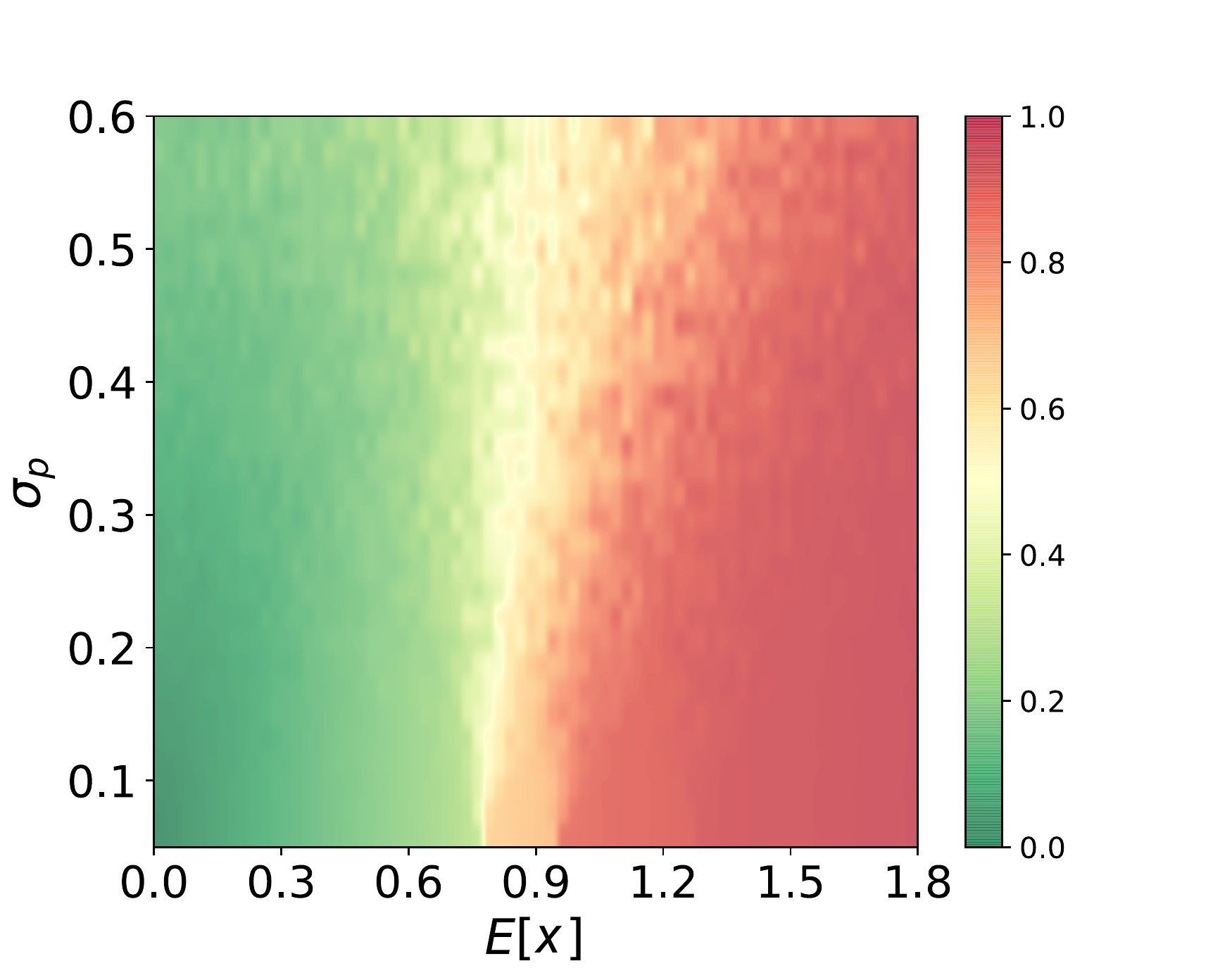}} 
	\caption{Phase diagrams of the mean density $\langle\rho_\infty\rangle$ in a lattice with $10\times10$ nodes under a homogeneous injection rate $ x$, with respect to different properties. The settings are the same with Figure \ref{fig:phaseDiag_congest}.}
	\label{fig:densityplot}
\end{figure}

\subsection{Impact of parameter $\kappa$}\label{sec:imapact_kappa}
\begin{figure}[!htp]
	\centering
	\subfloat[$\chi$]
	{\includegraphics[width=0.26\textheight]{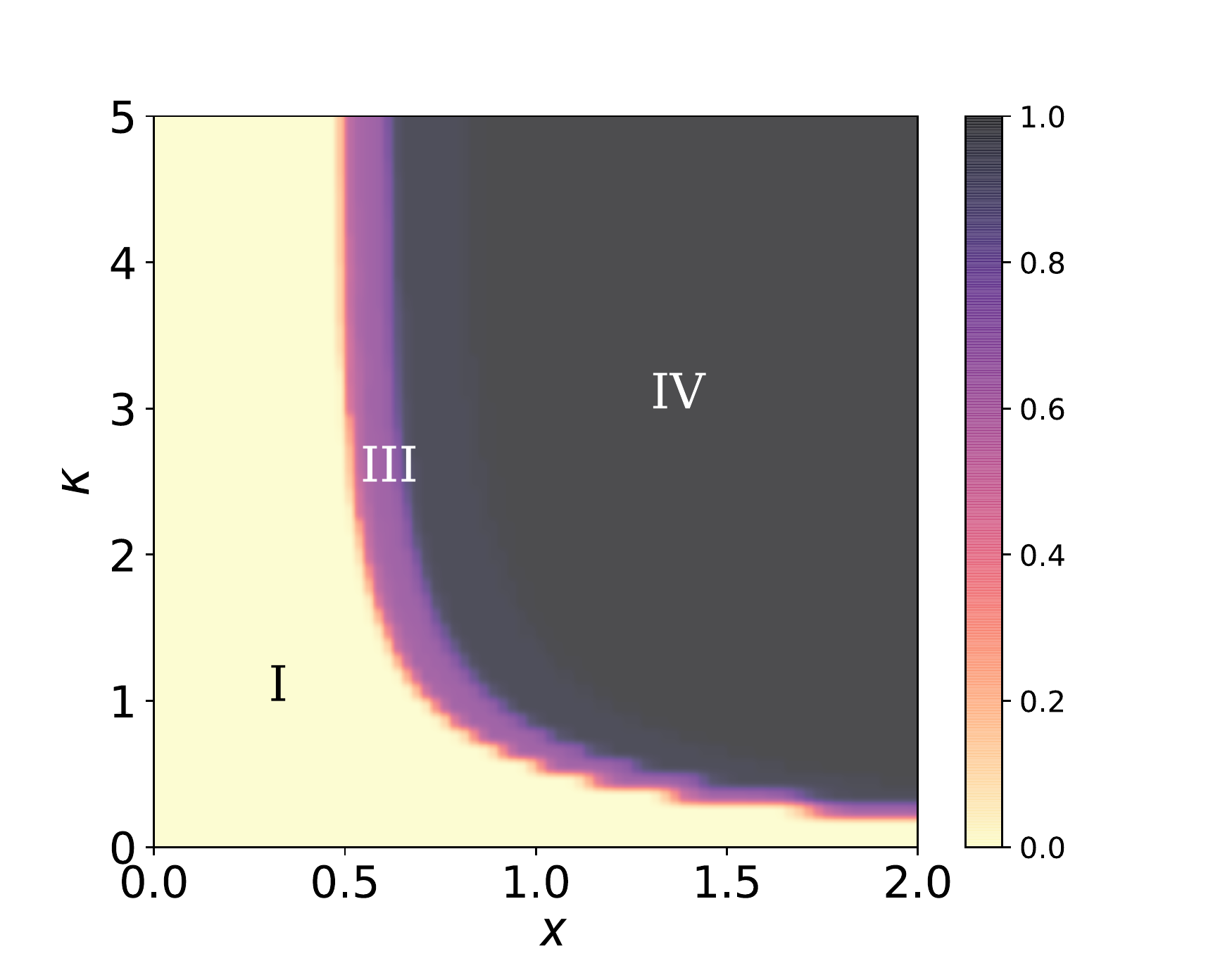}}
	\subfloat[$\langle\rho_\infty\rangle$]
	{\includegraphics[width=0.26\textheight]{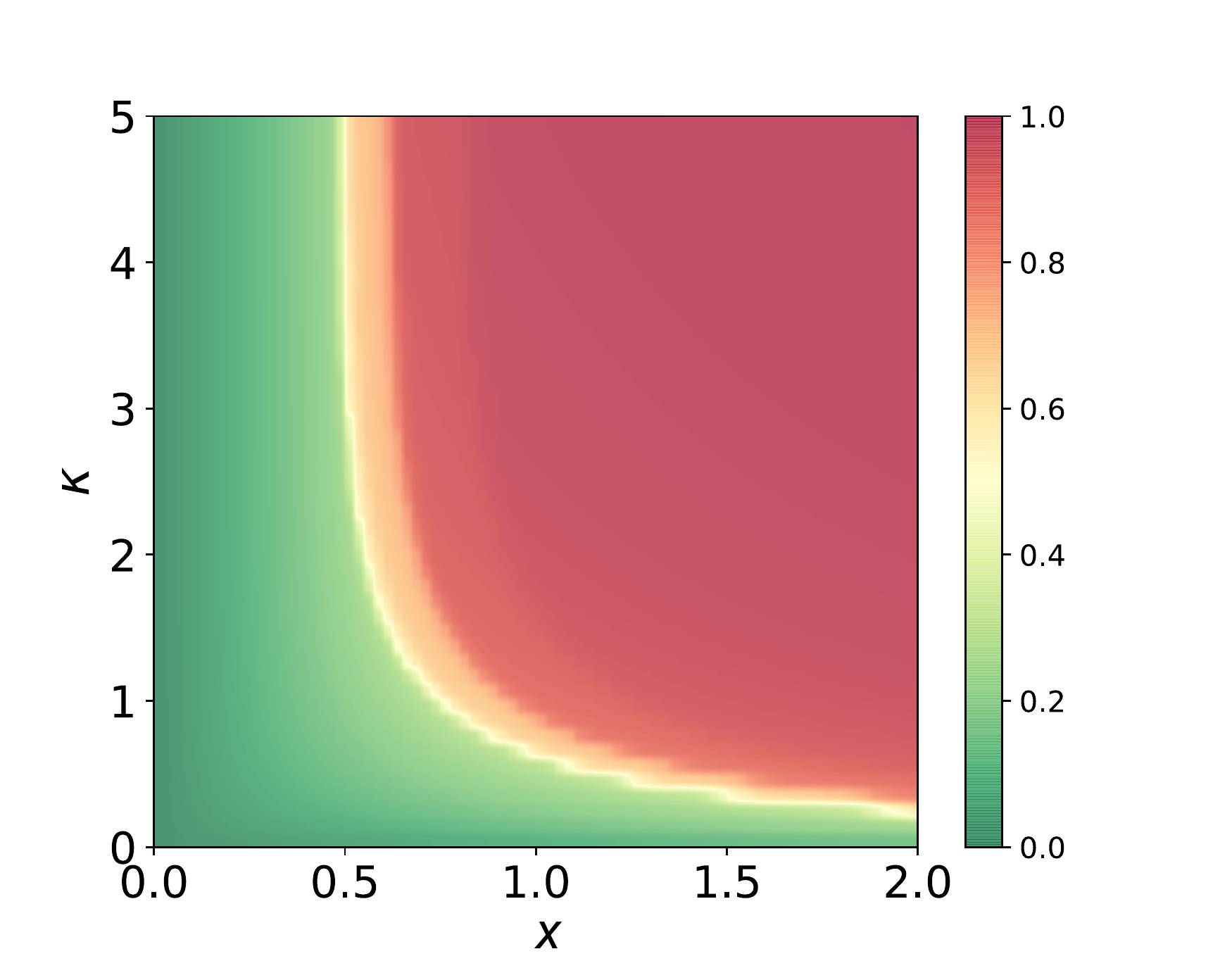}} 
	\caption{Phase diagrams of the mean density $\langle\rho_\infty\rangle$ in a lattice with $10\times10$ nodes under a homogeneous injection rate $ x$, with respect to the constant $\alpha$. The settings of numerical results are the same with Figure \ref{fig:phaseDiag_congest}.}
	\label{fig:phasediag_kappa}
\end{figure}

Figure \ref{fig:phasediag_kappa} shows the phase diagram with respect to the parameter $\kappa$ and the homogeneous injection rate $ x$. The different $\kappa>0$ does not alter the existence of the localization phase, but affect the congestion threshold $ x_c$. Specifically, the constant $\kappa<1$ increases the congestion threshold, while the constant $\kappa>1$ influences little on both the phases and the congestion threshold.

\end{appendices}

\section*{Acknowledgment}

\bibliographystyle{IEEEtran}
\bibliography{bibl}

\begin{thebibliography}{10}
\providecommand{\url}[1]{#1}
\csname url@samestyle\endcsname
\providecommand{\newblock}{\relax}
\providecommand{\bibinfo}[2]{#2}
\providecommand{\BIBentrySTDinterwordspacing}{\spaceskip=0pt\relax}
\providecommand{\BIBentryALTinterwordstretchfactor}{4}
\providecommand{\BIBentryALTinterwordspacing}{\spaceskip=\fontdimen2\font plus
\BIBentryALTinterwordstretchfactor\fontdimen3\font minus
  \fontdimen4\font\relax}
\providecommand{\BIBforeignlanguage}[2]{{%
\expandafter\ifx\csname l@#1\endcsname\relax
\typeout{** WARNING: IEEEtran.bst: No hyphenation pattern has been}%
\typeout{** loaded for the language `#1'. Using the pattern for}%
\typeout{** the default language instead.}%
\else
\language=\csname l@#1\endcsname
\fi
#2}}
\providecommand{\BIBdecl}{\relax}
\BIBdecl

\bibitem{hirokawa2005molecular}
N.~Hirokawa and R.~Takemura, ``Molecular motors and mechanisms of directional
  transport in neurons,'' \emph{Nature Reviews Neuroscience}, vol.~6, no.~3, p.
  201, 2005.

\bibitem{mieghem2006data}
P.~Van~Mieghem, \emph{Data communications networking}.\hskip 1em plus 0.5em
  minus 0.4em\relax Delft, 2011.

\bibitem{daganzo2011macroscopic}
C.~F. Daganzo, V.~V. Gayah, and E.~J. Gonzales, ``Macroscopic relations of
  urban traffic variables: {B}ifurcations, multivaluedness and instability,''
  \emph{Transportation Research Part B: Methodological}, vol.~45, no.~1, pp.
  278--288, 2011.

\bibitem{helbing2001self}
D.~Helbing, P.~Moln{\'a}r, I.~J. Farkas, and K.~Bolay, ``Self-organizing
  pedestrian movement,'' \emph{Environment and Planning B: Planning and
  Design}, vol.~28, no.~3, pp. 361--383, 2001.

\bibitem{anderson2018economy}
P.~W. Anderson, \emph{The economy as an evolving complex system}.\hskip 1em
  plus 0.5em minus 0.4em\relax CRC Press, 2018.

\bibitem{helbing2001traffic}
D.~Helbing, ``Traffic and related self-driven many-particle systems,''
  \emph{Reviews of Modern Physics}, vol.~73, no.~4, p. 1067, 2001.

\bibitem{chowdhury2000statistical}
D.~Chowdhury, L.~Santen, and A.~Schadschneider, ``Statistical physics of
  vehicular traffic and some related systems,'' \emph{Physics Reports}, vol.
  329, no. 4-6, pp. 199--329, 2000.

\bibitem{mendes2012traffic}
G.~Mendes, L.~Da~Silva, and H.~J. Herrmann, ``Traffic gridlock on complex
  networks,'' \emph{Physica A: Statistical Mechanics and its Applications},
  vol. 391, no. 1-2, pp. 362--370, 2012.

\bibitem{daganzo2007urban}
C.~F. Daganzo, ``Urban gridlock: {M}acroscopic modeling and mitigation
  approaches,'' \emph{Transportation Research Part B: Methodological}, vol.~41,
  no.~1, pp. 49--62, 2007.

\bibitem{lighthill1955kinematic}
M.~J. Lighthill and G.~B. Whitham, ``On kinematic waves {II}. {A} theory of
  traffic flow on long crowded roads,'' \emph{Proceedings of the Royal Society
  of London. Series A. Mathematical and Physical Sciences}, vol. 229, no. 1178,
  pp. 317--345, 1955.

\bibitem{newell1993simplified}
G.~F. Newell, ``A simplified theory of kinematic waves in highway traffic, part
  {I}: {G}eneral theory,'' \emph{Transportation Research Part B:
  Methodological}, vol.~27, no.~4, pp. 281--287, 1993.

\bibitem{daganzo1994cell}
C.~F. Daganzo, ``The cell transmission model: {A} dynamic representation of
  highway traffic consistent with the hydrodynamic theory,''
  \emph{Transportation Research Part B: Methodological}, vol.~28, no.~4, pp.
  269--287, 1994.

\bibitem{schreckenberg1995discrete}
M.~Schreckenberg, A.~Schadschneider, K.~Nagel, and N.~Ito, ``Discrete
  stochastic models for traffic flow,'' \emph{Physical Review E}, vol.~51,
  no.~4, p. 2939, 1995.

\bibitem{bressloff2013stochastic}
P.~C. Bressloff and J.~M. Newby, ``Stochastic models of intracellular
  transport,'' \emph{Reviews of Modern Physics}, vol.~85, no.~1, p. 135, 2013.

\bibitem{newman2018networks}
M.~Newman, \emph{Networks}.\hskip 1em plus 0.5em minus 0.4em\relax Oxford
  University Press, 2018.

\bibitem{wu2006congestion}
J.~Wu, Z.~Gao, H.~Sun, and H.~Huang, ``Congestion in different topologies of
  traffic networks,'' \emph{EPL (Europhysics Letters)}, vol.~74, no.~3, p. 560,
  2006.

\bibitem{zhao2005onset}
L.~Zhao, Y.-C. Lai, K.~Park, and N.~Ye, ``Onset of traffic congestion in
  complex networks,'' \emph{Physical Review E}, vol.~71, no.~2, p. 026125,
  2005.

\bibitem{li2015percolation}
D.~Li, B.~Fu, Y.~Wang, G.~Lu, Y.~Berezin, H.~E. Stanley, and S.~Havlin,
  ``Percolation transition in dynamical traffic network with evolving critical
  bottlenecks,'' \emph{Proceedings of the National Academy of Sciences}, vol.
  112, no.~3, pp. 669--672, 2015.

\bibitem{neri2011totally}
I.~Neri, N.~Kern, and A.~Parmeggiani, ``Totally asymmetric simple exclusion
  process on networks,'' \emph{Physical Review Letters}, vol. 107, no.~6, p.
  068702, 2011.

\bibitem{mones2014shock}
E.~Mones, N.~A. Ara{\'u}jo, T.~Vicsek, and H.~J. Herrmann, ``Shock waves on
  complex networks,'' \emph{Scientific Reports}, vol.~4, p. 4949, 2014.

\bibitem{sole2016congestion}
A.~Sol{\'e}-Ribalta, S.~G{\'o}mez, and A.~Arenas, ``Congestion induced by the
  structure of multiplex networks,'' \emph{Physical Review Letters}, vol. 116,
  no.~10, p. 108701, 2016.

\bibitem{zeng2019switch}
G.~Zeng, D.~Li, S.~Guo, L.~Gao, Z.~Gao, H.~E. Stanley, and S.~Havlin, ``Switch
  between critical percolation modes in city traffic dynamics,''
  \emph{Proceedings of the National Academy of Sciences}, vol. 116, no.~1, pp.
  23--28, 2019.

\bibitem{moretti2013griffiths}
P.~Moretti and M.~A. Mu{\~n}oz, ``Griffiths phases and the stretching of
  criticality in brain networks,'' \emph{Nature Communications}, vol.~4, p.
  2521, 2013.

\bibitem{parmeggiani2004totally}
A.~Parmeggiani, T.~Franosch, and E.~Frey, ``Totally asymmetric simple exclusion
  process with {L}angmuir kinetics,'' \emph{Physical Review E}, vol.~70, no.~4,
  p. 046101, 2004.

\bibitem{van2014performance}
P.~Van~Mieghem, \emph{Performance analysis of complex networks and
  systems}.\hskip 1em plus 0.5em minus 0.4em\relax Cambridge University Press,
  2014.

\bibitem{white2006viscous}
F.~M. White and I.~Corfield, \emph{Viscous fluid flow}.\hskip 1em plus 0.5em
  minus 0.4em\relax McGraw-Hill New York, 2006, vol.~3.

\bibitem{bonacich2001eigenvector}
P.~Bonacich and P.~Lloyd, ``Eigenvector-like measures of centrality for
  asymmetric relations,'' \emph{Social Networks}, vol.~23, no.~3, pp. 191--201,
  2001.

\bibitem{ghosh2012rethinking}
R.~Ghosh and K.~Lerman, ``Rethinking centrality: the role of dynamical
  processes in social network analysis,'' \emph{arXiv preprint
  arXiv:1209.4616}, 2012.

\bibitem{van2017pseudoinverse}
P.~Van~Mieghem, K.~Devriendt, and H.~Cetinay, ``Pseudoinverse of the
  {L}aplacian and best spreader node in a network,'' \emph{Physical Review E},
  vol.~96, no.~3, p. 032311, 2017.

\bibitem{gayah2011clockwise}
V.~V. Gayah and C.~F. Daganzo, ``Clockwise hysteresis loops in the macroscopic
  fundamental diagram: an effect of network instability,'' \emph{Transportation
  Research Part B: Methodological}, vol.~45, no.~4, pp. 643--655, 2011.

\bibitem{dafermos1969traffic}
S.~C. Dafermos and F.~T. Sparrow, ``The traffic assignment problem for a
  general network,'' \emph{Journal of Research of the National Bureau of
  Standards B}, vol.~73, no.~2, pp. 91--118, 1969.

\bibitem{geroliminis2008existence}
N.~Geroliminis and C.~F. Daganzo, ``Existence of urban-scale macroscopic
  fundamental diagrams: {S}ome experimental findings,'' \emph{Transportation
  Research Part B: Methodological}, vol.~42, no.~9, pp. 759--770, 2008.

\bibitem{roberts2007targeting}
P.~J. Roberts and C.~J. Der, ``Targeting the {R}af-{MEK}-{ERK}
  mitogen-activated protein kinase cascade for the treatment of cancer,''
  \emph{Oncogene}, vol.~26, no.~22, p. 3291, 2007.

\bibitem{portes2001information}
R.~Portes, H.~Rey, and Y.~Oh, ``Information and capital flows: {T}he
  determinants of transactions in financial assets,'' \emph{European Economic
  Review}, vol.~45, no. 4-6, pp. 783--796, 2001.

\bibitem{lokhov2017optimal}
A.~Y. Lokhov and D.~Saad, ``Optimal deployment of resources for maximizing
  impact in spreading processes,'' \emph{Proceedings of the National Academy of
  Sciences}, vol. 114, no.~39, pp. E8138--E8146, 2017.

\bibitem{he2018optimal}
Z.~He and P.~Van~Mieghem, ``Optimal induced spreading of {SIS} epidemics in
  networks,'' \emph{IEEE Transactions on Control of Network Systems}, 2018.

\bibitem{cabral2011role}
J.~Cabral, E.~Hugues, O.~Sporns, and G.~Deco, ``Role of local network
  oscillations in resting-state functional connectivity,'' \emph{Neuroimage},
  vol.~57, no.~1, pp. 130--139, 2011.

\bibitem{liu2018network}
Q.~Liu and P.~Van~Mieghem, ``Network localization is unalterable by infections
  in bursts,'' \emph{IEEE Transactions on Network Science and Engineering},
  2018.

\bibitem{de2006self}
L.~de~Arcangelis, C.~Perrone-Capano, and H.~J. Herrmann, ``Self-organized
  criticality model for brain plasticity,'' \emph{Physical Review Letters},
  vol.~96, no.~2, p. 028107, 2006.

\bibitem{gu2017optimal}
S.~Gu, R.~F. Betzel, M.~G. Mattar, M.~Cieslak, P.~R. Delio, S.~T. Grafton,
  F.~Pasqualetti, and D.~S. Bassett, ``Optimal trajectories of brain state
  transitions,'' \emph{NeuroImage}, vol. 148, pp. 305--317, 2017.

\bibitem{munoz2010griffiths}
M.~A. Mu{\~n}oz, R.~Juh{\'a}sz, C.~Castellano, and G.~{\'O}dor, ``Griffiths
  phases on complex networks,'' \emph{Physical Review Letters}, vol. 105,
  no.~12, p. 128701, 2010.

\bibitem{andrade2009fracturing}
J.~S. Andrade~Jr, E.~Oliveira, A.~Moreira, and H.~J. Herrmann, ``Fracturing the
  optimal paths,'' \emph{Physical Review Letters}, vol. 103, no.~22, p. 225503,
  2009.

\bibitem{ccolak2016understanding}
S.~{\c{C}}olak, A.~Lima, and M.~C. Gonz{\'a}lez, ``Understanding congested
  travel in urban areas,'' \emph{Nature Communications}, vol.~7, p. 10793,
  2016.

\bibitem{schmittmann1998driven}
B.~Schmittmann and R.~Zia, ``Driven diffusive systems. {A}n introduction and
  recent developments,'' \emph{Physics reports}, vol. 301, no. 1-3, pp. 45--64,
  1998.

\bibitem{sutera1993history}
S.~P. Sutera and R.~Skalak, ``The history of {P}oiseuille's law,'' \emph{Annual
  Review of Fluid Mechanics}, vol.~25, no.~1, pp. 1--20, 1993.

\bibitem{caffrey2013use}
M.~A. Caffrey and S.~P. Horn, ``The use of lithium heteropolytungstate in the
  heavy liquid separation of samples which are sparse in pollen,''
  \emph{Palynology}, vol.~37, no.~1, pp. 143--150, 2013.

\bibitem{tsuzuki2018effect}
S.~Tsuzuki, D.~Yanagisawa, and K.~Nishinari, ``Effect of self-deflection on a
  totally asymmetric simple exclusion process with functions of site
  assignments,'' \emph{Physical Review E}, vol.~97, no.~4, p. 042117, 2018.

\bibitem{popkov2003localization}
V.~Popkov, A.~R{\'a}kos, R.~D. Willmann, A.~B. Kolomeisky, and G.~M.
  Sch{\"u}tz, ``Localization of shocks in driven diffusive systems without
  particle number conservation,'' \emph{Physical Review E}, vol.~67, no.~6, p.
  066117, 2003.

\bibitem{kolomeisky1998phase}
A.~B. Kolomeisky, G.~M. Sch{\"u}tz, E.~B. Kolomeisky, and J.~P. Straley,
  ``Phase diagram of one-dimensional driven lattice gases with open
  boundaries,'' \emph{Journal of Physics A: Mathematical and General}, vol.~31,
  no.~33, p. 6911, 1998.

\bibitem{denisov2015totally}
D.~Denisov, D.~Miedema, B.~Nienhuis, and P.~Schall, ``Totally asymmetric simple
  exclusion process simulations of molecular motor transport on random networks
  with asymmetric exit rates,'' \emph{Physical Review E}, vol.~92, no.~5, p.
  052714, 2015.

\bibitem{raguin2013role}
A.~Raguin, A.~Parmeggiani, and N.~Kern, ``Role of network junctions for the
  totally asymmetric simple exclusion process,'' \emph{Physical Review E},
  vol.~88, no.~4, p. 042104, 2013.

\bibitem{pastor2015epidemic}
R.~Pastor-Satorras, C.~Castellano, P.~Van~Mieghem, and A.~Vespignani,
  ``Epidemic processes in complex networks,'' \emph{Reviews of Modern Physics},
  vol.~87, no.~3, p. 925, 2015.

\bibitem{saberi2019simple}
M.~Saberi, M.~Ashfaq, H.~Hamedmoghadam, S.~A. Hosseini, Z.~Gu, S.~Shafiei,
  D.~J. Nair, V.~Dixit, L.~Gardner, S.~T. Waller \emph{et~al.}, ``A simple
  contagion process describes spreading of traffic jams in urban networks,''
  \emph{arXiv preprint arXiv:1906.00585}, 2019.

\end{thebibliography}

\end{document}